\documentclass[aps,pre, superscriptaddress]{revtex4-1}
\usepackage[dvips]{graphicx}
\usepackage{amsmath}
\usepackage{amssymb}
\usepackage{natbib}
\usepackage{color}
\usepackage{epsfig}
\usepackage{subfigure}
\usepackage{setspace}

\begin{document}
\title{Noncrossing partition flow and random matrix models}
 \author{Mario Pernici}
\email{mario.pernici@mi.infn.it}
\affiliation{ Istituto Nazionale di Fisica Nucleare, Sezione di Milano,\\ 16 Via Celoria, 20133 Milano, Italy}

\begin{abstract}
We study a generating function flowing from the one enumerating a
set of partitions to the one enumerating the corresponding
set of noncrossing partitions; numerical simulations indicate that
its limit in the Adjacency random matrix model on bipartite Erd\"os-Renyi
graphs gives a good approximation of the spectral distribution 
for large average degrees.
This model and a Wishart-type random matrix model are described
using congruence classes on $k$-divisible partitions.

 We compute, in the $d\to \infty$ limit with $\frac{Z_a}{d}$ fixed,
 the spectral distribution of an Adjacency and of a Laplacian random block 
matrix model, on bipartite Erd\"os-Renyi
graphs and on bipartite biregular graphs with degrees $Z_1, Z_2$;
the former is the approximation previously mentioned;
the latter is a mean field approximation of the Hessian
of a random bipartite biregular elastic network;
it is characterized by an isostatic 
line and a transition line between the one- and the two-band regions.

 \vskip 0.3cm
Keywords: random matrix theory, random trees, block matrix, moments method,
noncrossing partitions
\end{abstract}
\maketitle

\section{Introduction}
The study of random matrix models has been started by Wigner and Dyson
to investigate the spectra of hamiltonians of complex systems
\cite{wigner, dyson}. One of the ways 
to examine them is the computation of the moments of the spectral density.

Recurrence equations for 
the moments of the Adjacency random matrix  on Erd\"os-Renyi (ER)
graphs \cite{Erd},
which can be used to study conductors with impurities,
have been found in \cite{bau,khor}; an effective medium (EM) approximation 
for this model has been given in \cite{SC}.

In \cite{parisi} a Laplacian random block matrix model on random
regular graphs has been introduced, in which the $d$-dimensional blocks
are projectors corresponding to random independent versors uniformly
distributed on a sphere in $d$ dimensions.
The Laplacian random block matrix represents a mean field approximation of the Hessian 
of a random elastic network in $d$ dimensions.
In $d=3$, near the isostatic point of the mechanical system \cite{maxwell},
it can be interpreted as an approximate description
of the low frequency modes in jammed soft spheres, 
which have been used to model the Boson peak in glasses \cite{ohern,wnw}.
At the isostatic point the spectral distribution $\rho(\lambda)$
diverges at $\lambda \to 0$; this behavior
is typical for a broad class of continuous constraint satisfaction problems
at the SAT-UNSAT transition \cite{fpuz}.
For $d\to \infty$ with $\frac{Z}{d}$ fixed, $Z$ being the coordination 
number, it has been shown in \cite{parisi} that 
the spectral distribution of the Laplacian random block matrix
is the Marchenko-Pastur distribution 
\cite{marcpas}. The $d=3$ model has been further studied in \cite{benetti}.

In \cite{CZ} a random block matrix model in $d$ dimensions on
ER graphs has been studied.  In $d=1$ it reduces to the
Adjacency and Laplacian random matrix models on ER
graphs.
As in \cite{parisi}, the Laplacian random block matrix
represents a mean field approximation of the Hessian of a random elastic 
network in $d$ dimensions.
From the analytic form of the first $5$ moments of the spectral
distribution, it has been conjectured in \cite{CZ} that,
for $d\to \infty$ with $t=\frac{Z}{d}$ fixed, the spectral
distribution is the EM approximation \cite{SC}
in the Adjacency case, the Marchenko-Pastur distribution
in the Laplacian case.
In \cite{PC} it has been observed that, in this limit,
a closed walk on a tree graph contributes to a moment
of the spectral distribution only if its sequence of steps forms a
noncrossing partition; as in the Adjacency random matrix model on 
ER graphs, the contribution
of a walk gets a factor $t$ for each distinct edge in the walk.
Using the noncrossing property and the decomposition of closed 
walks in primitive walks, algebraic equations for
the resolvent of the Adjacency and Laplacian random block
matrices have been obtained, giving respectively the EM and
the Marchenko-Pastur distribution, proving the conjectures in \cite{CZ}.
Another proof of these conjectures has been given recently in \cite{dkl}.

In \cite{bose} it has recently been shown that the set of closed walks 
contributing to the moments of the Adjacency random matrix model on 
ER graphs in the large-$N$ limit is isomorphic to a set
of $2$-divisible partitions, whose properties are investigated.

There is a relationship between free probability, random matrices
and noncrossing partitions.
The relationship between free probability and noncrossing partitions
is discussed in \cite{speicher}.
In \cite{voic91} it has been shown that Hermitian Gaussian random matrices 
are asymptotically free.
The Fuss-Narayana polynomials enumerate $NC^{(k)}(l)$, the set of
noncrossing $k$-divisible  partitions of size $kl$,
according to the number of blocks; they appear in the free Bessel laws 
\cite{bani,bani2} 
in the context of free probability theory \cite{voi}.
In \cite{alex} a random matrix model with the product of $k$ independent 
square Gaussian random matrices has been considered, in which the moments are
Fuss-Narayana polynomials.
In \cite{lecz,leczsa} they showed that the limit moments of 
a Wishart-type product $B(n)B(n)^*$ \cite{wishart},
where $B(n)$ is a product of $k$ independent rectangular Gaussian 
random matrices, are the generating functions enumerating a set of noncrossing
pair partitions according to the pairing types; they are
generalized Fuss-Narayana polynomials; see also \cite{leczsa2}.

In \cite{krew} the number of noncrossing partitions of $n$ elements with 
$h$ blocks has been obtained.  
In \cite{proding} it has been shown that there is a correspondence between
ordered trees and noncrossing partitions.
In \cite{edel} it has been computed the number of $k$-divisible noncrossing 
partitions of $n$ elements with $h$ blocks.

Given a set ${\cal A}$ of partitions irreducible under noncrossing
and the set ${\cal P}$ of all the partitions whose irreducible components 
are isomorphic to partitions in ${\cal A}$, 
in \cite{speich94,beiss,callan} it has been found an equation, 
called in \cite{callan} the noncrossing partition transform,
relating the generating function (g.f.) 
of the number of partitions in  ${\cal P}$
and the g.f. of the number of partitions in ${\cal A}$.
The noncrossing partition transform has the
same form as the free R-transform introduced in \cite{voi86}, of which a
combinatorial interpretation has been given in \cite{speich94}.

In \cite{stein, stein2} it has been shown that, in the case of the $2$-equal
partitions, the number of irreducible partitions (linked chords)
divided by the number of partitions tends to $e^{-1}$ in the limit
of infinite partition size.

We start studying a g.f. flowing from the one enumerating a
set of partitions to the one enumerating the corresponding
set of noncrossing partitions.
Enumerating a set of partitions according to the number of elements,
of blocks and of irreducible components under noncrossing,
one can define a g.f. with a parameter $s$, which flows from $s=1$,
in which it enumerates all the partitions of the set,
to $s\to \infty$, in which it enumerates the 
corresponding noncrossing partitions; let us call this flow
'noncrossing partition flow'.
Expanding this flowing g.f.
in powers of $\frac{1}{s}$ around $s=\infty$, the coefficient of
$\frac{1}{s^g}$ enumerates the partitions in which the difference between the
number of blocks and the number of irreducible components is $g$;
the leading term is the g.f. enumerating the
noncrossing partitions.

We say that a set of partition is nc-closed if it is the set generated by
the set of its irreducible partitions. These sets and the noncrossing
partition transform (NCPT) have been first studied in \cite{beiss}.
For a nc-closed set of partitions,
one can compute the $\frac{1}{s^g}$ term of the flowing g.f.
in terms of the leading order term and the g.f. enumerating partitions
till $g+1$ blocks; we give expressions for the terms with $g \le 3$.

For a set of partitions whose size is multiple of $k$, we define
as congruence class (CC) of a block of a partition its minimum element 
taken modulo $k$;
let us call CC type of a partition the list of numbers of blocks with given CC.
If a nc-closed set of partitions has all partitions of size multiple of $k$,
then one can enumerate them according to their
CC type; we write an extended NCPT for this set of partitions.

Define a $k$-gon tree as a directed connected graph formed by oriented 
$k$-gons joined at vertices and with no edge in common,
such that no other loops apart from the $k$-gons are present;
$k$-gon trees are a generalization of trees, the $k=2$ case.  
There is an isomorphism between the set of length-$kl$ closed walks covering 
rooted $k$-gon trees
and the set $P^{(k)}_k(l)$ of $k$-divisible partitions with $kl$ elements,
in which for any two blocks $B_i$ and $B_j$ of a partition,
the number of elements of $B_j$ between two elements of $B_i$ is multiple
of $k$. In the case $k=2$ this set of partitions has been studied in 
\cite{bose}, where it is shown its relation with the moments of the
Adjacency random matrix on the ER graphs.
From this isomorphism it follows that there is an isomorphism between the 
set  of closed walks covering rooted $k$-gon trees, with noncrossing 
sequence of steps, and the set $NC^{(k)}(l)$ of noncrossing 
$k$-divisible partitions with $kl$ elements.
$P^{(k)}_k$ and its subset $NC^{(k)}$ are both nc-closed;
there is a noncrossing partition flow between the generating functions
enumerating them.

We give recursive equations, generalizing the recursion relations obtained in
\cite{bau} for the case $k=2$, to enumerate $P^{(k)}_k$ according to the
number of elements and of blocks.

In the simpler case in which all the blocks of $P^{(k)}_k$ have size $2 k$,
these recursive equations can be solved for $P^{(k)}_k(l)$ with fairly large
$l$; a fit with Pade' approximants in the cases
$k=2,3$ and $4$ indicates that
the ratio $q_l$ between the number of irreducible partitions of size
$kl$ and the number of partitions of size $kl$ tends to $e^{-k}$
in the infinite size limit;
in the case $k=1$, describing partitions with all blocks of size $2$ (chords),
this has been proven in \cite{stein,stein2}.

The recursive equations for $P^{(k)}_k$ can be solved at low orders 
expanding in the number $h$ of blocks; we solve them till $h=2$ for
any $k$; for $k=2$ we solve them till $h=10$.

We extend the above recursive equations to compute the number of partitions in 
$P^{(k)}_k(l)$ with given CC types;
the powers of parameters $t_1,\cdots,t_k$  specify the number of blocks with
the respective CC $1,\cdots,k$.

Using the $k$-parameter extension of the NCPT for 
a nc-closed set of partitions of size multiple of $k$,
we give, in the case of $NC^{(k)}(l)$, the explicit form of the g.f.
enumerating $k$-divisible noncrossing partitions of $kl$ elements
according to the CC types.
It is a generalization of their enumeration according to the number of
blocks, studied in \cite{bani}, reducing to it by replacing its parameters
$t_1,\cdots,t_k$ with $t$; in the latter case, we give a formula
for the $\frac{1}{s}$ term in the noncrossing partition flow from the g.f.
enumerating $P^{(k)}_k$  to the g.f. enumerating $NC^{(k)}$.

The g.f. enumerating $NC^{(k)}(l)$ according to the CC types
is simply related to the g.f. of the $l$-th moment
of the Wishart-type random matrix
$X_1\cdots X_k (X_1\cdots X_k)^*$, in the limit $n\to \infty$, where
$X_a$ is a Gaussian random $N_{a-1}(n) \times N_a(n)$ matrix 
\cite{lecz,leczsa,leczsa2}.
The set of pair partitions associated to the contractions contributing to 
the $l$-th moment of this Wishart-type random matrix model
is isomorphic to $NC^{(k)}(l)$. We construct a $1$-to-$1$ correspondence
between the partitions in these two sets, mapping the CC type and the rank of
a $k$-divisible partition to the pairing type of the corresponding pair partition.

The $2l$-th moment of the Adjacency random matrix model on ER
graphs with average degree $Z$ is the g.f. enumerating $P^{(2)}_2(l)$
according to the number of blocks.
In the $\frac{1}{s}$-expansion of the flow between the g.f. enumerating
$P^{(2)}_2$ and the one enumerating
$NC^{(2)}$, the leading order gives the EM approximation;
we compute the first four terms of the $\frac{1}{s}$-expansion
of the g.f. of the moments.

The $2l$-th moment of the Adjacency random matrix model on bipartite
ER graphs with average degrees $Z_1,Z_2$ is the
g.f. enumerating $P^{(2)}_2(l)$ according to the CC types.
We compute the leading order and the first order term in the 
$\frac{1}{s}$-expansion of the g.f. of the moments;
the leading order approximation is the g.f. enumerating $NC^{(2)}$ according 
to the number of elements and the CC types.
Numerical simulations indicate that, for large degrees, the leading order 
gives a good approximation of the spectral distribution.

We study a generalization of the Adjacency random block matrix model
introduced in \cite{parisi,CZ}, to the case of 
random bipartite biregular (BB) graphs with degrees $Z_1$ and $Z_2$,
and of bipartite ER graphs with average degrees $Z_1$ and $Z_2$.
For $d=1$, in the case of bipartite ER graphs, this is
the Adjacency random matrix model on bipartite ER graphs;
in the case of random BB graphs, it is the Adjacency random matrix model on
BB graphs, whose spectral distribution has been computed in \cite{gm}, see
Appendix C.A .
In the limit $d\to \infty$ with
$t_1 = \frac{Z_1}{d}$ and $t_2 = \frac{Z_2}{d}$ fixed, 
the g.f. of the moments of this Adjacency random block matrix
enumerates $NC^{(2)}$ according to the number of elements
and CC types; the same limit holds for the
Adjacency random block matrix model on random BB graphs 
with degrees $Z_1$ and $Z_2$.
It is the same as the leading order in the $\frac{1}{s}$-expansion of the
g.f. of the moments of the Adjacency random matrix 
on bipartite ER graphs which we discussed above.

We study similarly the Laplacian random block matrix model,
a mean field approximation of the Hessian of a random 
BB $d$-dimensional elastic network.
In the limit $d \to \infty$ with $t_1=\frac{Z_1}{d}$ and $t_2=\frac{Z_2}{d}$ 
fixed, only walks with
noncrossing sequence of moves contribute to the moments of the Laplacian
block matrix, giving the same contribution in the case of bipartite
ER graphs
and in the case of random BB graphs, so the moments are
the same in these two models.
We generalize the sequence of moves on trees,
giving a closed walk contributing
to the moments of the Laplacian random block matrix on ER graphs,
to a sequence of moves on $k$-gon trees.  
There is a $1$-to-$1$ correspondence between these sequences of moves
on $k$-gon trees and a set of bicolored partitions, called ${\cal C}^w_{(k)}$,
with a $k$-divisibility
restriction on the elements of the first color; ${\cal C}^w_{(k)}$ is nc-closed.
Let us call ${\cal C}_{(k)}$ the corresponding set of irreducible
partitions.
An algebraic equation for the g.f. enumerating the partitions
in ${\cal C}_{(k)}$ according to the number of blocks can be obtained
using the NCPT;
in the case $k=2$ this is the g.f. of
the moments of the Laplacian random block matrix model on random regular 
graphs, or on ER graphs, in the limit $d \to \infty$ with $\frac{Z}{d}$ 
fixed; it follows the Marchenko-Pastur distribution,
as found in respectively in \cite{parisi} and \cite{PC};
it is the leading order in the $\frac{1}{s}$-expansion in the
noncrossing partition flow from the g.f. enumerating ${\cal C}^w_{(2)}$ to 
the one enumerating ${\cal C}_{(2)}$.

One can define CC types on these bicolored partitions;
we could not find an extension of the NCPT for ${\cal C}_{(k)}$ with
CC types, so we cannot compute the g.f. of ${\cal C}_{(k)}$
with CC types from the g.f. of the corresponding irreducible partitions.
In the case $k=2$, corresponding to the Laplacian random block matrix model
on random BB graphs or bipartite ER graphs,
in the limit $d \to \infty$ with $t_a=\frac{Z_a}{d}$ fixed,
we used instead the decomposition of a closed walk in primitive walks, 
as in \cite{PC},
to write a set of algebraic equations for the g.f. enumerating them;
we solve these equations, obtaining in this limit
the resolvent of the Laplacian random block matrix on random BB graphs
or on bipartite ER graphs;
the spectral distribution is characterized
by two curves in the $t_1, t_2$ plane: the isostatic line
of the mechanical system, and a line separating
the region in which the spectrum has one band from the region in
which it has two bands.
The isostatic line is an hyperbola,
reducing for $t_1=t_2$ to the isostatic point of the regular graph
case; on an isostatic point the spectral distribution diverges as
$\rho(\lambda) \sim \lambda^{-\frac{1}{2}}$, as in the 
case of the random regular graph case, with Marchenko-Pastur distribution.

Numerical simulations with Laplacian random block matrices on
BB graphs indicate that for $d \ge 2$ the spectral distribution 
diverges at isostatic points for $\lambda \to 0$ and that
the region in the $t_1-t_2$ plane
in which there is a single spectral band is approximately delimited
by the transition line found in the $d\to \infty$ case.
The spectral distribution of the Laplacian random matrix
on random BB graphs is reported in Appendix C.B; for $Z_1 \neq Z_2$
there are always two bands.

Numerical simulations in the Laplacian random matrix model on bipartite 
ER graphs indicate that, for average degrees $t_1$ and $t_2$ 
within the transition line of the $d\to \infty$ model, there is a single band, 
while outside of
that transition line, the band tend to split in two bands, forming a quasi-gap 
between them far away from it. 

Unlike in the case of the Adjacency random matrix models on bipartite
ER graphs, we find that for large average degrees $t_1$ and $t_2$
the spectral distribution of the Laplacian random block matrix model,
in the limit $d\to \infty$ with $t_a = \frac{Z_a}{d}$ fixed,
does not approximate well the spectral distribution of
the Laplacian random matrix model on bipartite ER graphs, especially
when $t_2$ is close to the transition line or beyond it, where
the latter has wide oscillations.

In Section II we review the noncrossing partition transform;
for a set of partitions we define a g.f. flowing from the g.f. 
enumerating all partitions and the one enumerating the corresponding 
noncrossing partitions; for nc-closed sets of partitions we compute the first
four terms in its $\frac{1}{s}$-expansion.
We give a multi-parameter extension of the noncrossing partition transform
in the case of partitions with size multiple of $k$.

In Section III we prove the isomorphism of $P^{(k)}_k(l)$ with the set
of closed walks of length $kl$ covering rooted $k$-gon trees and
some other properties of $P^{(k)}_k(l)$.

In Section IV we study the enumeration of $P^{(k)}_k(l)$ and some related 
sets.

In Section V we enumerate partitions in $NC^{(k)}(l)$ with CC types and
give  a $1$-to-$1$ correspondence between these partitions
and the pair partitions contributing
to the $l$-moment of a Wishart-type random matrix model
with the product of $k$ rectangular Gaussian random matrices.

In Section VI we study the Adjacency random block matrix model on 
BB graphs and on bipartite ER graphs, 
finding a cubic equation for its resolvent in the limit
$d\to \infty$ with $t_a=\frac{Z_a}{d}$ fixed;
 we study the spectral distribution for this model.
 We give the first four terms in the
 $\frac{1}{s}$-expansion of the g.f. of the moments
 of the random matrix model on  ER graphs and compare the
 leading order approximation with numerical simulations.

In Section VII we consider the Laplacian random block matrix model on 
BB graphs, finding a cubic equation for its resolvent 
in the same limit;
we study the set of bicolored partitions corresponding to the walks
contributing to the moments of the Laplacian random block matrix
and give the spectral distribution for this model, exhibiting an
isostatic line and one- and two-banded regions in the $t_1$,$t_2$ plane.
We compare this spectral distribution with some numerical simulation.

In Appendix A we illustrate at low orders some set of partitions considered
in the paper.

In Appendix B we give the first few moments for the Adjacency and the
Laplacian random block matrices on BB graphs, for generic $d$.

In Appendix C we review the spectral distribution of the Adjacency random
matrix model on random BB graphs and we report the spectral
distribution of the Laplacian random matrix model on these graphs.

\section{Noncrossing partition transform and noncrossing partition flow}
Let us review some definitions concerning partitions; 
see e.g. \cite{sim, arm, klazar}.

A partition of a set of $n$ elements is a family
of non-empty, pairwise disjoined sets, called blocks, whose union is
this set. Unless specified, we refer to a partition as a partition of 
$[n] = \{1,\cdots, n\}$, where $n$ is the size of the partition;
let $P(n)$ be the set of partitions of $[n]$.
The elements of the blocks are listed in increased order,
and the blocks are listed ordered according to their lowest element.

The sequential form $(s_1,\cdots,s_n)$ of a partition is given assigning to 
the $i$-th block $({b_1^{(i)}},\cdots, {b_{m_i}}^{(i)})$
the number $i$ in the positions which are elements of the block,
$s_{b_j^{(i)}} = i$, for $j=1,\cdots, m_i$.
For instance the partition $(1,2,5,6,7)(3,4)$ has sequential form
$(1,1,2,2,1,1,1)$.

Two distinct blocks cross each other if $a,c$ belong to the first block, 
$b,d$ belong to the second block, and $a < b < c < d$.
A partition is noncrossing if none of its blocks are crossing; equivalently
its sequential form has no subsequence
$a,b,a,b$, with $a \neq b$. $NC(n)$ is the set of noncrossing
partitions of $[n]$; the number of these partitions with
$b$ blocks is the Narayana number \cite{krew}.
\begin{equation}
N(n, b) = \frac{1}{n}\binom{n}{b} \binom{n}{b-1}.
\label{nara}
\end{equation}
$NC^{(k)}(l)$ is the set of $k$-divisible noncrossing
partitions, i.e. the noncrossing partitions
of $[kl]$ in which all its blocks have size multiple of $k$;
the number of noncrossing $k$-divisible partitions of $[kl]$ 
with given number of blocks has been computed in \cite{edel}.

$NC_k(l)$ is the set of noncrossing partitions of $[kl]$ with all blocks of size $k$.

\subsection{Noncrossing partition transform}

A partition $p$ is called {\it irreducible} under noncrossing if one cannot
separate its blocks into two non-empty subsets $S_1$ and $S_2$,
such that none of the blocks of $S_1$ crosses a block of $S_2$.
A {\it component} of a partition $p$ is a subset of blocks of $p$,
none of which crosses blocks not in this subset; an irreducible component 
of $p$ is an irreducible partition on the set of its elements.
For instance the partition $(1, 2, 7)(3, 4, 8)(5, 6)$ has irreducible components
$(1,2,7)(3,4,8) \equiv (1,2,5)(3,4,6)$ and $(5,6) \equiv (1,2)$,
where the equivalence is obtained reducing a partition to a partition
of $[n]$, where $n$ is its size,
preserving the ordering of the elements and the block structure.

Given a set ${\cal A}$ of irreducible partitions, define $\cal P_{\cal A}$
as the set of all the partitions whose irreducible components
are equivalent to partitions in ${\cal A}$. Let us say that $\cal P_{\cal A}$
is the set generated by ${\cal A}$ and that a partition in $\cal P_{\cal A}$
is generated by a subset of partitions of ${\cal A}$.
For instance, let ${\cal A}$ consist of the single partition $(1,2)$;
then ${\cal P}_{\cal A} = NC_2$, the union of the sets $NC_2(l)$, $l \ge 1$.
Let us say that a set of partitions $\cal P$ is {\it nc-closed} if it is the
set generated by the set of its irreducible partitions.

In \cite{beiss} it has been studied the relation between a set ${\cal A}$ of
irreducible partitions and the nc-closed set $\cal P_{\cal A}$.
An equation, called in \cite{callan}
noncrossing partition transform, relates the g.f. 
enumerating $\cal P_{\cal A}$
and the g.f. enumerating ${\cal A}$.
The case in which the partitions in ${\cal A}$ are noncrossing
had been studied before in \cite{speich94}.

Let us review the derivation of the noncrossing partition transform in 
\cite{callan}, adding to it a parameter to obtain the number of irreducible
components in a partition, given in \cite{beiss}.

Let ${\cal P}$ be a nc-closed set of partitions, generated by
the set of irreducible partitions ${\cal A}$.
Let $U(n)$ be the subset of ${\cal P}$ with size-$n$ partitions; 
it is a subset of the set of partitions of $[n]$.
Given a partition $p$ of $U(n)$, let $c$ be the irreducible component
of $p$ containing $n$, with size $h$ and elements  $c_1,\cdots, c_h=n$,
listed in increasing order.
On one of the $h$ intervals $[c_{i-1}+1, c_i-1]$ (letting $c_0=0$)
there is a partition $p_i$, formed by blocks of $p$
(some of these partitions can be trivial); if $p_i$ had elements belonging
to a block with elements also in another interval, then $p_i$ would cross $c$,
so $c$ would not be an irreducible component of $p$; therefore a nontrivial
partition $p_i$ is a component of the partition $p$.

Let us consider an example with $2$-equal partitions (chords). The partition
$p = (1,2)(3,9)(4,10)(5,6)(7,8)$ has the irreducible component $c=(3,9)(4,10)$
with the highest element $n=10$.
Between $c_0=0$ and the elements $3,4,9,10$ of $c$ there are the $4$ intervals:
$I_1=[1,2]$; $I_2$ is empty; $I_3=[5,8]$, $I_4$ is empty.
$p_1=(1,2)$ and $p_3=(5,6)(7,8)$ are components of $p$ defined
on these intervals; $p_3$ is isomorphism to $(1,2)(3,4)$.

Define the g.f. of the number $a_h$ of partitions in ${\cal A}$
with $h$ elements
\begin{equation}
a(x) = \sum_{h\ge 1} a_h x^h
\label{Aeq}
\end{equation}
Let $U(n,h)$ be the subset of $U(n)$
in which $n$ occurs in an irreducible component of size $h$.
Let $u(n,h,s)$ be the g.f. of the number $[s^c]u(n,h,s)$
of elements of $U(n, h)$ with $c$ irreducible components
(here and in the following $[x^n] f(x)$ denotes the coefficient of $x^n$ in $f(x)$);
the parameter $s$ has been introduced to count the number of irreducible
components of a partition.

Define 
\begin{equation}
u(n, s) = \sum_{h\ge 1} u(n,h,s)
\end{equation}
for $ n \ge 0$ and $u(0,s) = 1$.

The g.f. enumerating partitions in ${\cal P}$ 
according to the number of elements and of irreducible components is
\begin{equation}
f(x,s) = \sum_{n \ge 0} u(n, s) x^n
\label{fxs}
\end{equation}
$f(x,1)$ is the g.f. of the number of partitions
with given number of elements, $[x^n]f(x, 1) = |U(n)|$.

As discussed above,
for a partition $p$ in $U(n,h)$ the elements $c_1,\cdots, c_{h-1},n$ of the 
irreducible component $c$, containing the element $n$,
separate the $n-h$ elements of $[n]$, which are not in $c$, in $h$ intervals 
of $[n]$, that is $[1, c_1-1]$, $[c_1+1, c_2-1]$, $\cdots$,
$[c_{h-1}+1,n - 1]$ respectively of sizes $m_1,\cdots,m_h$,
with $m_1 + \cdots + m_h = n-h$ and $m_i \ge 0$; on each non-empty 
interval a component $p_i$ of $p$ is defined,  with $m_i$ elements,
so $p_i$ is equivalent to a partition in $U(m_i)$; therefore
\begin{equation}
u(n,h,s) = s a_h \sum_{m_1+\cdots +m_h = n-h} u(m_1,s)\cdots u(m_h,s)
\end{equation}
where the factor $s$ counts the irreducible partition $c$.
Summing over $h$ one gets for $n \ge 1$
\begin{eqnarray}
[x^n]f(x,s) &=&  s \sum_{h \ge 1} a_h \sum_{m_1+\cdots +m_h = n-h} u(m_1,s)\cdots u(m_h,s) = \nonumber \\
    &&   s[x^{n-h}] \sum_{h\ge 1} a_h f(x,s)^h =
s[x^n] \sum_{h\ge 1} a_h (x f(x,s))^h
\label{nctran0}
\end{eqnarray}
from which one gets the noncrossing partition transform (NCPT)
\begin{equation}
f(x,s) = 1 + s a(x f(x,s))
\label{nctran}
\end{equation}
The case $s=1$ of Eq. (\ref{nctran}) has been named in \cite{callan}
noncrossing partition transform; Eq. (\ref{nctran}) is a one-parameter
extension of it.

Using the Lagrange inversion theorem one gets
\begin{equation}
[x^n]f(x, s) = \frac{1}{n+1}[y^n](1 + s a(y))^{n+1} \nonumber \\
\label{nctran1aa}
\end{equation}
and
\begin{equation}
[s^h x^n]f(x, s) = \frac{1}{n+1}\binom{n+1}{h} [y^n] a(y)^h
= \frac{1}{h}\binom{n}{h-1} [y^n] a(y)^h
\label{nctran1}
\end{equation}
This equation gives the number of partitions with $n$ elements
and $h$ irreducible components, obtained in \cite{beiss}.

One can compute $[x^n]a(x)$ from $f(x, 1)$; $[x]a(x) = [x]f(x, 1)$;
for $n > 1$ \cite{beiss}
using the inverse NCPT
\begin{equation}
[x^n] a(x) = -\frac{1}{n-1} [y^n] f(y,1)^{1-n}
\label{nctraninv}
\end{equation}

\subsection{Noncrossing partition flow}
Let ${\cal P}$ be a set of partitions; let $f(x, s, t)$ be the
g.f. enumerating ${\cal P}$ according to the number of elements,
of blocks and of irreducible components; the parameters $s$ and
$t$ are used to count respectively the number of irreducible components
and the number of blocks of a partition:
$[x^n t^b s^c]f(x,s, t)$ is the number of partitions with $n$ elements,
$b$ blocks and $c$ irreducible components.
Let $b(p)$ be the number of blocks and
$c(p)$ be the number of irreducible components of a partition $p$;
$c(p) - b(p) = \sum_{i=1}^{c(p)} (1 - b(p_i)) \le 0$, where $p_i$ is an 
irreducible component of $p$. The condition $c(p)-b(p) = 0$ implies 
$b(p_i) = 1$ for each irreducible component $p_i$ of $p$; it
is equivalent to the fact that $p$ is noncrossing. One has

\begin{equation}
f(x,s,t) = \sum_{p \in {\cal P}} x^{|p|}t^{b(p)} s^{c(p)} = 
1 + \sum_{b,c \ge 1} t^b s^c [t^b s^c] f(x, s, t)
\label{fbc}
\end{equation}
where $|p|$ is the size of $p$, so that
\begin{equation}
f(x,s,\frac{t}{s}) = 1 + \sum_{n \ge 1} x^n \sum_{b,c = 1}^n
t^b s^{c-b} [x^n t^b s^c] f(x, s, t) =
1 + \sum_{n \ge 1} x^n \sum_{b = 1}^n t^b \alpha_{n,b}(s)
\label{fsexp}
\end{equation}

Since $c - b \le 0$, $\lim_{s\to \infty} f(x,s,\frac{t}{s})$ 
exists and contains only the terms with $c = b$, so there
is a noncrossing partition flow from $s=1$ to
$s \to \infty$, in the sense that $f(x,s,\frac{t}{s})$
interpolates between the limit $s=1$, in which it
enumerates the partitions according to the number of blocks,
and the limit $s \to \infty$, in which it enumerates the corresponding
noncrossing partitions; the
$\frac{1}{s^g}$-th term of its $\frac{1}{s}$-expansion
enumerates the partitions with $b(p)-c(p) = g$.

There is an infinite number of sets of partitions flowing to 
$\lim_{s\to \infty} f(x,s,\frac{t}{s})$; the partitions which are not 
noncrossing are irrelevant in the limit $s \to \infty$.

While the notion of noncrossing partition flow can be defined for any set
of partitions, it is of little use unless one can compute exactly at least
the leading term in the $\frac{1}{s}$-expansion.

Let ${\cal P}$ be a nc-closed set of partitions.
Since the number of blocks in a
component partition $p_i$ with $m_i$ elements is equal to the number
of blocks in the equivalent partition in $U(m_i)$, Eq. (\ref{nctran})
still holds, with $f$ and $a$ depending also on the parameter $t$
\begin{equation}
f(x,s,t) = 1 + s a(x f(x,s,t), t)
\label{nctrant}
\end{equation}
$[x^n t^b] a(x, t)$ is the number of irreducible partitions with $n$ elements
with $b$ blocks.
The inverse NCPT becomes
\begin{equation}
[x^n] a(x, t) = -\frac{1}{n-1} [y^n] f(y,1, t)^{1-n}
\label{nctraninvt}
\end{equation}

Define $\sigma = \frac{1}{s}$ and 
$\psi(x, \sigma, t) = f(x, s,\frac{t}{s})$; Eq. (\ref{nctrant}) becomes
\begin{equation}
\sigma \psi(x, \sigma, t) - \sigma = a(x \psi(x, \sigma, t), t\sigma)
\label{ntransig}
\end{equation}
Expanding this equation around $\sigma=0$ one can compute
$[\frac{1}{s^g}] f(x, s, \frac{t}{s}) = 
[\sigma^g]\psi(x, \sigma, t) \equiv \psi_g(x, t)$.

Define 
$a_{(j)}(x) = [t^j] a(x, t) = \frac{1}{j!}\frac{\partial^j a(x, t)}{\partial t^j}|_{t=0}$
and
$f_{(j)}(x) = [t^j] f(x, 1, t) = \frac{1}{j!}\frac{\partial^j f(x, 1, t)}{\partial t^j}|_{t=0}$ for $j \ge 1$.
From Eq. (\ref{nctrant}) with $s=1$ one gets at the first four orders
\begin{eqnarray}
&&a_{(1)}(x) = f_{(1)}(x); \qquad 
a_{(2)}(x) = f_{(2)}(x) - x f_{(1)}(x)\frac{d f_{(1)}(x)}{d x} \nonumber \\
&& a_{(3)} = f_{(3)} - x\frac{d(f_{(1)}f_{(2)})}{d x} + 
x f_{(1)}^2 \frac{d f_{(1)}}{d x} +
x^2 f_{(1)} \big(\frac{d f_{(1)}}{d x}\big)^2 +
\frac{x^2}{2} f_{(1)}^2 \frac{d^2 f_{(1)}}{d x^2} \label{a12} \\
&& a_{(4)} = f_{(4)} - x f_{(3)}\frac{d a_{(1)}}{d x} -
x^2 f_{(1)} f_{(2)}\frac{d^2 a_{(1)}}{d x^2} -
\frac{x^3}{6}f_{(1)}^3 \frac{d^3 a_{(1)}}{d x^3} - 
x f_{(2)} \frac{d a_{(2)}}{d x} -
\frac{x^2}{2}f_{(1)}^2 \frac{d^2 a_{(2)}}{d x^2} - 
x f_{(1)}\frac{d a_{(3)}}{d x}
\nonumber
\end{eqnarray}
Expanding Eq. (\ref{ntransig}) around $\sigma=0$ one gets at order $0$
\begin{equation}
\lim_{s\to \infty} f(x, s,\frac{t}{s}) = \psi(x, 0, t) \equiv
\psi_0(x,t) = 1 + t a_{(1)}(x \psi_0(x, t))
\label{ntransig0}
\end{equation}
which is the NCPT for the g.f. enumerating the noncrossing
partitions according to the number of blocks.

Observe that $a_{(1)}(x)$ is universal, in the sense that it is the same
for nc-closed sets having the same noncrossing partitions;
$a_{(j)}(x)$ for $j > 1$ enumerates the irreducible partitions
with $j > 1$ blocks, which are irrelevant in the limit $s \to \infty$.

At first order one gets
\begin{equation}
[\frac{1}{s}] f(x, s, \frac{t}{s}) = 
\psi_1(x, t) = \frac{t^2 a_{(2)}(y)}
{1-t x \frac{d a_{(1)}(y)}{d y}} =
t^2 \frac{f_{(2)}(y) - y f_{(1)}(y) \frac{d f_{(1)}(y)}{d y} }
{1-t x \frac{d f_{(1)}(y)}{d y}}
\label{ntransig1}
\end{equation}
where we set
\begin{equation}
y = x \psi_0(x, t)
\label{ypsi}
\end{equation}
At the next two orders we get
\begin{eqnarray}
&&\psi_2(x, t) = \frac{t^3 a_{(3)}(y) + 
t^2 x \psi_1(x, t)\frac{d a_{(2)}(y)}{d y} +
t \frac{x^2}{2}\psi_1(x, t)^2 \frac{d^2 a_{(1)}(y)}{d y^2 }}
{1-t x \frac{d a_{(1)}(y)}{d y}} \label{ntransig2} \\
&&\psi_3(x, t) =  \big[t^4 a_{(4)}(y) + 
t x^2 \psi_1(x, t) \psi_2(x, t) \frac{d^2 a_{(1)}(y)}{d y^2 } +
\frac{t}{6} x^3 \psi_1(x, t)^3 \frac{d^3 a_{(1)}(y)}{d y^3} +
t^2 x \psi_2(x, t) \frac{d a_{(2)}(y)}{d y} + \nonumber \\
&&\frac{t^2}{2} x^2 \psi_1(x, t)^2 \frac{d^2 a_{(2)}(y)}{d y^2 } +
t^3 x \psi_1(x, t) \frac{d a_{(3)}(y)}{d y}\big]/(1-t x \frac{d a_{(1)}(y)}{d y})
\label{ntransig3}
\end{eqnarray}
In the expressions for $\psi_2$ and $\psi_3$, $a_{(j)}(y)$ are given 
in Eq. (\ref{a12}). In general
$\psi_g(x, t)$ can be expressed in terms of
$y = x \psi_0(x, t)$ and $[t^r] f(x,1,t)$ for $r=1,\cdots,g+1$.

For instance $P = \{P(n), \, n \ge 1\}$ is nc-closed; for $s=1$, the 
g.f. $f(x,1,t)$, 
enumerates the partitions according to the number of blocks and it
has as coefficients the Stirling numbers of the second kind,
$\alpha_{n,b}(1) = S(n, b)$; 
$\psi_0(x,t) = \lim_{s \to \infty}f(x,s,\frac{t}{s})$ enumerates
the set $NC$ of the noncrossing partitions according to the number of blocks.
The g.f. enumerating the noncrossing irreducible partitions 
is
$a_{(1)} = \frac{x}{1-x}$ and Eq. (\ref{ntransig0}) gives the equation for
the g.f. enumerating the noncrossing partitions
\begin{equation}
x \psi_0(x,t)^2 + (t x - x - 1)\psi_0(x,t) + 1 = 0
\label{ntransig0a}
\end{equation}
Its coefficients are the Narayana numbers
$\lim_{s \to \infty} \alpha_{n,b}(s) = N(n,b)$.
For example one gets
$\alpha_{8,4}(s) = 490 + \frac{812}{s} + \frac{372}{s^{2}} + \frac{27}{s^{3}}$,
which interpolates between
$\alpha_{8,4}(1) = 1701 = S(8, 4)$ and
$\lim_{s \to \infty} \alpha_{8,4}(s) = 490 = N(8, 4)$.
Eq. (\ref{ntransig1}) gives
\begin{eqnarray}
\psi_1(x, t) = \frac{t^{2} y^{4}}{(1 - 2 y)(1 - y)(1 - 2 y + y^2 - x t)}
\label{ntransig1P}
\end{eqnarray}
We checked till order $x^{10}$ Eq. (\ref{ntransig1P}) and 
Eqs. (\ref{ntransig2}, \ref{ntransig3})
listing partitions of $[n]$ till $n=10$.

If all the partitions in the nc-closed set have all
the blocks with the same size $m$, for a partition $p$ with $n$ elements
$b(p) = \frac{n}{m}$, so that one can obtain the number of irreducible
partitions from Eq. (\ref{nctran}) without introducing the parameter $t$;
this is done in the case of the
chords, the partitions of $n$ with block size $m=2$ \cite{flanoy}, where
the notation $I(x^2) = f(x, t)$, $I(x^2,s) = f(x,s,t)$ and $C(x^2) = a(x, t)$ 
is used; see also \cite{beiss}.
In this case one has
\begin{equation}
f_{(n)}(x) = (2n-1)!! x^{2n}; \qquad a_{(n)}(x) = C_n x^{2n}
\label{chfa}
\end{equation}
where $C_n$ satisfies the recursion relation
$C_n = (n-1) \sum_{i=1}^{n-1} C_i C_{n-i}$, with $C_1=1$ \cite{stein};
one has $C_2=1, C_3=4$.
Eq. (\ref{ntransig0}) gives
\begin{equation}
\psi_0(x, t) = 1 + t x^2 \psi_0(x, t)^2
\label{ch0}
\end{equation}
so $\psi_0(x, t)$ is the g.f. enumerating $NC_2$ according
to the number of blocks.
Eq. (\ref{ntransig1}) gives
\begin{equation}
\psi_1(x,t) = \frac{t^2 y^4}{1 - 2 t x y}
\label{ch1}
\end{equation}
where $y$ is given by Eqs. (\ref{ypsi}, \ref{ch0}).
From Eq. (\ref{ntransig2}) one gets
\begin{equation}
\psi_2(x,t) = \frac{4 t^3 y^6 + 4 t^2 x y^3 \psi_1(x,t) + t x^2 \psi_1(x,t)^2}
{1 - 2 t x y} 
\label{ch2}
\end{equation}

\subsection{CC types and extended NCPT for nc-closed sets of partitions of size multiple of $k$}
Consider a nc-closed set ${\cal P}$ of partitions with size multiple of $k$.
Let us say that a block has congruence class (CC) $a=1,\cdots,k$
if its minimum element is equal to $a$ modulo $k$.
We want to count the number of partitions in ${\cal P}$ with given number
of elements and of blocks with given CC.
Parameters $t_1,\cdots,t_k$ are used to count the number of partitions 
with blocks in the congruence classes $1,\cdots,k$.
A partition $p$ has $j_1(p)$ blocks with CC $1$, $\cdots$,
$j_k(p)$ blocks with CC $k$; let us call $(j_1(p), \cdots, j_k(p))$
the CC type of the partition $p$, with $j_1(p) \ge 1$ and
$j_a(p) \ge 0$ for $a \ne 1$; it has the corresponding monomial
$t_1^{j_1(p)}\cdots t_k^{j_k(p)}$.
Partitions on an interval starting with $a$ have CC type $j_a \ge 1$ and
$j_i \ge 0$ for $i \ne a$.
Let $f_a(x,s,t_1,\cdots,t_k)$ be the g.f. for the number of
partitions with sequential form starting at position $a$ modulo $k$, 
with given numbers of elements, components and CC types.
$[x^n s^c t_1^{j_1}\cdots,t_k^{j_k}]f_a(x, s,t_1,\cdots,t_k)$ is the number
of partitions starting with $a$, with $c$ irreducible components, $j_b$
blocks with CC $b$, with $b=1,\cdots,k$.
The g.f. $f_a$ is obtained from $f_1$ by a 
cyclic permutation of the $t_1,\cdots,t_k$ parameters,
\begin{equation}
f_a(x,s,t_1,\cdots,t_k) = f_1(x,s,t_a,t_{a+1\mod k}, \cdots,t_{a-1\mod k})
\label{faf1}
\end{equation}
Let $a(x,t_1,\cdots,t_k)$ be the g.f. for the corresponding
irreducible partitions (starting at position $1$), with given CC types.

Following the derivation of the NCPT in subsection II.A,
a size-$m_i$ irreducible component $p_i$ of a partition $p$
has blocks with lowest element $1$ plus the number of elements preceding 
them in $p$, so its CC type is in general
different from the one of the equivalent partition in $U(m_i)$; for instance
in the example in Appendix A.A, the partition $(1,2,5)(3,4,8)(6,7)$,
having size multiple of $k=2$, has
the irreducible component $(6,7)$, consisting of a single block with CC $2$, 
while the equivalent irreducible partition $(1,2)$ has a single block with 
$CC$ 1.
Therefore Eq. (\ref{nctran}) cannot be trivially extended to the enumeration
of partitions according to their CC types.

We saw in the subsection II.A that a partition $p$ of $[n]$
can be decomposed as
$p_1 c_1 \cdots p_h c_h$, where $c_1,\cdots,c_h=n$ are the elements
of the irreducible partition $c$ containing $n$, while $p_1,\cdots, p_h$
are components of $p$ defined on the $h$ intervals between these elements.
Since $p_i$ has size multiple of $k$, while $c_i$
is a single element, it follows that $p_i$ starts at position $i$ modulo $k$.
Let $u_a(n,s, t_1,\cdots,t_k) = [x^n] f_a(x,s,t_1,\cdots,t_k)$
and $a_h(t_1,\cdots,t_k) = [x^h] a(x,t_1,\cdots,t_k)$.
Instead of Eq. (\ref{nctran0}) one has in this case
\begin{eqnarray}
f_1(x,s,t_1,\cdots,t_k) &=&  1 + 
s \sum_{n \ge k} x^n  \sum_{h \ge k} a_h(t_1,\cdots,t_k) \sum_{m_1+\cdots +m_h = n-h} 
u_1(m_1,s, t_1,\cdots,t_k)\cdots u_h(m_h,s,t_1,\cdots,t_k) = \nonumber \\
&& 1 + s \sum_{j \ge 1} a_{jk}(t_1,\cdots,t_k) (x^k \prod_{a=1}^k f_a(x,s,t_1,\cdots,t_k))^j
\end{eqnarray}
where $n, h$ and $m_1,\cdots,m_h$ are divisible by $k$; from this equation
one gets the extended NCPT
\begin{equation}
f_1(x,s,t_1,\cdots,t_k) = 1 + s a\Big(x \prod_{a=1}^k f_a(x,s,t_1,\cdots,t_k)^{\frac{1}{k}}, t_1,\cdots,t_k\Big)
\label{nctrank}
\end{equation}

In Appendix A.A we give an example of the use of this equation in
the case of a nc-closed set of partitions of size multiple of $2$, but
which are not $2$-divisible partitions (we dealt only with the case $s=1$).

We do not have a close formula for the inverse of Eq. (\ref{nctrank}),
like Eq. (\ref{nctraninvt}); $a(x,t_1,\cdots,t_k)$ can be computed iteratively
from Eqs. (\ref{faf1}, \ref{nctrank}).

Like in the previous subsection,
$f_1(x,s,\frac{t_1}{s},\cdots,\frac{t_k}{s})$ can be expanded in powers
of $\frac{1}{s}$: 
$[\frac{1}{s^g}]f_1(x,s,\frac{t_1}{s},\cdots,\frac{t_k}{s})$
is the g.f. enumerating the partitions according to
the CC types and $b(p) - c(p) = g$, where $b(p)$ is the number of blocks
and $c(p)$ is the number of irreducible components.
Define $\sigma = \frac{1}{s}$ and
\begin{equation}
\psi_1(x, \sigma, t_1,\cdots, t_k) = 
\sum_{j \ge 0} \sigma^j \psi_{1,j}(x, t_1,\cdots, t_k) = 
f_1(x,s,\frac{t_1}{s},\cdots,\frac{t_k}{s})
\end{equation}
which satisfies
\begin{equation}
\sigma \psi_1(x, \sigma, t_1,\cdots, t_k) - \sigma =
a\Big(x \prod_{a=1}^k \psi_a(x,\sigma,t_1,\cdots,t_k)^{\frac{1}{k}}, t_1\sigma,
\cdots,t_k\sigma\Big)
\label{psik1}
\end{equation}
The lowest order in $\sigma$ of this equation gives
\begin{equation}
\psi_{1,0}(x, t_1,\cdots, t_k) - 1 = 
t_1 \alpha(
x\prod_{b=1}^k \psi_{b,0}(x,t_1,\cdots,t_k)^{\frac{1}{k}})
\label{psik2}
\end{equation}
where $\alpha(x) = [t_1] a(x, t_1,0,\cdots,0)$, from which, using 
Eq. (\ref{faf1})
\begin{equation}
\psi_{a,0}(x, t_1,\cdots, t_k) - 1 = t_a \alpha(x\prod_{b=1}^k \psi_{b,0}(x,t_1,\cdots,t_k)^{\frac{1}{k}})
\label{psik0}
\end{equation}
From Eq. (\ref{psik0}) it follows that 
$\frac{\psi_a(x,t_1,\cdots, t_k)-1}{\psi_b(x,t_1,\cdots, t_k)-1} = \frac{t_a}{t_b}$, so one has
\begin{equation}
\psi_{a,0}(x,t_1,\cdots, t_k) = 1 + t_a \Phi(x,t_1,\cdots, t_k)
\label{psik0a}
\end{equation}
where $\Phi(x,t_1,\cdots, t_k)$ is totally symmetric in $t_1,\cdots, t_k$.
From Eq. (\ref{psik0}) one gets
\begin{equation}
\Phi(x,t_1,\cdots, t_k) = \alpha(x \prod_{a=1}^k(1 + t_a \Phi(x,t_1,\cdots, t_k))^{\frac{1}{k}})
\label{psik0b}
\end{equation}
For example, in the case of a flow from the g.f. enumerating a nc-closed set 
${\cal{P}}$  and the g.f. enumerating 
$NC_k$, one has $\alpha(x,t_1,\cdots, t_k) = t_1 x^k$
and Eq. (\ref{psik0b}) becomes
\begin{equation}
\Phi(x,t_1,\cdots, t_k) = 
x^k \prod_{a=1}^k(1 + t_a \Phi(x,t_1,\cdots, t_k))
\label{psik0b1}
\end{equation}
In Section V we will consider the case of $NC^{(k)}$ with CC types and rank.

Let us compute the $\frac{1}{s}$ term of $\frac{1}{s}$-expansion of $f_1$ 
in the case $k=2$.
Defining 
\begin{eqnarray}
&& y = x \sqrt{\psi_{1,0}(x,t_1,t_2) \psi_{1,0}(x,t_2,t_1)} \nonumber \\
&& \eta = \frac{x}{2}
\frac{\psi_{1,1}(x,t_1,t_2)\psi_{1,0}(x,t_2,t_1) + 
\psi_{1,0}(x,t_1,t_2)\psi_{1,1}(x,t_2,t_1)}
{\sqrt{\psi_{1,0}(x,t_1,t_2)\psi_{1,0}(x,t_2,t_1)}}
\label{yeta}
\end{eqnarray}
Eq. (\ref{psik2}) becomes
\begin{equation}
\sigma (\psi_{1,0}(x,t_1,t_2) - 1) + \sigma^2 \psi_{1,1}(x,t_1,t_2) =
\sigma t_1 a_{(1,0)}(y + \sigma \eta) + \sigma^2 t_1^2 a_{(2,0)}(y) +
\sigma^2 t_1 t_2 a_{(1,1)}(y) + O(\sigma^3)
\end{equation}
so that

\begin{equation}
\psi_{1,1}(x,t_1,t_2) = t_1^2 a_{(2,0)}(y) + 
t_1 t_2 a_{(1,1)}(y) + t_1 \eta \frac{d a_{(1,0)}(y)}{d y}
\label{psik2cc}
\end{equation}
This equation and the one with $t_1,t_2$ exchanged give a 
system of linear equations for $\psi_{1,1}(x,t_1,t_2)$
and $\psi_{1,1}(x,t_2,t_1)$, from which one gets $\psi_{1,1}(x,t_1,t_2)$
\begin{equation}
\psi_{1,1}(x,t_1,t_2) = t_1 \frac{t_2(t_2 - t_1)\chi(t_2,t_1)
\frac{d a_{(1,0)}(y)}{d y}(a_{(2,0)}(y) - a_{(1,1)}(y)) +
t_1 a_{(2,0)}(y) + t_2 a_{(1,1)}(y)}
{1 - (t_1 \chi(t_1,t_2) + t_2 \chi(t_2,t_1))\frac{d a_{(1,0)}(y)}{d y}}
\label{psik2cc2}
\end{equation}
where
\begin{equation}
\chi(t_1,t_2) = \frac{x}{2} \sqrt{\frac{\psi_{1,0}(t_2,t_1)}{\psi_{1,0}(t_1,t_2)}}
\label{psik2cc3}
\end{equation}
We checked this Eq. (\ref{psik2cc2}) in the case of chords, listing partitions
till size $14$.

In the rest of the paper we will set $s=1$ and call $f(x,1,t)$ and
$f_1(x,1,t_1,\cdots,t_k)$ respectively $f(x,t)$ and $f_1(x,t_1,\cdots,t_k)$,
unless explicitly stated.

\section{Walks on $k$-gon trees and the corresponding partitions}

Let us call $P^{(k)}(l)$ the set of $k$-divisible partitions of $kl$;
define $P^{(k)}_h(l)$ as the set of $k$-divisible partitions of $kl$
in which, between any two distinct blocks $B_i$ and $B_j$,
the number of elements in $B_j$ between two elements of $B_i$ is a multiple
of $h$. 

$P^{(1)}_1(l) = P(l)$ is the set of partitions of $[l]$.
One has trivially $P^{(1)}_h(kl) \supset P^{(k)}_h(l)$ and
$P^{(k)}(l) = P^{(k)}_1(l) \supset P^{(k)}_h(l)$
for $h, k \ge 1$.

{\bf Observation $1$.}
The set of noncrossing partitions in $P^{(k)}_k(l)$ is $NC^{(k)}(l)$.

For $k=1$, $P^{(1)}_1(l)$ is the set of the partitions of $[l]$,
so that the set of noncrossing partitions in $P^{(k)}_k(l)$ is $NC(l)$.
Let $k \ge 2$.  Both sets are sets of $k$-divisible partitions
and if an element of a block $B_i$ of  $p \in NC^{(k)}(l)$ is
between two elements of $B_j$, with $j\neq i$, then the noncrossing condition 
implies that all the elements of $B_i$ are between these two elements,
so that $NC^{(k)}(l)$ is included in the set of noncrossing
partitions in $P^{(k)}_k(l)$. 
Viceversa, $P^{(k)}_k(l) \subset P^{(k)}(l)$, and by definition 
$NC^{(k)}(l)$ is the subset of noncrossing partitions of $P^{(k)}(l)$,
so that the set of noncrossing partitions of $P^{(k)}_k(l)$ is a subset
of $NC^{(k)}(l)$.

{\bf Observation $2$.} 
The set of partitions in $P^{(k)}_k(l)$ with all blocks 
of length $k$ is $NC_k(l)$.

In the case $k=1$, the only partition in both sets in $(1)\cdots(l)$.
Let $k > 1$. If there is a crossing between two distinct blocks $B_i$ and 
$B_j$ in a partition in $P^{(k)}_k(l)$, say an ordered sequence 
$a_{i_1}\cdots b_{j_1} \cdots a_{i_2} \cdots b_{j_2}$, where $a_{i_r}$
belongs to $B_i$ and $b_{j_r}$ belongs to $B_j$, then
there must be $mk \ge k$ elements of $B_j$ between $a_{i_1}$ and
$a_{i_2}$, so there are at least $k+1$ elements of $B_j$; similarly for $B_i$.
So $B_i$ and $B_j$ must have at least $2k$ elements each.
Therefore a partition in $P^{(k)}_k(l)$ which has all blocks of length $k$
is noncrossing, hence it belongs to $NC_k(l)$.
Conversely let $p$ be a partition in $NC_k(l)$; 
given two distinct blocks $B_i$ and $B_j$ of $p$, if there is an element 
of a block $B_j$ between two elements of $B_i$, then by the noncrossing
condition all the $k$ elements of $B_j$ must be there; hence 
$NC_k(l)$ is a subset of the set of partitions in $P^{(k)}_k(l)$
with all blocks of length $k$, which ends the proof.

\vspace{4 mm}

For $k \ge 2$, 
define a {\it $k$-gon tree} as a connected directed graph which is formed 
by oriented $k$-gons, having no edges in common and joined at vertices,
in such a way that the graph has no other loop, apart from the $k$-gons.

A rooted $k$-gon tree is a $k$-gon tree in which one of the vertices
is the root.

A walk on a rooted $k$-gon tree is said to have a noncrossing sequence
of steps if the sequence of edges which it traverses does not have a 
subsequence $a\cdots b \cdots a \cdots b$, with $a \neq b$.

Let us say that a walk loops around a $k$-gon $j$ times if it traverses
each of its edges exactly $j$ times.

{\bf Lemma $1$.} In a closed walk traversing at least one edge in each
$k$-gon of a $k$-gon tree, each $k$-gon is looped around one or more times.

Let us consider a closed walk $w$ traversing at least one edge in each
$k$-gon of a $k$-gon tree $T$.
The shortest closed walk is a loop around a $k$-gon tree consisting of
a single $k$-gon, so the minimum length of $w$ is $k$; in fact if it were
shorter, $T$ would have another loop, apart from the $k$-gons.
Assume by induction that the closed walks of length less or equal to $j k$
loop around each $k$-gon one or more times, having therefore length multiple 
of $k$.
Consider a closed walk $w$ of length $n$ between $jk+1$ and $jk+k$.
The walk starts along the first $k$-gon $G_1$; if it never leaves it,
it loops $j+1$ times around it; otherwise, let $v$ be the node at which
$w$ leaves $G_1$ for the first time. The walk must return to $v$, otherwise
there would be a loop in $T$ which is not a $k$-gon. Consider the walk $w_2$
which is the part of $w$ starting at $v$ and returning to it for the first
time after leaving it.
Let $w_1$ be the closed walk obtained from $w$ subtracting $w_2$.
If $n=jk+1$, $w_1$ and $w_2$ have length less or equal to $jk$, so by
induction they would have length multiple of $k$; this is not possible,
so there is no such walk. Proceeding in the same way for $n=jk+2,\cdots,jk+k$,
one sees that for $n=jk+2,\cdots,jk+k-1$ there are no such walks; 
by induction, for $n=jk+k$, $w_1$ and $w_2$ loop around the $k$-gons
one or more time, therefore $w$ loops around the $k$-gons of 
$T$ one or more time, and $w$ has length $jk+k$.

From this lemma it follows that a closed walk, traversing at least one edge 
in each $k$-gon of a $k$-gon tree $T$,
covers $T$ and has length multiple of $k$.

Given a length-$n$ walk on a $k$-gon tree,
use $i=1,\cdots,n$ as a label for the edge through which the walk passes at
the $i$-th step; each edge has a set of these labels.
Order the $k$-gons according to the smallest element of their label set, 
that is according to which $k$-gon appears first in the walk.
Associate to the $k$-gon $G_j$ the label set $B_j$
which is the union of the sets of labels of its edges.
In this way one gets a partition of $[n]$ corresponding to the walk on
the $k$-gon tree.

{\bf Lemma $2$.} For $k \ge 2$, the partition corresponding to a length-$kl$ 
closed walk covering a $k$-gon tree $T$ belongs to $P^{(k)}_k(l)$.

Let $p$ be the
partition formed by the label sets of the $k$-gons associated to the closed 
walk $w$ covering a $k$-gon tree $T$.
By Lemma $1$ the label sets have length multiple of $k$, so $p$
belongs to $P^{(k)}(l)$.

If $w$ has length $k$, then $p$ is $(1,\cdots,k)$, which  belongs to 
$P^{(k)}_k(1)$.
Let $w$ have length $k l > k$.
If $T$ consists of a single $k$-gon, then its label set is $(1,\cdots,kl)$,
giving a partition in $P^{(k)}_k(l)$.
If $T$ has more than one $k$-gon,
let $i$ and $j$ be two consecutive elements of a block $B_a$, corresponding
to two edges of the $k$-gon $G_a$ joined at the node $v$.
If $j=i+1$, there are no elements of other blocks between these two elements.
If $j > i+1$, there are $j-i-1$ elements of other blocks between them.
The steps $i+1,\cdots,j-1$
of $w$ form a closed walk $w_1$ rooted at $v$, so by Lemma $1$
$w_1$ loops one or more
time around the $k$-gons it enters. Let $G_b$ be one of these $k$-gons;
then the number of elements of its label set $B_b$ which are in the interval
$[i+1,j-1]$ is a multiple of $k$.
Therefore $p$ belongs to $P^{(k)}_k(l)$.

\vspace{4 mm}
Let us remark that a length-$n$ walk $w$ (not necessarily closed)
on a $k$-gon tree $T$, going through 
one or more edges of its $k$-gons, has the corresponding partition belonging 
to $P^{(1)}_k(n)$.
In fact, suppose that $w$ at step $i_1$ ends in $v$, that the edge at step $i_1$
belongs to the $k$-gon $G_a$ and the next edge to another $k$-gon,
and that at step $i_2-1$ it returns to $v$, entering at step $i_2$ again in $G_a$;
on the steps $i_1+1,\cdots, i_2-1$ the walk is not on $G_a$.
Then by Lemma $1$ the closed walk $w_1$ rooted in $v$, made by these steps,
loops one or more times around the $k$-gons which have at least one edge with
one of these labels; reasoning as in Lemma $2$ it follows
that between the elements $i_1$ and $i_2$, belonging to the block $B_a$,
the number of elements in the interval $[i_1+1,i_2-1]$ belonging to another
block is a multiple of $k$.
Therefore the set of label sets $\{B_j\}$
associated to $w$ on $T$ is a partition in $P^{(1)}_k(n)$.

{\bf Theorem $1$.} The set of length-$kl$ closed walks, such that a walk $w$ 
covers a rooted $k$-gon tree, is isomorphic to $P^{(k)}_k(l)$ for $k \ge 2$.
The set of label sets of the $k$-gons associated to $w$
is the set of blocks of the corresponding partition.

Consider a length-$kl$ closed walk $w$ covering a rooted $k$-gon tree $T$. 
By Lemma $2$ the partition $p$ formed by the
set of label sets of the $k$-gons associated to $w$ belongs to $P^{(k)}_k(l)$.

Let us construct viceversa a length-$kl$ closed walk $w$ covering a rooted 
$k$-gon tree $T$ from a partition $p \in P^{(k)}_k(l)$, where $k \ge 2$.
Let $B_1,\cdots, B_h$ be the blocks of $p$.

If $l=1$, there is a single partition, having one block, $B_1 = (1,\cdots,k)$,
to which the closed walk of length $l$ looping once around the rooted
$k$-gon $G_1$ corresponds.

Assume by induction that from a partition in $P^{(k)}_k(l')$, for $l' < l$
one can construct a closed walk covering a rooted $k$-gon tree.
Starting from the root, the first $k$-gon $G_1$
corresponds to the block $B_1$ of length $k i$, with elements
$b_1=1,\cdots, b_{ki}$.
In the intervals $[b_j+1,b_{j+1}-1]$, $j=1,\cdots,i-1$ and $[b_{ki}+1,kl]$ 
the elements
must be assigned to other $k$-gons. 
If all these intervals are empty, then $B_1=(1,\cdots,kl)$, to which
the closed walk rooted in $R$, looping around $G_1$ $l$ times is
associated.
Let instead $I_1 = [b_j+1,b_{j+1}-1]$ be the first not empty interval. 
Then the elements $1,\cdots,b_j$ belong to $B_1$; they correspond to a walk 
$w_1$ of length $b_j$, moving around $G_1$ starting from $R$ and ending 
at node $v$.
Consider the set $S$ of elements constructed in this way:
$I_1$ is a subset of $S$; for each element in $I_1$, the block containing it
is a subset of $S$; the intervals similarly defined on this block have elements 
belonging to other blocks, which are added to $S$, unless they belong to $B_1$;
and so on.
The set of blocks $B_i$ in $S$ forms a partition of $S$.
It is clear that relabeling indices, these blocks form a partition
in $P^{(k)}_k(l')$, with $l' = \frac{|S|}{k} < l$. Using the induction
hypothesis, associate to it a closed walk $w_2$ on a $k$-gon tree rooted at $v$.

Consider the set of blocks of $p$ which are not in $S$.
Relabeling indices, they form a partition of $P^{(k)}_k(l-l')$.
By induction hypothesis it gives a closed walk on a $k$-gon tree, rooted at 
$R$. It is formed by the walk $w_1$ going from $R$ to $v$ constructed before,
and a walk $w_3$ from $v$ to $R$. The desired closed walk $w$ is obtained 
concatenating $w_1$, $w_2$ and $w_3$.

\vspace{4 mm}

\begin{figure*}[h]
\begin{center}
\epsfig{file=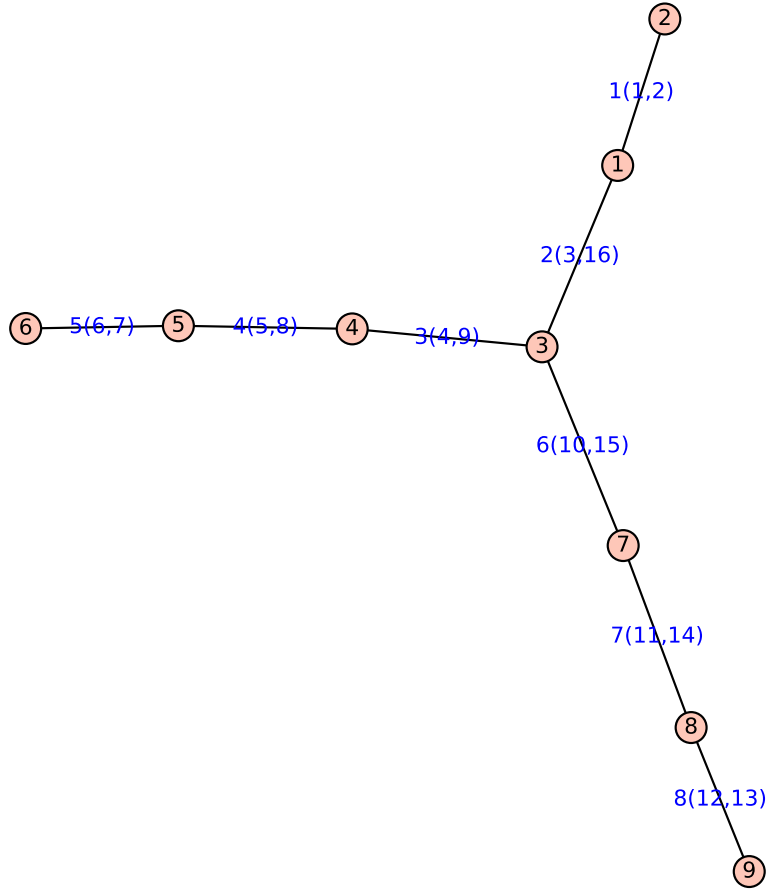, width=12.00cm  }
\caption{Graphical representation of the $2$-divisible noncrossing partition 
    $(1,2)(3,16)(4,9)(5,8)(6,7)(10,15)(11,14)(12,13)$, as a closed walk 
    covering a rooted tree with root in node $1$.
    The edges have a label
    $n(b_1,\cdots,b_j)$ where $n$ is the block number, and $b_1,\cdots,b_j$
    are the block elements.
}
\label{gpartk2}
\end{center}
\end{figure*}

In the case $k=2$, the $2$-gon is an edge, running first away from the root
of the $2$-gon tree, then towards it; the closed walk on a 
rooted $2$-gon tree is a closed walk on a rooted tree.
An example is in Fig. \ref{gpartk2}.

\begin{figure*}[h]
\begin{center}
\epsfig{file=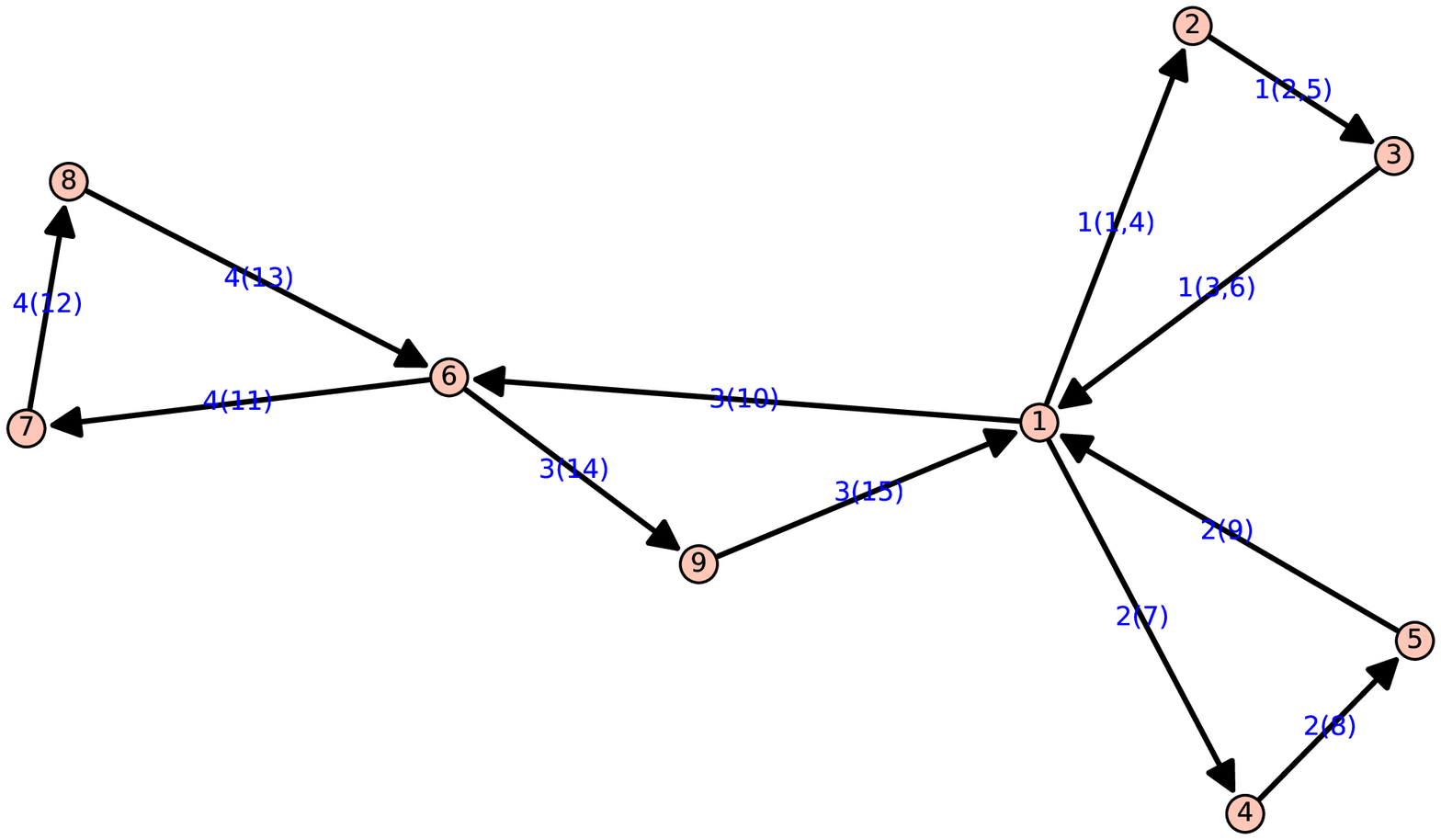, width=12.00cm  }
\caption{Graphical representation of the $3$-divisible noncrossing partition 
    $(1,2,3,4,5,6) (7,8,9) (10,14,15) (11,12,13)$, as a closed walk
    covering a rooted $3$-gon tree.  The edges have a label
    $n(b_1,\cdots,b_j)$ where $n$ is the block number, and $b_1,\cdots,b_j$
    are the block elements. The closed walk is represented by the sequence of
    edges with block elements in the interval $[1,15]$, with
    sequence of nodes $(1, 2, 3, 1, 2, 3, 1, 4, 5, 1, 6, 7, 8, 6, 9, 1)$.
}
\label{gpart}
\end{center}
\end{figure*}

An example of closed walk covering a rooted $3$-gon tree 
is given in Fig. \ref{gpart}.

There is a cyclic symmetry group of transformations in $P^{(k)}_k(l)$. 
Consider a length-$kl$ closed walk $w$
covering a $k$-gon tree $T$, starting from a node in $T$; 
label the sequence of
edges with $1,\cdots, kl$ and consider the corresponding partition $p$ on
these elements, with blocks given by the label sets of the $k$-gons.
Consider the closed walk $\rho(w)$ starting at the second step of $w$
and ending with its first step. The corresponding partition $\rho(p)$
is obtained from $p$ by the cyclic permutation $(1,\cdots,kl) \to 
(2,\cdots,kl,1)$. $\rho^{kl} = 1$; by Lagrange's theorem,
the orbits under $\rho$ have size dividing $kl$.
This symmetry relates closed walks covering $T$ with different roots.

{\bf Corollary $1$.} The set of length-$kl$ closed walks covering rooted $k$-gon
trees, with noncrossing sequence of steps, is isomorphic to $NC^{(k)}(l)$,
for $k \ge 2$.

In fact by Theorem $1$ the set of walks covering rooted $k$-gon
trees is isomorphic to $P^{(k)}_k(l)$; the fact that the sequence of
steps is noncrossing corresponds to consider the subset of $P^{(k)}_k(l)$
with noncrossing partitions, so by Observation $1$ the corollary follows.

{\bf Corollary $2$.} The set of closed length-$kl$ walks covering rooted 
$k$-gon trees and looping exactly 
once around each $k$-gon is isomorphic to $NC_k(l)$, for $k \ge 2$.

By Theorem $1$ the set of walks covering rooted $k$-gon
trees is isomorphic to $P^{(k)}_k(l)$;
the condition that the walks loop exactly once around each $k$-gon corresponds
to the condition that all the blocks be of size $k$. From Observation $2$
it follows that it is equal to $NC_k(l)$.

In the case of $NC(l) = NC^{(1)}(l)$ there is no direct interpretation as walks 
on $k$-gon trees; however using the $1$-to-$1$ correspondence between
noncrossing partitions and labeled rooted trees \cite{proding}, and the 
$1$-to-$1$ correspondence between the latter and closed walks on rooted trees
going exactly twice on each edge of the tree, one gets a
$1$-to-$1$ correspondence between the latter and noncrossing partitions.

{\bf Observation} $3$: In an irreducible partition of $P^{(k)}_k(l)$ each
block has size equal or larger than $2k$, apart from the partition 
$(1,\cdots,k)$. 

Suppose that there are only $k$ elements in a block $B_i$ in an
irreducible partition $p$ of $P^{(k)}_k(l)$ and that there are elements
of other blocks in $p$. If there are no elements
of other blocks between the elements of the block $B_i$, then $p$ is reducible.
In order that $p$ be an irreducible partition of $P^{(k)}_k(l)$,
there must be elements of other blocks between two non-empty groups of elements
of $B_i$, having less than $k$ elements. Some elements of the blocks containing
these elements must also be outside the two groups of elements of $B_i$,
otherwise $p$ would be reducible. But then one of the groups of $B_i$
should have a nonvanishing number of elements multiple of $k$,  so that
$B_i$ would have size larger than $k$,
which is a contradiction; hence the observation follows.

This observation leads to a way of listing irreducible partitions of
$P^{(k)}_k(l)$.
Since by Observation $3$ 
all the irreducible partitions of $P^{(k)}_k(l)$, apart from $(1,\cdots, k)$,
have at most $\frac{l}{2}$ blocks, one can list
all partitions in $P^{(k)}_k(l)$ with at most $\frac{l}{2}$ blocks and
select the irreducible partitions.
More efficiently,
one can consider the set $Q_k(l')$ of the partitions of $P^{(k)}_k(l')$
with at most $\frac{l'}{2}$ blocks and
in which there are not more than $k$ consecutive elements in a block;
from the irreducible partitions of $Q_k(l')$, with $l' \le l$, one obtains
the irreducible partitions of $P^{(k)}_k(l)$ in the following way.

One can represent the sequential form $v$ of a partition $p$ as a monomial 
$X_{i_i}^{j_1}\cdots X_{i_m}^{j_m}$, where $X_i$ are
noncommuting variables, by replacing each element $i$ in the sequential form
of $p$ with $X_i$.
The indices $i_h$ are ordered according to the first appearence of an
element of a block.
Let $\pi = X_{i_i}^{j_1}\cdots X_{i_m}^{j_m}$ be an irreducible partition in 
$Q_k$; $j_s \in \{1,\cdots, k\}$ for $s=1,\cdots,m$; $\pi$ has $m$ block parts.
Then $X_{i_i}^{j_1'}\cdots X_{i_m}^{j_m'}$, 
with $j_s' = j_s \,mod\, k$ in $[k]$, is a irreducible partition in $P^{(k)}_k$;
the series of partitions thus obtained from $\pi$ is
$\frac{X_{i_1}^{j_1}}{1-X_{i_1}^k}\cdots \frac{X_{i_m}^{j_m}}{1-X_{i_m}^k}$.
The subset of the partitions of size $kl$ of the
union of the sets of partitions obtained in this way from the sets of
irreducible partitions of $Q_k(l')$, with $l' \le l$,
is the set of irreducible partitions of $P^{(k)}_k(l)$.
For instance the partition $p=(1,4,5,8)(2,3,6,7)$ has sequential form
$(1,2,2,1,1,2,2,1)$ and monomial $X_1 X_2^2 X_1^2 X_2^2 X_1$.
$p$ is an irreducible partition in $Q_2(4)$; then
$\frac{X_1}{1-X_1^2} \frac{X_2^2}{1-X_1^2} \frac{X_1^2}{1-X_1^2} \frac{X_2^2}{1-X_1^2}\frac{X_1}{1-X_1^2}$ is the series of irreducible partitions of
$P^{(2)}_2(l')$, with $l' \ge 4$, obtained from $p$.
For $l'=5$, it gives $5$ partitions, $X_1^3 X_2^2 X_1^2 X_2^2 X_1$,
$X_1 X_2^4 X_1^2 X_2^2 X_1$,
$\cdots$, $X_1 X_2^2 X_1^2 X_2^2 X_1^3$. Similarly from $X_1^2 X_2^2 X_1^2 X_2^2$
one obtains $4$ partitions.
Therefore from the three irreducible partitions $X_1 X_2^2 X_1^2 X_2^2 X_1$,
$X_1^2 X_2^2 X_1^2 X_2^2$ and $X_1^2 X_2^2 X_1^2 X_2^2 X_1^2$ of $Q_2$
one obtains $10$ irreducible partitions of $P^{(2)}_2(5)$; they belong to
the same orbit.

{\bf Observation $4$}.
$P^{(k)}_k \equiv \{P^{(k)}_k(l), l \ge 1 \}$ is nc-closed.

Let ${\cal A}$ be the set of the irreducible partitions of $P^{(k)}_k$
and ${\cal P}_{\cal A}$ be the set of partitions generated 
by ${\cal A}$.
An irreducible component $p_i$ of a partition $p \in P^{(k)}_k(l)$
has a set of blocks with size multiple of $k$; between two elements
of a block in $p_i$, the number of elements belonging to another block
is multiple of $k$; so $p_i$ is equivalent to an irreducible
partition in ${\cal A}$, belonging to $P^{(k)}_k(m_i)$, where $m_i$
is the number of elements of $p_i$, so $p \in {\cal P}_{\cal A}$.
Consider viceversa a partition $p$ in ${\cal P}_{\cal A}$.
Let $i_1$ and $i_2$, with $i_1 < i_2$ be consecutive elements in a 
block $B$ of an irreducible partition $p_1$, component of $p$.
Consider the elements in $[i_1+1, i_2-1]$ which belong
to a block $B'$; if $B'$ belong to $p_1$, then the number of elements
of  $B'$ in this interval is a multiple of $k$, because $p_1$
is isomorphic to a partition in ${\cal A}$; if $B'$ does not belong
to $p_1$, it belongs to an irreducible component $p_2$ nested
in this interval; since $p_2$ belongs to ${\cal A}$, it is $k$-divisible,
hence $B'$ has size multiple of $k$, with all its elements between $i_1$
and $i_2$. Therefore $p \in P^{(k)}_k$, ending the proof.

Since a partition $p$ in $P^{(k)}_k(l)$ has size multiple of $k$,
one can define CC types $j_a(p)$, with $a=1,\cdots, k$, as defined in
the previous Section.
The number of elements between 
two consecutive elements $i$ and $j$ of a block of $p$ 
is multiple of $k$, hence $j = i + 1\,mod\,k$;
therefore in a block of size $jk$ there are $j$ elements for each CC.

{\bf Observation $5$.}
The last block of a partition in $P^{(k)}_k(l)$ is the union of sets of 
consecutive integers, each with size multiple of $k$.

Let $i_r$ and $i_{r+1}$ be two consecutive elements of the last block $B_h$,
with $i_{r+1} > i_r + 1$. 
Let $B_j$ be a block with some elements between $i_r$ and $i_{r+1}$;
$B_j$ must have elements less than $i_1$, the lowest element of $B_h$,
since $B_h$ is the last block; then the number of elements $i_1,\cdots,i_r$
must be a multiple of $k$. This is true for any such pair of elements
$i_r$ and $i_{r+1}$, so that all the sets of consecutive integers in 
$B_h$, apart from the last one, have size multiple of $k$.
Since $B_h$ has size multiple of $k$, the Observation is proven.

This property of the last block is taken as part of the definition
of the special symmetric partition set $SS(2l)$ in \cite{bose}, 
which is $P^{(2)}_2(l)$. In the case $k=2$ the CC type of an element
of a partition in $P^{(2)}_2(l)$ is $1$ or $2$, i.e. the element
is odd or even; as remarked above, consecutive elements in a block $i$
and $j$ in a block have $j = i+1\,mod\,k$, so that even and odd numbers
alternate in a block in the case $k=2$, which is also taken as
part of the definition of $SS(l)$ in \cite{bose}.

\section{Enumeration of partitions in $P^{(k)}_k$ for $k \ge 2$}

In \cite{bau}  recursion relations have been found to enumerate
the closed walks of length $2j$ on rooted trees with $l$ edges,
in the study of the moments of the Adjacency random matrix model
in the $N \to \infty$ limit,
on ER graphs with average degree $t$.

These recursion relations are easily generalizable to the enumeration of 
length-$jk$ closed walk on rooted $k$-gon trees.
Let us consider a length-$jk$ closed walk $w$ covering a rooted $k$-gon tree 
$T$; denote the root by $r_1$.
Let $m$ be the number of times the walk returns to its root $r_1$.
The first edge of the walk is $(r_1, r_2)$, belonging to the $k$-gon
$\bar G$; the vertices of $\bar G$ are $r_1,\cdots, r_k$.
Let $T_a$ be the $k$-gon tree attached to the vertex $r_a$ of $\bar G$,
not including edges in $\bar G$; it can be trivial, consisting of a
single node.
Let $w_a$ be the part of $w$ on $T_a$,
with $kj_a$ steps and $m_a$  returns to $r_a$.
The walk loops $\bar m = j - \sum_{a=1}^k j_a \ge 1$ times around $\bar G$
and returns $m = m_1 + \bar m$ times to $r_1$.

Let $J_{j,l,m}$ be the number of closed walks with $kj$ steps, 
$l$ $k$-gons and $m$ returns to the root.
Then the number of closed walks with $kj$ steps and
$l$ $k$-gons is given by
\begin{equation}
J_{j,l} = \sum_{m=0}^j J_{j,l, m}
\end{equation}

By Theorem $1$ we can equivalently talk of the corresponding partitions.
The g.f. for the number of partitions in $P^{(k)}_k(j)$
with $jk$ elements and $l$ blocks is
\begin{equation}
f(x, t) = \sum_{j,l \ge 0} J_{j,l} x^{kj} t^l
\label{f1J}
\end{equation}
Define
\begin{equation}
\psi_{j,m}(t) = \sum_{l \ge 0} J_{j,l,m} t^l
\label{psijm}
\end{equation}
$\psi_{0,0} = 1$ corresponds to the trivial walk; a nontrivial walk
returns at least once to the root, so $\psi_{j,0} = 0$ for $j > 0$;
a walk with length $jk$ can return to the root at most $j$ times, so
$\psi_{j, m} = 0$ for $m > j$.

Following \cite{bau}, let us count the number of ways in which the walk
can arrive to $r_i$.
It can arrive to $r_i$ from $T_i$ or from $\bar G$; the first time it arrives
from $\bar G$. Using $\bar G$
and $T_i$ as objects, one has strings of $m_i + \bar m$ objects
with $\bar m$ $\bar G$'s, the rest with $T_i$; 
the first entry of a string is $\bar G$.
There are $\bar m - 1$ $\bar G$ to place in $m_i + \bar m - 1$ places,
so there is a combinatorial factor $\binom{m_i + \bar m - 1}{\bar m - 1}$
associated to $T_i$; one gets, for $j > 0$
\begin{equation}
\psi_{j,m}(t) = t \sum_{\bar m=1}^m
\sum_{\substack{j_1,\cdots, j_k \ge 0 \\ \sum_1^k j_i = j - \bar m}}
\psi_{j_1,m-\bar m}(t) \binom{m - 1}{\bar m - 1}
\prod_{b=2}^k \sum_{m_b=0}^{j_b} \psi_{j_b,m_b}(t) \binom{m_b + \bar{m} - 1}{\bar{m} - 1}
\label{psijm0}
\end{equation}

Define
\begin{equation}
\psi_j(y, t) = \sum_{m=0}^j y^m \psi_{j, m}(t)
\label{psijt}
\end{equation}
where the parameter $y$ is used to count the number of times the walk returns
to the root.
For $j > 0$, $\psi_j(y, t)$ can be written, using Eq. (\ref{psijm0})
and the fact that $m, \bar m \le j$, as
\begin{equation}
\psi_j(y, t) = t \sum_{m=1}^j \sum_{\bar m=1}^j \theta(m - \bar m)
\sum_{\substack{j_1,\cdots, j_k \ge 0 \\ \sum_1^k j_i = j - \bar m}}
\sum_{m_1=0}^{j_1} \delta_{m, \bar m + m_1} \psi_{j_1,m_1}(t)
\binom{m_1 + \bar{m} - 1}{\bar{m} - 1} y^m 
\prod_{b=2}^k \sum_{m_b=0}^{j_b} \psi_{j_b,m_b}(t) \binom{m_b + \bar{m} - 1}{\bar{m} - 1}
\nonumber
\end{equation}
where $\theta(n) = 1$ for $n \ge 0$ and zero otherwise, $\delta_{i,j}$
is the Kronecker function.
$\theta(m - \bar m)$ can be dropped from this sum, since 
$\delta_{m, \bar m + m_1}$ guarantees that $m \ge \bar m$; then the sum
over $m$ can be moved to the right of the sums over $\bar m$,
$j_1,\cdots,\j_k$ and $m_1$; hence one gets
\begin{eqnarray}
&&\psi_j(y, t) = t \sum_{\bar m = 1}^j y^{\bar m}
\sum_{\substack{j_1,\cdots, j_k \ge 0 \\ \sum_1^k j_i = j - \bar m}}
H_{j_1,\bar m}(y,t)H_{j_2,\bar m}(1,t)\cdots H_{j_k,\bar m}(1,t) 
\label{psiH1} \\
&&H_{j,\bar m}(y, t) = \sum_{m=1}^j y^m [y^m]\psi_{j}(y,t) \binom{m+\bar m - 1}{\bar m - 1};
\qquad \psi_j(y, t) = H_{j, 1}(y, t)
\label{psiH}
\end{eqnarray}
and $\psi_0(y, t) = 1 = H_{0,\bar m}(y, t)$.

One can further simplify these equations defining
$\psi(z, y, t) = \sum_{j\ge 0} z^j \psi_j(y, t)$ and
$H_{\bar m}(z, y, t) = \sum_{j\ge 0} z^j H_{j,\bar m}(y, t)$.

One gets
\begin{eqnarray}
&&\psi(z, y, t) = 1 + t \sum_{\bar m \ge 1} (y z)^{\bar m}
H_{\bar m}(z,y,t)H_{\bar m}(z,1,t)^{k-1}
\label{psiH1z} \\
&&H_{\bar m}(z, y, t) = 1 + \sum_{m \ge 1} y^m [y^m] \psi(z,y,t) \binom{m+\bar m - 1}{\bar m - 1};
\qquad \psi(z, y, t) = H_{1}(z, y, t)
\label{psiHz}
\end{eqnarray}
From Eqs. (\ref{f1J},\ref{psijm},\ref{psijt},\ref{psiHz}) one has
\begin{equation}
f(x, t) = \psi(x^k,1,t) = H_1(x^k,1,t)
\label{fpsiH}
\end{equation}
where $z = x^k$.

One can count similarly the closed walks on rooted $k$-gon trees, in which
there is a restriction on the number of times the walk can loop around
the $k$-gons.
Let us consider the subset $P^{M,(k)}_k(l)$ of $P^{(k)}_k(l)$
in which the partitions have block sizes divided by $k$ belonging to
a subset $M$ of $\{\bar m \ge 1\}$.
Eqs. (\ref{psiH1} - \ref{psiHz}) can be modified to enumerate the
number of partitions in $P^{M,(k)}_k(l)$ according to the number of blocks, 
by restricting the sum over $\bar m$ in Eqs. (\ref{psiH1}, \ref{psiH1z}) 
to this set.

Eqs. (\ref{psiH1z}, \ref{psiHz}) become
\begin{eqnarray}
&&\psi_M(z, y, t) = 1 + t \sum_{\bar m \in M} (y z)^{\bar m}
H_{M,\bar m}(z,y,t) H_{M,\bar m}(z,1,t)^{k-1}
\label{psiM} \\
&&H_{M,\bar m}(z, y, t) = 1 + \sum_{m \ge 1} y^m [y^m] \psi_M(z,y,t) \binom{m+\bar m - 1}{m};
\qquad \psi_M(z, y, t) = H_{M, 1}(z, y, t)
\label{psiMH}
\end{eqnarray}
with
\begin{equation}
f_M(x, t) = \psi_M(x^k,1,t) = H_{M,1}(x^k,1,t)
\label{fpsiHM}
\end{equation}
By Observation $4$, $P^{(k)}_k$ is nc-closed, so one can
use the inverse NCPT Eq. (\ref{nctraninvt}) to get the g.f.
$a(x, t)$, enumerating the irreducible partitions of $P^{(k)}_k$ according to
the number of elements and of blocks, from $f(x, t)$
(actually we found it faster to compute $a(x, t)$ solving iteratively
Eq. (\ref{nctrant}) ).
It is easy to see that similarly $P^{M, (k)}_k$, in which the block
sizes divided by $k$ are restricted to be in $M$, is nc-closed.
Splitting the set $M=\{\bar m \ge 1\}$ in two disjoint sets $M_1$ and $M_2$,
one gets in this way the g.f.s $a_{M_1}$ and $a_{M_2}$ 
of the irreducible partitions
with block lengths divided by $k$ respectively in set $M_1$ and $M_2$.
Calling $a$ the g.f. corresponding to $\{\bar m \ge 1\}$,
$a - a_{M_1} - a_{M_2}$ is the g.f. of the irreducible
partitions having block lengths divided by $k$ in both sets $M_1$ and $M_2$.
The latter g.f. has in general infinite terms; the only
exception is $M_1 = \{1\}$ and $M_2 = \{\bar m \ge 2\}$, in which case
$a = a_{M_1} + a_{M_2}$, since by Observation $3$ the only irreducible 
partition having a
block length equal to $k$ is $(1,\cdots,k)$, in which case
$a_{M_1}(x, t) = t x^k$.

Let $f(x, t)$ be the g.f. enumerating
$P^{(k)}_k$ according to the number of elements and blocks,
$f_{M_2}(x, t)$ the g.f. enumerating the partitions with 
block size $2k$ or larger.
From $a(x, t) = t x^k + a_{M_2}(x, t)$ and the NCPT Eq. (\ref{nctrant}) one gets
\begin{eqnarray}
&&f_{M_2}(x, t) = 1 + a_{M_2}(x f_{M_2}(x, t), t) \label{faM2a} \\
&&f(x, t) = 1 + t(x f(x, t))^k + a_{M_2}(x f(x, t), t)
\label{faM2}
\end{eqnarray}

$f(x) = A(x f(x))$ implies $f(\frac{x}{A(x)}) = A(x)$ \cite{speich94};
using this identity in Eq. (\ref{faM2a}), replacing $x$ with $x f(x, t)$
and then using Eq. (\ref{faM2}) one gets
\begin{equation}
f_{M_2}\Big(\frac{x}{1 - tx^k f(x,t)^{k-1}}, t\Big) = f(x,t) - tx^k f(x,t)^k
\label{ffM2}
\end{equation}
Let us remark that, while
we have obtained Eq. (\ref{ffM2}) in the case in which $f(x,t)$ is the
g.f. for $P^{(k)}_k$, in which $k \ge 2$,
this equation can be applied with $k=1$, in the case of a nc-closed 
set of partitions, containing the singleton partition $(1)$.
$a_{M_2}$ is the g.f. for an irreducible set of partitions
not including the singleton partition $(1)$, and $a_1(x) = t x$.
For $k=1$ Eq. (\ref{ffM2}) becomes
\begin{equation}
f(x, t) = \frac{1}{1-t x}f_{M_2}(\frac{x}{1-tx})
\label{ffM2chords}
\end{equation}
so that $f$ is the generalized binomial transform of $f_{M_2}$,
\begin{equation}
f(x, t) = \sum_{n \ge 0} x^n \sum_{h=0}^n \binom{n}{h} t^{n-h} [x^h]f_{M_2}(x,t)
\label{ffM2chords1}
\end{equation}
which can be obtained also using the Lagrange inversion formula.
Let us consider in particular the set of partitions with blocks of size $1$
or $2$.
In this case $f_{M_2}$ is the g.f. enumerating
partitions with blocks of size $2$, i.e. $M_2 = \{2\}$,
according to the number of elements and blocks;
$f$ is the g.f. enumerating partitions with blocks of size $1$ or $2$.
This equation has been given, for $t=1$, in \cite{gil},
with the identifications $1 + P(x^2) = f_{M_2}(x)$ and $Y(x) = f(x)$.

Returning to the enumeration of $P^{(k)}_k$,
one can compute $f(x,t)$ computing first $f_{M_2}(x, t)$
using Eqs. (\ref{psiM}, \ref{psiMH})
and then $f(x,t)$ from it using Eqs. (\ref{faM2a}, \ref{faM2}).
By Observation $3$, 
the polynomials $[x^{2j}]f_{M_2}(x, t)$ have degree $j$ in $t$,
while the polynomials $[x^{2j}]f(x, t)$ have degree $2j$, so it is faster
to enumerate $P^{(k)}_k$ in this way than using Eqs. (\ref{psiH1}, \ref{psiH}).

Using Sage \cite{sage} with polynomials over rational numbers, 
for $k=2$ we computed $a(x, t)$ and $f(x, t)$
till $j=120$ in half an hour on a desktop computer, 
storing $H_{j,\bar m}(y, t)$  
for $\bar m \le 10$, to avoid the repetition of their computation many times.
Till order $j=6$ one gets
\begin{equation}
a(x, t) = t x^{2} + t x^{4} + t x^{6} + (2t^{2} + t) x^{8} + (10t^{2} + t) x^{10} + (11t^{3} + 32t^{2} + t) x^{12}
\end{equation}
from which one sees that the lowest order in which it differs from
the g.f. $t \frac{x^2}{1-x^2}$ enumerating the
irreducible partitions of $NC^{(2)}(j)$ is $x^8$; at that order,
beyond the one-block partition $(1, 2, 3, 4, 5, 6, 7, 8)$, there are two
other irreducible partitions, the crossing partitions
$(1, 4, 5, 8)(2, 3, 6, 7)$ and $(1, 2, 5, 6)(3, 4, 7, 8)$, 
belonging to the same orbit in $P^{(2)}_2(4)$, so that
one gets the term $(2t^{2} + t) x^{8}$ in $a(x, t)$.
At order $x^{10}$ there are $10$ irreducible partitions 
in the orbit of $(1, 2, 5, 6, 7, 8) (3, 4, 9, 10)$ in $P^{(2)}_2(5)$, 
as we saw in last section; for higher orders there
is more than one orbit for each number of blocks.

In the simpler case in which one wants to enumerate partitions
irrespective of the number of blocks
one sets $t=1$ in Eqs. (\ref{psiH1}, \ref{psiH});
for $k=2$ we computed $\psi_j(x, 1)$ till $j=250$
in about the same time in which we computed $\psi_j(x, t)$ till order $j=120$ ; 
here are the first $12$ terms
\begin{eqnarray}
    f(x,1) &&= 1 + x^{2} + 3x^{4} + 12x^{6} + 57x^{8} + 303x^{10} + 1747x^{12} + 10727x^{14} + 69331x^{16} + 467963x^{18} + 3280353x^{20} + \nonumber \\
    &&    23785699x^{22} + 177877932x^{24}
\nonumber
\end{eqnarray}
and the corresponding irreducible terms
\begin{equation}
a(x,1) =  x^{2} + x^{4} + x^{6} + 3x^{8} + 11x^{10} + 44x^{12} + 204x^{14} + 1029x^{16} + 5562x^{18} + 31994x^{20} + 194151x^{22} + 1236251x^{24}
\nonumber
\end{equation}
We find
that $\frac{[x^{2j}]a(x,1)}{[x^{2j}]f_{M_2}(x,1)}$ is about $0.87$ for
$j=250$; 
a plot of this ratio against $j$ suggests that it tends to a value
between $0.9$ and $1$ for $j \to \infty$,
while $\frac{[x^{2j}]a(x,1)}{[x^{2j}]f(x,1)}$ appears to go to zero faster 
than $1/j$.
Therefore $\frac{[x^{2j}]f_{M_2}(x,1)}{[x^{2j}]f(x,1)}$
seems to go faster than linearly to zero for $j \to \infty$.

In Section II.B we considered the g.f. $f(x, s, t)$, obtained
from $a(x, t)$ using the NCPT Eq. (\ref{nctrant}).
In the large-$s$ expansion of $f(x, s, s^{-1}t)$,
the leading term is the g.f. enumerating $NC^{(k)}$,
the first order in the $\frac{1}{s}$-expansion can be obtained enumerating the 
irreducible partitions with one and two blocks and then using 
Eq. (\ref{nctrant}); the exact expression for this term is given in
the next Section.
We give here the case $k=2$ till $j=6$
\begin{eqnarray}
&&f(x, s, \frac{t}{s}) = 1 + t x^2 + (2 t^2 + t)x^4 + (5 t^3 + 6 t^2 + t)x^6 
+ (14 t^4 + 28 t^3 + 12 t^2 + t + 2 \frac{t^2}{s})x^8 + \nonumber \\
&&\big(42 \, t^{5} + 120 \, t^{4} + 90 \, t^{3} + 20 \, t^{2} + t + \frac{10 \, {\left(2 \, t^{3} + t^{2}\right)}}{s} \big)x^{10} + \nonumber \\
&&\big(132 \, t^{6} + 495 \, t^{5} + 550 \, t^{4} + 220 \, t^{3} + 30 \, t^{2} + t 
+ \frac{4 \, {\left(33 \, t^{4} + 36 \, t^{3} + 8 \, t^{2}\right)}}{s} + 
\frac{11 \, t^{3}}{s^{2}}\big)x^{12} + O(x^{14})
\label{f1s2a}
\end{eqnarray}

\subsection{Partitions in $P^{(k)}_k$ with block size $k \bar m$}

Let us consider the case in which all the blocks have
the same size $k \bar m$. In this case one can neglect the $t$ variable,
since the number of blocks in a partition of size $j k$ is $\frac{j}{\bar m}$.
Then Eq. (\ref{psiM}) with $M = \{ \bar m \}$ gives, for $k \ge 2$,
\begin{equation}
\psi(z,y) = 1 + (z y)^{\bar m} H_{\bar m}(z, y)H_{\bar m}(z, 1)^{k-1}
\label{psiH1mb}
\end{equation}
where $H_{\bar m}$ is given in Eq. (\ref{psiMH}).
For $y=1$ Eq. (\ref{psiH1mb}) gives, using Eq. (\ref{fpsiH})
\begin{equation}
f(x) = 1 + z^{\bar m} H_{\bar m}(z,1)^k
\label{psiH1mby1}
\end{equation}
For $\bar m=1$, from Eqs. (\ref{fpsiH}, \ref{psiH1mby1}) one gets 
\begin{equation}
f(x) = 1 + x^k f(x)^k
\label{feqk}
\end{equation}

This can be seen also from Corollary $2$:
the case $\bar m = 1$, in which the blocks have size $k$, 
describes the noncrossing partitions of $kl$ with blocks of size $k$,
whose set is $NC_k(l)$;
in $NC_k(l)$ the only irreducible partition is $(1,\cdots,k)$,
so that the g.f. for its irreducible partitions is 
$a(x) = x^k$ and Eq. (\ref{feqk}) is the corresponding NCPT.

The case $k=2$ is well known: Eq. (\ref{feqk}) is
the g.f. of the Catalan numbers,
$[x^{2i}]f^{(2)}(x) = C_i$, the $i$-th Catalan number.
$f(x)$ counts closed walks on rooted trees,
in which each edge is traversed exactly twice (as in depth-first search). 
These are isomorphic
to ordered trees, which are isomorphic to noncrossing partitions \cite{proding};
the number of size-$n$ noncrossing partitions is the $n$-th Catalan number.

Doing a Taylor expansion around $y=0$ in Eq. (\ref{psiH1mb}) one gets
\begin{equation}
\psi(z,y) = 1 + \sum_{i \ge 1} (y z)^{\bar{m}i} H_{\bar m}(z,1)^{i(k-1)}
\prod_{j=1}^i \binom{j\bar m - 1}{\bar m - 1}
\label{psiH1mb2}
\end{equation}
from which, using Eq. (\ref{psiH1mby1}) one obtains
\begin{equation}
f(x) = 1 + \sum_{i \ge 1} x^{\bar{m} i} \big(f(x)-1)^{i\frac{k-1}{k}}
\prod_{j=1}^i \binom{j\bar m - 1}{\bar m - 1}
\label{psiH1mb2a}
\end{equation}

Let us see how the ratio 
$q_j = \frac{[x^{\bar m kj}]a(x)}{[x^{\bar m kj}]f(x)}$ 
between the number of irreducible partitions of size $\bar m k j$ and
the number of partitions of the same size varies with $\bar m$.
Let us consider the case $k=2$.
For $\bar m = 1$, there is a single irreducible partition of size $2$ for 
$j=1$
while from Eq. (\ref{feqk}) one sees that there are $C_j$ partitions of 
size $2j$, where $C_j$ is
the $j$-th Catalan number, so $q_j$ is zero for $j > 1$.
For $\bar m = 2$ Eq. (\ref{psiH1mb2a}) becomes
\begin{equation}
f(x) = 1 + \sum_{i \ge 1} x^{2i} (f(x)-1)^{i/2} (2i-1)!!
\label{psifmq2}
\end{equation}
which can be solved iteratively; however we found it faster to solve
iteratively Eq. (\ref{psiH1mb}).

Using Eq. (\ref{psiH1mb}) we computed $[x^{4j}]f(x)$ till $j=1000$.
Using the Pade' approximants
in the interval $[100, 1000]$
with polynomials of degree $(3,3),(4,4),(5,5)$ we get that
$\lim_{j\to \infty} |e^{-2} - q_j| < 4 \times 10^{-10}$

For $\bar m = 3$, we computed $[x^{6j}]f(x)$ till $j=400$.
Using the Pade' approximants
in the interval $[100,400]$ with polynomials of degree $(3,3),(4,4),(5,5)$ we get that
$\lim_{j\to \infty} |q_j - 1| < 2 \times 10^{-8}$.

In the case $k=3$, $q_j=0$ for $\bar m = 1$ and $j > 1$.
For $\bar m = 2$, using the Pade' approximants in the interval $[100, 500]$
with polynomials of degree $(3,3),(4,4),(5,5)$ we get that
$\lim_{j\to \infty} |e^{-3} - q_j| < 3 \times 10^{-9}$.
For $\bar m = 3$ the Pade' approximants indicate that $q_j \to 1$.

In the case $k=4$, again $q_j=0$ for $\bar m = 1$, $j > 1$;
Pade' approximants indicate 
that $q_j \to 1$ for $\bar m = 3$ for $j\to \infty$.
For $\bar m = 2$ we computed $[x^{8j}] f$ and $[x^{8j}] a$
till $j=175$; using Pade' approximants in the interval $[100, 175]$ 
with polynomials of degree $(3,3),(4,4),(5,5)$
we get $0.0183157(3)$, close to
$e^{-4} = 0.0183156388\cdots$.

These numerical results suggest that
$q_j \to e^{-k}$ for $k \ge 2$ and $\bar m = 2$,
while, for $\bar m > 2$, $q_j \to 1$.
We have however for $k > 3$ too little data to check this possibility
with good accuracy.

Observe that an analogous result is
obtained in the case $k=1$, $\bar m=2$; the set of partitions
in $P^{(1)}_1$ with blocks of size $2$ is the set of chords; the irreducible
partitions correspond to the linked chords;
$q_j$ is the ratio of
the number of irreducible $2$-equal partitions of size $2 j$ and that of
$2$-equal partitions of size $2 j$;
in that case it has been conjectured \cite{klei}, 
and then proved \cite{stein, stein2}, that $\lim_{j\to \infty} q_j = e^{-1}$.
In \cite{klei} it has been argued that more in general the proportion of
$k$-connected chords goes as $e^{-k}$.

\begin{figure*}[h]
\begin{center}
\epsfig{file=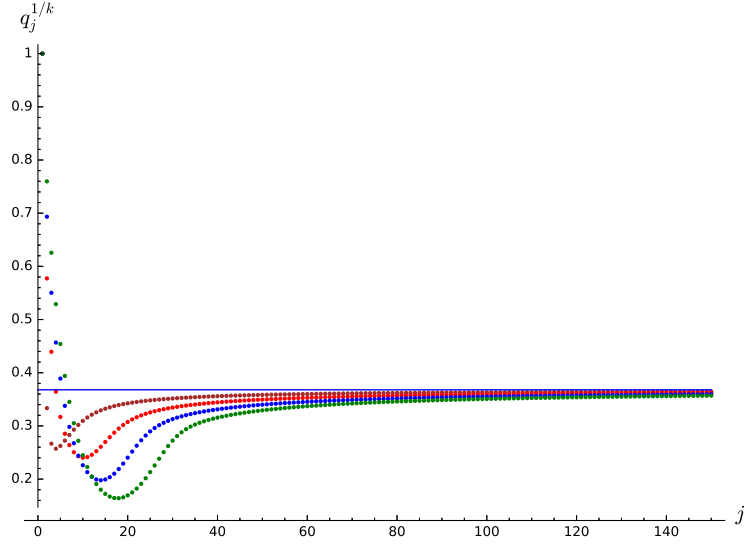, width=10.00cm}
    \caption{Plot of $q_j^{\frac{1}{k}}$ against $j$, 
    where $q_j$ is the ratio
    between the number of irreducible partitions of size $2 k j$ and
    the number of partitions of the same size
    for the set of partitions $P^{(k)}_k(2 j)$ with partitions
    having block size
    equal to $2 k$, for $k=1$ (brown), $2$ (red), $3$ (blue) and $4$ (green).
    The blue horizontal line has height $\frac{1}{e}$.
}
\label{FigLink}
\end{center}
\end{figure*}

$q_j^{\frac{1}{k}}$ is plotted in Fig. \ref{FigLink} against $j$ for
$k=1,2,3$ and $4$ till $j=150$.
In all cases $q_j^{\frac{1}{k}}$ decreases to a minimum, then it
approaches $e^{-1}$. In making the fits with the Pade' approximants
we used $q_j$ for $j \ge 100$, to avoid the region around the minimum.

\subsection{Enumeration of $P^{(k)}_k$ with given number of blocks. }
For $h$ small, one can compute exactly the generating
function enumerating the partitions in $P^{(k)}_k$ with $h$ blocks
according to the number of elements and returns to the root of 
the corresponding closed walks.
Let us expand Eq. (\ref{psiH1z}) in powers of $t$, defining
$H_{\bar m,h}(z,y) = [t^h]H_{\bar m}(z,y,t)$ for $h > 0$
and $H_{\bar m,0}(z,y) = 1$.
We use $z = x^k$.
One gets, for $h \ge 1$,
\begin{equation}
H_{1,h}(z,y) = \sum_{\bar m\ge 1} (zy)^{\bar m} 
\sum_{\substack{h_1,\cdots, h_k \ge 0 \\  h_1+\cdots+h_k=h-1}}
H_{\bar m,h_1}(z,y) H_{\bar m,h_2}(z,1)\cdots H_{\bar m,h_k}(z,1)
\label{bgx1a}
\end{equation}
$H_{1,h}(x^k,y)$ is the g.f. enumerating the closed walks covering rooted
$k$-gon trees with $h$ $k$-gons, according to the length of the walk
and the number of returns to the root.

Eq. (\ref{psiHz}) becomes
\begin{equation}
H_{\bar m,h}(z,y) = \sum_{m\ge 1} y^m [y^m] H_{1,h}(z,y) \binom{m+\bar m-1}{m}
\label{bgxma}
\end{equation}
and Eq. (\ref{fpsiH}) gives
\begin{equation}
[t^h]f(x,t) = H_{1,h}(z,1)
\label{fH1h}
\end{equation}

One has
\begin{equation}
H_{1,1}(z, y) = \sum_{\bar m \ge 1}(zy)^{\bar m} = \frac{z y}{1-z y}
\label{H11}
\end{equation}
so that 
\begin{equation}
[t]f(x, t) = H_{1,1}(z, 1) = \frac{z}{1-z}
\label{tfk}
\end{equation}
Using the identity
\begin{equation}
\frac{1}{(1-x)^m} = \sum_{j \ge 0} x^j \binom{m+j-1}{j}
\label{idbin}
\end{equation}
from Eq. (\ref{H11}) one gets
\begin{equation}
H_{\bar m,1}(z, y) = \frac{1}{(1 - y z)^{\bar m}} - 1
\label{Hm1}
\end{equation}

For $h=2$ Eqs. (\ref{bgx1a}, \ref{Hm1}) give
\begin{equation}
H_{1,2}(z,y) = \sum_{\bar m \ge 1} (y z)^{\bar m}(H_{\bar m,1}(z,y)
+ (k-1)H_{\bar m,1}(z,1)) = 
\frac{y z}{1 - 2 y z} - \frac{k y z}{1 - y z} + \frac{(k-1) y z}{1 - z - yz}
\nonumber
\end{equation}
so that
\begin{equation}
[t^2]f(x, t) = H_{1,2}(z, 1) = \frac{k z^2}{(1 - 2 z)(1 - z)}
\label{t2fk}
\end{equation}
From now on we will consider only the case $k=2$.
Let us prove by induction that
\begin{equation}
H_{1,h}(z,y) = \sum_{i > 0} \frac{p_{h,i}(z)y}{a_{h,i}(z) - b_{h,i}(z) y}
\label{pab}
\end{equation}
where $a_{h,i}(z)$ and $b_{h,i}(z)$ are polynomials in $z$.
This is true at first order, with $p_{1,1}=z$, $a_{1,1}=1$ and
$b_{1,1}= z$.

From Eqs. (\ref{bgxma},\ref{idbin},\ref{pab}) one gets
\begin{equation}
H_{\bar m,h}(z,y) = \sum_{i > 0} \frac{p_{h,i}(z)}{b_{h,i}(z)}\big( 
\big(\frac{a_{h,i}(z)}{a_{h,i}(z)-b_{h,i}(z) y}\big)^{\bar m} - 1\big)
= \sum_{i \ge 0} c_{h,i}(z) \big(\frac{a_{h,i}(z)}{a_{h,i}(z)-b_{h,i}(z) y}\big)^{\bar m}
\label{Hcab}
\end{equation}
where in the last sum $c_{h,i}(z) = \frac{p_{h,i}(z)}{b_{h,i}(z)}$
for $i > 0$ and
$c_{h,0}(z) = - \sum_i \frac{p_{h,i}(z)}{b_{h,i}(z)}$, with
$a_{h,0}=1$ and $b_{h,0}=0$.

Substituting this equation in Eq. (\ref{bgx1a}) one obtains
\begin{equation}
H_{1,h}(z, y) = \sum_{i,j} \sum_{h_1+h_2=h-1}
    \frac{c_{h_1,i} c_{h_2,j} a_{h_1,i}(z) a_{h_2,j}(z)z y}{a_{h_1,i}(z)(a_{h_2,j}(z)-b_{h_2,j}(z)) -
\big(b_{h_1,i}(z) (a_{h_2,j}(z)-b_{h_2,j}(z)) + a_{h_1,i}(z)a_{h_2,j}(z)z\big)y}
\label{H1cab}
\end{equation}
which has again the form of Eq. (\ref{pab}), with
\begin{eqnarray}
&&p_I(z) = a_{h_1,i}(z) a_{h_2,j}(z) c_{h_1,i}(z) c_{h_2,j}(z) z \label{gr1a} \\
&&a_I(z) = a_{h_1,i}(z)(a_{h_2,j}(z) - b_{h_2,j}(z)) \label{gr1b} \\
&&b_I(z) = b_{h_1,i}(z)(a_{h_2,j}(z) - b_{h_2,j}(z)) + za_{h_1,i}(z) a_{h_2,j}(z)
\label{gr1c}
\end{eqnarray}
ending the proof by induction.
In this way we computed $H_{1,h}$ till $h=10$,
as a sum of terms in the representation of Eq. (\ref{pab}),
each term represented by a triplet $[p(z), a(z), b(z)]$ corresponding to 
$\frac{p(z) y}{a(z)-b(z) y}$. At the end, one sets $y=1$, so the
poles are in $a(z) - b(z) = 0$ or in a factor $b_j(z)$ in $p(z)$.

In all the cases considered, i.e. till $h=10$, we found that, 
while $\frac{p_I(z)}{b_I(z)}$
is generally $z$-dependent, after collecting terms one gets that, 
in a term $[p,a,b]$,
$c = \frac{p(z)}{b(z)}$ is a rational number, so that from Eq. (\ref{gr1a})
it follows that also $p(z)$ is a polynomial in $z$; therefore in all these
cases the poles of $[t^h]f(x, t) = H_{1,h}(x^2, 1)$ are in the terms
$a(z)-b(z)$ in the addends $\frac{p(z)}{a(z)-b(z)}$ of $H_{1,h}(z,1)$.

Let us indicate terms $\frac{p y}{a-b y}$ 
of $H_{1,h}$ in Eq. (\ref{pab}) by $[p,a,b]$,
the terms of $c\big(\frac{a}{a-b}\big)^{\bar m}$ of $M_h \equiv H_{\bar m, h}$
in Eq. (\ref{Hcab}) by $(c,a,b)$.
One has $M_0 = (1,1,0)$; $H_{1,1} = [z,1,z]$ and $M_1 = (1,1,z) + (-1,1,0)$.

Eq. (\ref{H1cab}) can be written symbolically as
$H_{1,h} = \sum_{h_1+h_2=h-1} M_{h_1}\cdot M_{h_2}$, where the dot operation
is, according to Eqs. (\ref{gr1a}-\ref{gr1c})
\begin{equation}
(c_1,a_1,b_1)\cdot (c_2,a_2,b_2) = [a_1a_2c_1c_2 z,
a_1(a_2-b_2), b_1(a_2-b_2) + z a_1 a_2]
\label{gr1d}
\end{equation}
One gets 
\begin{equation}
H_{1,2} = M_0\cdot M_1 + M_1\cdot M_0 = [z,-z+1,z] + [-z,1,z] + [z,1,2z] + 
[-z,1,z] = [z,-z+1,z] + [-2z,1,z] + [z,1,2z] 
\nonumber
\end{equation}
Here the single contributions to $H_{1,2}$ have the form $[p_i, a_i, b_i]$
with $\frac{p_i}{b_i}$ $z$-independent even before collecting terms.
From $H_{1,2}$ one gets
\begin{equation}
M_2 = (1,-z+1,z) + (-2,1,z) + (\frac{-1}{2},1,2z) + (\frac{1}{2},1,0)
\nonumber
\end{equation}

$H_{1,3} = M_0\cdot M_2 + M_1\cdot M_1 + M_2\cdot M_0$; In this case there are
$12$ contributions; all of them apart from two have 
$\frac{p_i}{b_i}$ $z$-independent. The exceptions are 
$(1,1,z)\cdot (1,1,z) =[z,-z+1,-z^2+2z]$
coming from $M_1\cdot M_1$ and 
$(1,-z+1,z)\cdot (1,1,0) = [-z^2 + z,-z+1,-z^2+2z]$ coming from
$M_2\cdot M_0$; summing them one gets
$[p,a,b] = [-z^2 + 2z,-z+1,-z^2+2z]$ with $\frac{p}{b} = 1$.

In the following, instead of representing the terms 
$\frac{p(z) y}{a(z)-b(z) y}$ of $H_{1,h}(z,y)$ as
$[p(z),a(z),b(z)]$,
 we use shorter the representation $(c, a(z), b(z))$
where $c = \frac{p(z)}{b(z)}$ is a rational number for $h \le 10$.

We have computed $H_{1,h}(z,y)$ for $h \le 10$. 
We have already given $H_{1,1}$ in Eq. (\ref{tfk});
for $h = 2, 3, 4$ we list the triplets contributing to $H_{1,h}(z,y)$, the
representation of $[t^h]f(x, t) = H_{1,h}(z,1)$ as a ratio of polynomials and 
the first $10$ terms of the Taylor expansion, which correspond to the 
$h$-th column in Table $1$ in \cite{bau}.

  For $h=2$,
\begin{eqnarray}
&&(-2, 1, z), (\frac{1}{2}, 1, 2 z), (1, -z + 1, z) \nonumber \\
&&[t^2]f(x,t) =  \frac{2 z^{2}}{(2z - 1)(z - 1)} = \sum_{j\ge 1} (2^j-2) z^j =\label{t2fk2} \\
&&2z^{2} + 6z^{3} + 14z^{4} + 30z^{5} + 62z^{6} + 126z^{7} + 254z^{8} + 510z^{9} + 1022z^{10} + O(z^{11})
\end{eqnarray}
as in \cite{bau}.

For $h=3$,
\begin{eqnarray}
&&(1, -2 z + 1, -z^{2} + z), (\frac{1}{2}, -2 z + 1, z), (-3, -z + 1, z),
(\frac{1}{6}, 1, 3 z), (2, 1, z), (-\frac{3}{2}, 1, 2 z),
(1, -z + 1, -z^{2} + 2 z)   \nonumber \\
\nonumber \\
&&[t^3]f(x, t) = -\frac{{\left(8 \, z^{2} - 17 \, z + 5\right)} z^{3}}{{\left(z^{2} - 3 \, z + 1\right)} {\left(3 \, z - 1\right)} {\left(2 \, z - 1\right)} {\left(z - 1\right)}} = \nonumber \\
&& 5z^{3} + 28z^{4} + 110z^{5} + 375z^{6} + 1190z^{7} + 3628z^{8} + 10805z^{9} + 31740z^{10} + O(z^{11})
\label{t3fk2}
\end{eqnarray}
In \cite{bau} the following expression is given
\begin{equation}
[t^3 x^{2j}]f(x,t) = 3^{j-1} + \omega^j + {\bar \omega}^j - 3 \cdot 2^{j} + 2
\end{equation}
where $\omega$ and $\bar \omega$ are the roots of $z^2 - 3 z + 1$.
We checked agreement with our result for $[t^3]f(x, t)$ till order $100$
in powers of $z$.

For $h=4$,
\begin{eqnarray}
&& (-4, -2 z + 1, -z^{2} + z),
(\frac{1}{6}, -3 z + 1, z),
(\frac{9}{4}, 1, 2 z),
(-2, -2 z + 1, z),
(\frac{1}{2}, -3 z + 1, -2 z^{2} + z),
(\frac{9}{2}, -z + 1, z), \nonumber \\
&&(1, z^{2} - 3 z + 1, -2 z^{2} + z),
(\frac{1}{2}, -z + 1, -2 z^{2} + 3 z),
(1, z^{2} - 3 z + 1, -z^{2} + z),
(1, -2 z + 1, -3 z^{2} + 2 z), \nonumber \\
    &&(\frac{1}{2}, -2 z + 1, -2 z^{2} + 2 z),
(\frac{1}{24}, 1, 4 z),
(\frac{1}{2}, z - 1, -2 z),
(-\frac{4}{3}, 1, z),
(-\frac{2}{3}, 1, 3 z),
(-4, -z + 1, -z^{2} + 2 z) \nonumber \\
&&[t^4]f(x,t) = 
\frac{2 \, {\left(50 \, z^{4} - 158 \, z^{3} + 161 \, z^{2} - 59 \, z + 7\right)} z^{4}}{{\left(2 \, z^{2} - 4 \, z + 1\right)} {\left(z^{2} - 3 \, z + 1\right)} {\left(4 \, z - 1\right)} {\left(3 \, z - 1\right)} {\left(2 \, z - 1\right)} {\left(z - 1\right)}}
= \label{t4fk2} \\
&&14z^{4} + 120z^{5} + 682z^{6} + 3248z^{7} + 14062z^{8} + 57516z^{9} + 
227030z^{10} + O(z^{11})
\end{eqnarray}

All the poles of $[t^h]f(x, t) = H_{1,h}(z,1)$ have the form $a-b$,
where $(a,b)$ can be generated from $(1,0)$ using the product rule
\begin{equation}
(a_1,b_1)\dot (a_2, b_2) = (a_1(a_2-b_2), b_1(a_2-b_2) + z a_1 a_2)
\label{prod}
\end{equation}
obtained from Eq. (\ref{gr1d}) dropping the first entry of the triplet.
This product is neither associative nor commutative. 

For example, applying $(1,0)$ to the right $n$ times to $(a,b)$, one gets
$(a, b + n z)$, giving in particular for $a=1$ and $b=0$ the poles $1-n z$.

\subsection{Counting partitions of $P^{(k)}_k$ with CC types}
Let $J_{j,l_1,\cdots,l_k,m,a}$ be the number of closed walks on rooted
$k$-gon trees with $j k$ steps,
$l_1,\cdots, l_k$ edges with CC respectively $1,\cdots,k$, and
$m$ returns to the root with CC $a$. The CC type of a closed
walk is the CC type of the corresponding partition of $P^{(k)}_k$.
The number of closed walks with root of CC $a$ with $kj$ steps,
$l_1,\cdots, l_k$ edges with CC respectively $1,\cdots,k$ is given by
\begin{equation}
J_{j,l_1,\cdots,l_k,a} = \sum_{m=0}^k J_{j,l_1,\cdots,l_k, m,a}
\end{equation}

The g.f. for the number of partitions in $P^{(k)}_k(j)$
with first block of CC $a$ and with $l_b$ blocks having CC $b$ is
\begin{equation}
f_a(x, t_1,\cdots, t_k) = \sum_{j \ge 0} \sum_{l_1,\cdots,l_k \ge 0} 
t_1^{l_1}\cdots t_k^{l_k} J_{j,l_1,\cdots,l_k,a} x^{j k}
\label{f1Jc}
\end{equation}

Define
\begin{equation}
\psi_{j,m,a}(t_1,\cdots,t_k) = \sum_{l_1,\cdots, l_k \ge 0} J_{j,l_1,\cdots,l_k,m,a} t_1^{l_1}\cdots t_k^{l_k}
\end{equation}
$\psi_{0,0,a} = 1$, $\psi_{j,0,a} = 0$ for $j > 0$ and
$\psi_{j,m,a} = 0$ for $m > j$.

$\psi_{j,m,a}$ for $a \neq 1$ can be obtained from cyclic permutations
of the parameters from $\psi_{j,m,1}$
\begin{equation}
\psi_{j,m,a}(t_1,\cdots,t_k) = \psi_{j,m,1}(t_a, \cdots, t_{a-1\,mod\,k})
\end{equation}

A closed walk starts at $r_1$, with CC $a$, 
goes to $r_2$ with CC $a+1\,mod\,k$ along
the first edge of $\bar G$, there it can enter into the $k$-gon tree $T_2$, 
and so on; the last $k$-gon tree visited is $T_1$, at the node $r_1$.
The walk can arrive to $r_b$ from $T_b$ or from $\bar G$; using $\bar G$
and $T_b$ as objects, one has strings of $m_b + \bar m$ objects
with $\bar m$ $\bar G$'s, the rest with $T_b$;
the first entry of a string is $\bar G$.
The number of returns at $r_1$ is $m = m_1 + \bar m$.
There are $\bar m - 1$ $\bar G$ to place in $m_b + \bar m - 1$ places,
so there is a combinatorial factor $\binom{m_b + \bar m - 1}{\bar m - 1}$
associated to $T_b$, for $b=1,\cdots,k$. 
So one gets the recursion relations
\begin{eqnarray}
    \psi_{j, m, a}(t_1,\cdots,t_k) &&= t_a \sum_{\bar m=1}^m
\sum_{\substack{j_1,\cdots, j_k \ge 0 \\ \sum_1^k j_i = j - \bar m}}
\psi_{j_1,m-\bar{m},a}(t_1,\cdots,t_k)
\binom{m-1}{\bar{m} - 1} \nonumber \\
&&\prod_{b=2}^k \sum_{m_b=0}^{j_b}
\psi_{j_b,m_b,a+b-1 \,mod\,k}(t_1,\cdots,t_k) \binom{m_b + \bar{m}-1}{\bar{m}-1}
\label{psijc}
\end{eqnarray}
for $j > 0$.
In the case $k=2$ we solved on a desktop computer
 the recurrence equations Eq. (\ref{psijc}) at order $j=30$ in few minutes.
As discussed in Section II.C, after computing
$f_1(x,t_1,\cdots,t_k)$ till a low order $j$ with the recursion equations
Eq. (\ref{psijc}), one can compute the corresponding 
g.f. for the irreducible partitions, solving iteratively
Eqs.(\ref{faf1}, \ref{nctrank}) with $s=1$, and then obtaining from 
Eq. (\ref{nctrank}) the g.f.
$f_1(x,s,t_1,\cdots,t_k)$; from that one gets
$f_1(x,s, \frac{t_1}{s},\cdots,\frac{t_k}{s})$, which for $s\to \infty$
gives the g.f. enumerating $NC^{(k)}$ according
to the number of elements and the CC types, computed till order $j$;
in the next section we will compute it exactly.

Let us give, in the case $k=2$,
$f_1(x,s, \frac{t_1}{s},\frac{t_2}{s}) = \sum_j x^{2j} f_{1, 2j}$
till order $j=5$, where
$f_{1, 2j} = [x^{2j}] f_1(x,s,\frac{t_1}{s},\frac{t_2}{s})$
in which, as discussed in subsection II.B, the terms in $\frac{1}{s^g}$
correspond to partitions with $b(p) - c(p) = g$.
\begin{eqnarray}
&&f_{1,2} = t_{1} \nonumber \\
&& f_{1,4} = t_{1}^{2} + t_{1} t_{2} + t_{1} \nonumber \\
&& f_{1,6} = t_{1}^{3} + 3 \, t_{1}^{2} t_{2} + t_{1} t_{2}^{2} + 3 \, t_{1}^{2} + 3 \, t_{1} t_{2} + t_{1} \nonumber \\
&& f_{1,8} = t_{1}^{4} + 6 \, t_{1}^{3} t_{2} + 6 \, t_{1}^{2} t_{2}^{2} + t_{1} t_{2}^{3} + 6 \, t_{1}^{3} + 16 \, t_{1}^{2} t_{2} + 6 \, t_{1} t_{2}^{2} + 6 \, t_{1}^{2} + 6 \, t_{1} t_{2} + t_{1} + \frac{t_{1}^{2}}{s} + \frac{t_{1} t_{2}}{s} \nonumber \\
&& f_{1,10} = t_{1}^{5} + 10 \, t_{1}^{4} t_{2} + 20 \, t_{1}^{3} t_{2}^{2} + 10 \, t_{1}^{2} t_{2}^{3} + t_{1} t_{2}^{4} + 10 \, t_{1}^{4} + 
50 \, t_{1}^{3} t_{2} + 50 \, t_{1}^{2} t_{2}^{2} + 10 \, t_{1} t_{2}^{3} +
20 \, t_{1}^{3} + 50 \, t_{1}^{2} t_{2} + 20 \, t_{1} t_{2}^{2} + 
10 \, t_{1}^{2} + \nonumber \\
&&10 \, t_{1} t_{2} + t_{1} + 
\frac{5 \, t_{1}^{3}}{s} + \frac{10 \, t_{1}^{2} t_{2}}{s} + \frac{5 \, t_{1} t_{2}^{2}}{s} + \frac{5 \, t_{1}^{2}}{s} + \frac{5 \, t_{1} t_{2}}{s}
\label{fsk2bc}
\end{eqnarray}
We checked till order $x^{18}$ that listing the partitions of
$P^{(2)}_2(j)$ till $j=9$ and computing the CC types and number of
components one gets the same result for $f_1(x,s, t_1, t_2)$
as using Eq. (\ref{psijc}).

\section{$NC^{(k)}$ with CC types and a Wishart-type random matrix model}

In the first subsection we enumerate $NC^{(k)}$ according to the number of
elements and of CC types; in the second subsection we
prove an isomorphism between these partitions and the pair partitions
occurring in the Wishart-type random matrix model with 
the product of $k$ Gaussian random matrices introduced in \cite{lecz, leczsa}.

\subsection{Enumeration of $NC^{(k)}(h)$ according to the CC types.}

Using the noncrossing partition transform one can enumerate the
partitions of $NC^{(k)}(l)$ according to the number of blocks:
in this case the irreducible components are the partitions with one
block of size multiple of $k$, so that
$a_h=1$ for $h\, mod\, k = 0$, else $a_h = 0$; the generating
function of the irreducible partitions is
\begin{equation}
a(x, t) = \frac{t x^k}{1 - x^k}
\end{equation}
and Eq. (\ref{nctrant}) with $s=1$ gives
\begin{equation}
f(x, t) = 1 + x^k f^k(f + t - 1)
\label{ban0}
\end{equation}
obtained in \cite{bani2,bani}.

Consider the g.f. $f_1(x, t_1,\cdots,t_k)$ enumerating $NC^{(k)}(h)$ according 
to the CC types;
a $k$-divisible noncrossing partition of $[hk]$ having $j_1$ blocks
of CC $1$, $\cdots$, $j_k$ blocks of CC $k$
contributes a term $x^{hk} t_1^{j_1}\cdots t_k^{j_k}$ to it.

The g.f. of the irreducible partitions is
\begin{equation}
a(x, t_1,\cdots,t_k) = t_1\frac{x^k}{1 - x^k}
\label{a1}
\end{equation}
Since all irreducible
partitions in $NC^{(k)}(h)$ have a single block, the monomials in
$f_1(x,s,t_1,\cdots, t_k)$, computed with Eq. (\ref{nctrank}),
have degree in $s$ equal to the total degree
in $t_1,\cdots, t_k$, so we can neglect the parameter $s$.

The g.f. $f_a$ corresponds to 
partitions with lowest element with CC $a$;
it is obtained from $f_1$
by cyclic permutation of $t_1,\cdots, t_k$, see Eq. (\ref{faf1}).
From Eqs. (\ref{nctrank}, \ref{a1}) one has
\begin{eqnarray}
f_a(x, t_1,\cdots,t_k) &&= 1 + t_a\frac{g(x, t_1,\cdots,t_k)}{1-g(x, t_1,\cdots,t_k)} 
\nonumber \\
\qquad g(x, t_1,\cdots,t_k) &&= x^k \prod_{a=1}^k f_a(x, t_1,\cdots,t_k)
\label{f1nck}
\end{eqnarray}
so that one gets
\begin{equation}
g(x, t_1,\cdots,t_k) (f_a(x, t_1,\cdots,t_k) + t_a - 1) = f_a(x, t_1,\cdots,t_k) - 1 
\label{gfc1}
\end{equation}
From Eq. (\ref{gfc1}) one gets $(f_a - 1)t_b = (f_b - 1)t_a$ for any
$a,b=1,\cdots,k$.
Therefore $f_a$ can be parametrized as
\begin{equation}
f_a(x, t_1,\cdots,t_k) = 1 + t_a z \phi(z, t_1,\cdots,t_k)
\label{fphi0}
\end{equation}
where we set $z = x^k$.
One gets from Eqs. (\ref{gfc1}-\ref{fphi0})
\begin{equation}
\phi = (1+ t_0 z \phi)\cdots (1+ t_k z \phi)
\label{fphi}
\end{equation}
with $t_0 = 1$.
Eq. (\ref{fphi}) can be obtained directly from Eqs. (\ref{psik0b}, \ref{a1})
with $\Phi(x, t_1,\cdots, t_k) = z \phi(z, t_1,\cdots, t_k)$ and 
$t_1 \alpha(x) = a(x, t_1,\cdots, t_k)$.

For symmetry reasons let us extend Eq. (\ref{fphi}) to have $t_0$ as
 a new variable (it is related to the parameter $s$ we neglected);
so in Eq. (\ref{fphi}) $\phi = \phi(z, t_0,\cdots, t_k)$ and
Eq. (\ref{fphi0}) becomes
$f_a(x, t_0, \cdots,t_k) = 1 + t_a z \phi(z, t_0,\cdots,t_k)$.
Using the Lagrange inversion formula one gets, for $h > 0$,
\begin{eqnarray}
[z^h]\phi(z,t_0,\cdots,t_k) &=& \frac{1}{h+1} [z^h]((1+t_0 z)\cdots(1+t_k z))^{h+1} \nonumber \\
&=& \frac{1}{h+1} 
\sum_{\substack{j_1,\cdots, j_k \ge 0 \\ j_0+\cdots+j_k=h}}
    \binom{h+1}{j_0}\cdots
\binom{h+1}{j_k}t_0^{j_0}\cdots t_k^{j_k}
\label{phi}
\end{eqnarray}
From Eqs. (\ref{fphi0}, \ref{phi}) one gets
\begin{equation}
[z^h]f_a(x, t_0,\cdots,t_k) =
 \frac{t_a}{h} 
\sum_{\substack{j_1,\cdots, j_k \ge 0 \\ j_0+\cdots+j_k=h-1}}
    \binom{h}{j_0}\cdots
\binom{h}{j_k}t_0^{j_0}\cdots t_k^{j_k}
\label{fad}
\end{equation}
Setting $t_0=1$ one obtains
the g.f. $f_1(x,1,t_1,\cdots,t_k)$,
enumerating $NC^{(k)}(h)$ according to the CC types.

$[z^h] f_1(x, t_0,\cdots,t_k)$ is a homogeneous polynomial in
$t_0,t_1,\cdots,t_k$, of total degree $h$.
The total degree in $t_1,\cdots,t_k$ of $[z^h] f_a$ is, from Eq. (\ref{fad}),
$1 + \sum_{i=1}^k j_i$. In a $k$-divisible noncrossing partition $p$,
to a block of CC $a$ a factor $t_a$ is associated, so to $p$ is associated
a total degree $b$ in the $t_1,\cdots,t_k$ variables, where $b$ is the number of blocks; so
$\sum_{i=1}^k j_i = b - 1$. Since the sums in Eq. (\ref{fad})
satisfy $j_0 = h-1-\sum_{i=1}^k j_i$, it follows that
$j_0 = h-b$ is the rank of the $k$-divisible noncrossing partition \cite{edel}.
The CC type of $p$ is 
$(j_1(p),\cdots, j_k(p))$, contributing to the term in the sums in
Eq. (\ref{fad}) with $j_1 = j_1(p) - 1$ and $j_a = j_a(p)$ for $a=2,\cdots,k$;
$j_0 = j_0(p)$, the rank of $p$.

In Appendix A.B an example is given for $k=3$.

For $t_1=\cdots = t_k = t$ and $t_0=1$ one gets Eq. (\ref{ban0}) from 
Eqs. (\ref{fphi0},\ref{fphi}); $f(x, t) = f_1(x,t,\cdots,t)$.
In this case $[z^h]f$ is a Fuss-Narayana polynomial; we can obtain
it from Eq. (\ref{fad}) using the generalized Vandermonde identity,
\begin{equation}
[z^h]f =
\sum_{b=1}^h \frac{t^b}{h} \sum_{j_1+\cdots+j_{k}=b-1} \binom{h}{j_1}\cdots
\binom{h}{j_{k}}\binom{h}{b} = 
\sum_{b=1}^h \frac{t^b}{b} \binom{h-1}{b-1}\binom{k h}{b-1}
\label{fad1}
\end{equation}

$[t^b][z^h] f$ in Eq. (\ref{fad1})
is the number of $k$-divisible noncrossing partitions with
$h k$ elements and $b$ blocks \cite{edel}.

We have seen that, since $P^{(k)}_k$ is nc-closed,
the g.f. $f(x,s, \frac{t}{s})$,
enumerating $P^{(k)}_k$ according to the number of elements,
blocks and components of a partition, flows from the g.f.
enumerating $P^{(k)}_k$ for $s=1$, to the one enumerating
$NC^{(k)}$ for $s\to \infty$.
Using Eq. (\ref{ntransig1}) and Eqs. (\ref{tfk}, \ref{t2fk}) we get
the first order term in the $\frac{1}{s}$-expansion
\begin{equation}
[\frac{1}{s}]f(x, s, \frac{t}{s}) =
    \frac{k t^2 y^{4k}}{(1-y^k)(1-2y^k)\big((1-y^k)^2 - k t x y^{k-1}\big)}
\label{f1s2b}
\end{equation}
where $\frac{y}{x} = \lim_{s \to \infty} f(x, s, \frac{t}{s})$ 
is the g.f. enumerating $NC^{(k)}$ according to the
number of elements and of blocks; it satisfies Eq. (\ref{ban0}), 
with series expansion given in Eq. (\ref{fad1})
\begin{equation}
y = x \lim_{s \to \infty} f(x, s, \frac{t}{s}) =
x + x\sum_{h\ge 1} x^{2h} \sum_{b=1}^h \frac{t^b}{b} \binom{h-1}{b-1}\binom{k h}{b-1}
\label{f1s2b1}
\end{equation}
We checked Eq. (\ref{f1s2b}) using the recursive equations 
Eqs. (\ref{psiH1}, \ref{psiH}) and the NCPT Eq. (\ref{nctrant}) 
in the case $k=2$ till $x^{60}$;
in the cases $k=3,4$ and $5$ listing partitions,
respectively till $x^{21}$, $x^{24}$ and $x^{30}$.

\subsection{Moments of a Wishart-type random matrix model with the
product of $k$ Gaussian random matrices}

This subsection is based on the approach in \cite{leczsa}, in
which they discuss in detail the set of pair partitions used to compute
the limit moments;
we do not deal directly with the operator approach
used in \cite{lecz} and \cite{leczsa2} to compute them.

In \cite{lecz, leczsa} they considered a Wishart-type product 
$B(n)=X_1\cdots X_k$ 
of $k$ independent $N_{j-1} \times N_j$ complex Gaussian random matrices; 
the moments of $B(n)B(n)^*$ are evaluated in the limit $n\to \infty$,
with $d_j = \frac{N_j}{n}$ fixed; $d_0,d_1,\cdots,d_k$ are called 
asymptotic dimensions; the $h$-th moment is
\begin{equation}
P_h(d_0,\cdots,d_k) = \lim_{n\to \infty} \tau_0(n) (B(n)B(n)^*)^h
\label{GP}
\end{equation}
where $\tau_0(n)$ is the expectation of the trace over the set of $N_0(n)$ 
basis vectors.

They prove that 
the $h$-th limit moment $P_h(d_0,\cdots,d_k)$ is the generating polynomial for 
the number of noncrossing pair partitions of $[2hk]$, defined in this way:
consider the strings
\begin{eqnarray}
W_0 &&= 1\cdots k k^*\cdots 1^* \nonumber \\
W_i &&= i^*\cdots 1^* 1\cdots k k^*\cdots (i+1)^*, \quad i=1,\cdots,k
\label{Wi}
\end{eqnarray}
and the strings $W_i^h$, with $i=0,\cdots,k$, that is the string $W_i$
repeated $h$ times. Consider all the pairings
where all the elements $l$ in $W_i^h$ are paired with $l^*$ and viceversa; to a pairing
$(l, l^*)$ or $(l^*,l)$ associate a partition block $(a,b)$, where 
$a$ and $b$ are the positions of $l$ and $l^*$ in the string $W_i^h$.
These blocks form a pair partition of $[2hk]$. The set of the noncrossing
pair partitions thus defined is called $NC_{2hk}^2(W_i^h)$.
Denote by $r_l(\pi)$ the number
of pairings in $\pi \in NC_{2hk}^2(W_i^h)$ in which $l$ is on the right.
One has $h = r_0(\pi) \ge r_l(\pi) \ge r_{l+1}(\pi) \ge r_{k+1}(\pi) = 0$.
Define 
\begin{equation}
m_l(\pi) = r_{l+1}(\pi) - r_l(\pi) + h
\label{jr}
\end{equation}
which satisfies $0 \le m_l(\pi) \le h$, and
\begin{equation}
N_i(h, m_0,\cdots,m_k) = |\{\pi \in NC_{2hk}^2(W_i^h): m_l(\pi) = m_l, 
\quad l=0,\cdots,k \} |
\label{Ni}
\end{equation}
where $i=0,\cdots,k$.
The corresponding g.f.s are
\begin{equation}
{\cal N}_i(x, d_0,\cdots, d_k) = 1 + \sum_{h \ge 1} x^h 
\sum_{m_0,\cdots,m_k} N_i(h, m_0,\cdots,m_k) d_0^{m_0}\cdots d_k^{m_k} =
\sum_{h \ge 0} x^h \sum_{\pi \in NC^2_{2hk}(W_i^h)} d_0^{m_0(\pi)}\cdots d_k^{m_k(\pi)}
\label{Nigf}
\end{equation}
and one has
\begin{equation}
P_h(d_0,\cdots,d_k) = [x^h]{\cal N}_0(x, d_0,\cdots, d_k)
\label{GP1}
\end{equation}

In \cite{lecz, leczsa} they prove, for $i=0,\cdots,k$, that
\begin{eqnarray}
{\cal N}_i(x, d_0,\cdots,d_k) &&= 1 + \frac{1}{d_i}\sum_{h \ge 1} 
\frac{x^h}{h} \sum_{m_0 + \cdots + m_k = hk + 1}
\binom{h}{m_0}\cdots \binom{h}{m_k} d_0^{m_0}\cdots d_k^{m_k} \nonumber \\
&&= 1 + \frac{1}{d_i}\sum_{h \ge 1} \frac{(d_0\cdots d_k x)^h}{h}
\sum_{j_0 + \cdots + j_k = h-1}
\binom{h}{j_0}\cdots \binom{h}{j_k} d_0^{-j_0}\cdots d_k^{-j_k}
\label{pk2}
\end{eqnarray}
where $m_l = h - j_l$, so that $0 \le j_l \le h$.
Comparing Eqs. (\ref{fad}, \ref{pk2}) one gets
\begin{equation}
{\cal N}_a(t_0\cdots t_k x^k, t_0^{-1},\cdots,t_k^{-1}) = 
f_a(x, t_0,\cdots,t_k)
\label{pk2a}
\end{equation}
From Eqs. (\ref{Nigf}, \ref{pk2a})
\begin{equation}
[z^h] f_i(x,t_0,\cdots,t_k) = 
\sum_{\pi \in NC_{2hk}^2(W_i^h)} \prod_{b=0}^k t_b^{h-m_b(\pi)}
\nonumber
\end{equation}
Using Eq. (\ref{jr}) one gets
\begin{equation}
\sum_{p\in NC^{(k)}(h)} \prod_{i=0}^k t_i^{j_i(p)} = [z^h] f_1(x,t_0,\cdots,t_k) = \sum_{\pi \in NC_{2hk}^2(W_1^h)} 
\prod_{l=0}^k t_l^{r_l(\pi) - r_{l+1}(\pi)}
\label{far}
\end{equation}
where $j_0(p)$ and $(j_1(p),\cdots, j_k(p))$ are the rank and the CC type
of $p \in NC^{(k)}(h)$.
Eq. (\ref{far}) gives an isomorphism between $NC^{(k)}(h)$ and $NC_{2hk}^2(W_1^h)$.

In the following we construct a $1$-to-$1$ correspondence
between partitions in
$NC_{2hk}^2(W_1^h)$ and partitions in a set $S_{h,k} \subset NC_{2hk}^2(W_1^h)$, 
for $k \ge 2$,
with the following correspondence between the rank $j_0(p)$ and the CC type
$(j_1(p),\cdots,j_k(p))$ of $p \in NC^{(k)}(h)$
and the pairing type $r_i(\pi) - r_{i+1}(\pi)$, $i=0,\cdots, k$
of $\pi \in NC_{2hk}^2(W_1^h)$:
\begin{eqnarray}
&&r_i(\pi) - r_{i+1}(\pi) = j_i(p),  \quad i=0,1 \nonumber \\
&&r_i(\pi) - r_{i+1}(\pi) = j_{k+2-i\,mod\,k}(p) \quad i=2,\cdots,k
\label{paircc}
\end{eqnarray}
so that
\begin{equation}
\sum_{p\in NC^{(k)}(h)} \prod_{i=0}^k t_i^{j_i(p)} = 
 \sum_{\pi \in S_{h,k}} 
\prod_{i=0}^1 t_i^{r_i(\pi) - r_{i+1}(\pi)}
\prod_{l=2}^k t_{2+k-l\,mod\,k}^{r_l(\pi) - r_{l+1}(\pi)}
\label{far1}
\end{equation}
Comparing Eqs. (\ref{far}, \ref{far1}) and using the fact that
$f_1(x,t_0,\cdots,t_k)$ is symmetric in $t_2,\cdots,t_k$ it follows that
$S_{h,k} = NC_{2hk}^2(W_1^h)$, so that this is a $1$-to-$1$ correspondence
between partitions in $NC^{(k)}(h)$ and partitions in $NC_{2hk}^2(W_1^h)$.

While in the case of the $k$-divisible noncrossing partitions the
fundamental object is $f_1$, and the other $f_a$ are introduced to compute it,
in the case of $NC_{2hk}^2(W_a^h)$ the fundamental quantity is
${\cal N}_0$, the other ${\cal N}_i$ are introduced in its computation.

The existence of an isomorphism between $NC_{2hk}^2(W_1^h)$ and $NC^{(k)}(h)$
is not surprising; in \cite{alex} it has been shown that the limit moments
of the product of $k$ square matrices are Fuss-Catalan polynomials;
in \cite{bani} they showed that the one-parameter Fuss-Narayana polynomials
occur in the free Bessel laws; these polynomials enumerate $NC^{(k)}$.
It is reasonable that, in the case of the product of $k$ rectangular random
matrices, one obtains again limit moments which enumerate 
$NC^{(k)}(h)$ in the presence of $k$ parameters, the asymptotic
dimensions; these are expressed in terms of generalized 
Fuss-Narayana polynomials \cite{lecz, leczsa}.

Let us define the {\it list representation} $a$ of a partition  
$p \in NC^{(k)}(h)$   as the sequence
$(1,\cdots,k)$ repeated $h$ times, with a $[$ before the $i$-th element
of $a$, if $i$ is the first element of a block and a $]$ after the
 $i$-th element of $a$, if $i$ is the last element of a block.

For instance to the $2$-divisible noncrossing partition $(1,2,7,8)(3,6)(4,5)$
corresponds the list representation
$[1,2,[1,[2,1],2],1,2]$.

From the list representation $a$ one obtains the partition $p \in NC^{(k)}(h)$
assigning to a block the positions of the elements belonging to a sublist 
$b$ of $a$, which are not in a sublist nested in $b$.

Let us then construct the {\it intermediate string representation} $s'$ from the
list representation $a$ of a partition $p \in NC^{(k)}(h)$ in the following
way: 
to each $i = 2,\cdots,k$ substitute $(k+2-i)^*$ respectively; to each
$1$ substitute $1^*1\cdots k$, so that $1\cdots k$ is mapped to
$1^*1\cdots k k^* \cdots 2^* = W_1$;
to each $[$ substitute a $\langle$; to each $]$ substitute a $\rangle$.

Viceversa, given an intermediate string representation $s'$,
one obtains the list representation in this way:
substitute $\langle$ with $[$, $\rangle$ with $]$,
$i^*$ with $k+2-i$ for $i=2,\cdots,k$ 
and the substring $1^*1\cdots k$ with $1$.

Define a {\it string representation} $s$ as a string obtained from
an intermediate string representation $s'$, in which each occurrence of
the substring $1\cdots k k^*\cdots 1^*$
(neglecting nested sub-brackets and the elements within) is substituted by
$S_0 \equiv \langle 1\cdots k k^*\cdots 1^*\rangle$.
Viceversa, given a string representation, the corresponding
intermediate string representation is obtained substituting all the
substrings $\langle 1\cdots k k^*\cdots 1^*\rangle$ with
$1\cdots k k^*\cdots 1^*$ (also here nested sub-brackets are neglected).

Each well-parenthesised substring $\langle \cdots \rangle$ of a
string representation $s$
is consists in $2k$ elements at the outermost parenthesis level
and in nested $\langle \cdots \rangle$ substrings, with $2k$ elements
within each pair of parenthesis, apart the nested substrings; in fact
if there were more than $2k$ elements in such a substring, 
it would contain a subsequence 
$1\cdots k k^*\cdots 1^*$,
but the latter cannot occur in a string representation by definition.
For instance the string representation
$\langle 1^* 1 2 \langle 2^* 1^* 1 2\rangle \langle 2^* 1^* 
\langle 1 2 2^* 1^* \rangle 1 2 \rangle 2^*\rangle$ has, reading from the
left, the substrings $\langle1^* 1 2\langle \cdots \rangle 2^*\rangle$,
$\langle2^* 1^* 1 2\rangle$, $\langle2^* 1^* \langle \cdots \rangle 1 2\rangle$ 
and $\langle1 2 2^* 1^*\rangle$.

To a sublist $[i,\cdots]$ of the list representation (excluding
nested sublists)
corresponds $S_1 = \langle 1^* \cdots \rangle$ for $i=1$,
$S_{k+2-i\,mod\,k} = \langle (k+2-i\,mod\,k)^* \cdots \rangle
$ for $i =2,\cdots,k$; this sublist contributes
a monomial $t_i$ to the monomial giving the CC type;
if a sublist $[i,\cdots]$ contains $k m$ elements with $m > 1$, in
the string representation there is correspondingly a factor $S_0^{m-1}$.
Both the sublist and $S_0^{m-1}$ produce a factor $t_0^{m-1}$.
The  substring $S_i$ contributes a factor 
$\prod_{i=0}^1 t_i^{r_i(\pi) - r_{i+1}(\pi)}
\prod_{l=2}^k t_{2+k-l\,mod\,k}^{r_l(\pi) - r_{l+1}(\pi)}$, according to
Eq. (\ref{far1}).
The substrings $S_i$ and the corresponding monomials are
\begin{eqnarray}
&&S_0 = \langle 1\cdots k k^* \cdots 1^*\rangle \quad t_0 \nonumber \\
&&S_1 = \langle 1^* 1\cdots k  k^* \cdots 2^* \rangle \quad t_1 \nonumber \\
    &&S_i = \langle i^* \cdots k  k^* \cdots (i+1)^* \rangle \quad t_{k+2-i\,mod\,k},\quad 
i=2,\cdots,k-1 \nonumber \\
&&S_k = \langle k^*(k-1)^* \cdots 1^*1\cdots k \rangle  \quad t_2
\label{Ws}
\end{eqnarray}
Hence to a sublist $[i,\cdots]$, contributing $t_i$, corresponds,
for $i = 0, 1$,  the substring $S_i$ contributing the same;
for $2 \le i \le k$ it corresponds $S_{k+2-i\,mod\,k}$, contributing again
$t_i$.

Making these contractions in all the substrings of this kind,
all the elements of the string representation are contracted;
associate to the result a pair partition, in which each pair corresponds
to the positions of the elements forming a contraction.
The contractions in one of these substrings do not cross the contractions
of nested substrings. The contractions within one of these
substrings $S_i$ do not cross; this is obvious in the case of $S_0$;
for $i > 0$, $S_i$ is obtained from $S_{i-1}$ moving $i^*$ from the
right end to the left end of the substring; it is clear that this
change does not introduce crossings (a similar
bijection $\phi$ has been defined on $NC^2_{2hk}$ in \cite{leczsa}).
Therefore all these substrings
do not have crossings, hence the pair partition is noncrossing.
Let us call $S_{h,k}$ the set of pair partitions obtained from
$NC^{(k)}(h)$ using the above described correspondence;
it is a subset of $NC^2_{2kh}(W_1^h)$.

This sequence of representations defines a mapping $\iota$
from $NC^{(k)}(h)$ to $S_{h,k}$. To prove that it is an isomorphism
let us construct $\iota^{-1}$.

In the string representation, in the positions $i,i+1,\cdots,i+k-1$ 
for a given $i = 2\, mod\,2k$ are placed the elements $1,\cdots,k$
coming from the substitution of the element $1$ of the list representation,
which is in position $1\, mod\,k$ of the list representation,
with $1^* 1 \cdots k$. The pairings of these
$k$ elements identify the $2k$ elements within a pair of brackets in
the string representation (excluding nested brackets).

Therefore, to get $\iota^{-1}$
from a partition $\pi \in S_{h,k}$, group the pairs
containing the elements $i,i+1,\cdots,i+k-1$ for $i = 2\, mod\,2k$; 
these collections of pairs contain $2k$ elements.
Adding, for each of these groups of $2k$ elements,
$\langle$ to the string $W_1^l$ before the position given by the
smallest element of each collection of pairs, $\rangle$ after the
position of the largest element, the resulting string is the string
representation of $\pi$. From it one obtains $p \in NC^{(k)}(h)$.

As said after Eq. (\ref{far1}), from the fact that $\iota$ is a bijection
from $NC^{(k)}(h)$ to $S_{h,k}$ it follows that $S_{h,k} = NC^2_{2kh}(W_1^h)$.

For instance in the case $k=2$ we give a partition $p \in NC^{(2)}(4)$,
its list representation, the corresponding intermediate string representation,
the string representation
and the partition $\pi \in NC^2_{16}(W_1^4)$; from each of them
the following representation is immediately obtained
\begin{eqnarray}
&&p = (1,8)(2,3)(4,5,6,7); \quad [1,[2,1],[2,1,2,1],2] ; \quad
\langle 1^* 1 2 \langle 2^* 1^* 1 2\rangle \langle 2^* 1^* 1 2 2^* 1^* 1 2 \rangle 2^*\rangle;
    \nonumber \\
&&\langle 1^* 1 2 \langle 2^* 1^* 1 2\rangle \langle 2^* 1^* 
\langle 1 2 2^* 1^* \rangle 1 2 \rangle 2^*\rangle
; \quad \pi = (1,2)(3,16)(4,7)(5,6)(8,15)(9,14)(10,13)(11,12)
\nonumber
\end{eqnarray}
Let us see how one obtains the string representation from $\pi$ in this
example.
The elements $2,3$ of $\pi$ are paired with $1,16$ identifying
the positions of the elements of the substring
$\langle 1^* 1 2 \langle \cdots \rangle 2^*\rangle$ in the string 
representation; the elements $6,7$
of $\pi$ are paired with $4,5$ identifying the brackets for other $4$
elements: $\langle 1^* 1 2 \langle 2^* 1^* 1 2\rangle\cdots \rangle 2^*\rangle$
and so on. The monomial associated to $p$ is $t_0 t_1 t_2^2$, as can be
seen either counting the blocks of each $CC$ in $p$ or from the product of the
monomials associated to the substrings in Eq. (\ref{Ws}).
In Appendix A.C other examples of this $1$-to-$1$ correspondence are given.

Here is an example for $k=3$;
we give a partition $p \in NC^{(3)}(3)$,
its list representation, the corresponding intermediate string representation,
the string representation
and the partition $\pi \in NC^2_{18}(W_1^3)$; from each of them
the following representation is immediately obtained
\begin{eqnarray}
&&p = (1,2,3,4,5,9)(6,7,8) ; \quad [1,2,3,1,2,[3,1,2],3] \quad
\langle 1^* 1 2 3 3^* 2^* 1^* 1 2 3 3^* \langle 2^* 1^* 1 2 3 3^* \rangle
2^* \rangle; \quad
\langle 1^* \langle 1 2 3 3^* 2^* 1^* \rangle 1 2 3 3^* 
\langle 2^* 1^* 1 2 3 3^* \rangle 2^* \rangle \nonumber \\
&&\pi = (1,8)(2,7)(3,6)(4,5)(9,18)(10,11)(12,15)(13,14)(16,17)
\nonumber
\end{eqnarray}
$p$ has monomial $t_0 t_1 t_3$; the string representation has the
substrings $S_1$, $S_0$ and $S_2$, giving respectively $t_1$, $t_0$ and
$t_3$, so $\pi$ has the same monomial.
Starting from $\pi$, the first group is obtained grouping 
the pairs containing $2,3,4$; the second from the pairs containing $8,9,10$;
the third from the pairs containing $14,15,16$; these groups
identify the positions of the brackets in the string representation,
so $\iota^{-1}(\pi)$ is determined.

\section{Adjacency random matrix models}
In the first subsection we study a $d$-dimensional Adjacency random block 
matrix model on BB graphs with degrees $Z_1$ and $Z_2$
and on bipartite ER graphs with average degrees $Z_1$ and $Z_2$.
In the second subsection we study the noncrossing partition flow of the 
Adjacency random matrix model, which in the limit $s\to \infty$
has the same g.f. of the moments as the models in the first subsection,
in the limit $d\to \infty$ with $\frac{Z_a}{d}$ fixed.

\subsection{Adjacency random block matrix model}
In \cite{CZ} an Adjacency random block matrix model with $d$-dimensional
blocks has been introduced on ER graphs.

Let $X_{i,j}$ be an order-$d$ matrix, with $i \neq j$, defined for $d > 1$
as  $X_{i,j} = |\hat v_{i,j}><\hat v_{i,j}| = X_{j,i}$, where, for $i < j$,
$\hat v_{i,j}$ are $d$-dimensional versors, which are
random independent uniformly distributed on the unit sphere in 
$d$ dimensions; in $d=1$, $X_{i,j}$ reduces to $1$.

Define the Adjacency random block matrix
\begin{equation}
A_{i,j} = \alpha_{i,j} X_{i,j}
\label{AalX}
\end{equation}
for $i \neq j$ and $0$ otherwise;
$\alpha_{ij}$ is an element of the Adjacency matrix of a
sparse graph with $N$ nodes.
In the case of ER graphs, the elements $\alpha_{i,j}$ are 
i.i.d. random variables with the probability law
\begin{equation}
    P(\alpha) = \frac{Z}{N}\delta(\alpha - 1) +
(1 - \frac{Z}{N})\delta(\alpha)
\label{erlaw}
\end{equation}
In the large-$N$ limit only size-$2l$ closed walks on rooted tree 
graphs contribute to the $2l$-th moment of the Adjacency random block matrix, 
as in the $d=1$ case studied in \cite{bau}. 
As we saw in Section III, these walks are in $1$-to-$1$ 
correspondence with the set of partitions $P^{(2)}_2(l)$.
In $d=1$ these walks contribute with weight $Z^E$, where $E$ is the
number of distinct edges in the walk. Therefore in $d=1$ the $2l$-moment
is obtained enumerating the partitions in $P^{(2)}_2(l)$ with $E$ blocks, 
studied in \cite{bau} and in Section IV.
This result has been obtained recently in \cite{bose}, where $P^{(2)}_2(l)$
is called the set of special symmetric partitions $SS(2l)$.

For $d > 1$, beyond the factor $Z^E$, the weight of a closed walk contains
also the contribution given by the average of the trace of the blocks 
$X_{i,j}$ associated to the edges of the walk, divided by $d$.
In the limit $d\to \infty$ with $t = \frac{Z}{d}$ fixed, it has been
proven in \cite{PC} that only walks on trees with noncrossing sequence of steps
contribute to the moments of the Adjacency random block matrix,
and that a walk with $E$ distinct edges gives a contribution $t^E$.
From Corollary $1$ in Section IV it follows that the g.f. of the $2l$-th moment
is the g.f. enumerating the partitions in $NC^{(2)}(l)$ 
according to the number
of blocks, which satisfies Eq. (\ref{ban0}) for $k=2$, i.e. the
distribution of the Adjacency random matrix model in the EM approximation
on ER graphs \cite{SC}; this result has been proven in \cite{PC}
using the decomposition of the closed walks in primitive walks.

Consider now the Adjacency random block matrix Eq. (\ref{AalX}) on
random regular graphs with degree $Z$. In the large-N limit again
only length-$2l$ closed walks on rooted tree graphs contribute to the
$2l$-th moment of the Adjacency random block matrix.
In the limit $d\to \infty$ with $t = \frac{Z}{d}$ fixed, it is easy to see,
by an argument similar to the one used later in the BB case,
that the contribution of each closed walk to the moment of the Adjacency
random block matrix is also $t^E$, so the g.f.s of the moments
is the same as in the case of ER graphs.

We will consider two bipartite extensions of this model,
one on the bipartite ER graphs with average degrees
$Z_1$ and $Z_2$, the other on
random BB graphs with degrees $Z_1$ and $Z_2$.

Since only walks on tree graphs contribute to the moments in the large-$N$
limit, in the random block matrix models on ER
graphs with average degree $Z$ or on random regular graphs,
the graphs covered by these walks are naturally
bipartite; therefore these moments are the same as
in the corresponding models on
bipartite ER graphs or random BB graphs
with $Z_1 = Z_2 = Z$.
From this remark and \cite{PC} it follows that, 
in the limit $d\to \infty$ with $t = \frac{Z}{d}$ fixed,
the g.f. of the $2l$-th moment of the Adjacency random block matrix
on bipartite ER with average degree $Z$ or on random
BB graphs with $Z_1 = Z_2$ 
follows the EM distribution.

Consider the Adjacency matrix of a random BB graph with
two types of nodes,
with elements  $\alpha_{i,j}$, where $i=1,\cdots, N_1$ 
are the indices of the nodes of the first type with degree $Z_1$
and $j=N_1+1,\cdots, N$ those of the nodes of the second type with degree $Z_2$,
where $N=N_1 + N_2$.
The number of edges is $N_1 Z_1 = N_2 Z_2$.

Define $t_a = \frac{Z_a}{d}$, with $a=1,2$.

The g.f. for the moments of $A$ is defined as
\begin{eqnarray}
f(x) &&= \sum_{n \ge 0} x^n \mu_n = 
\lim_{N\to \infty}\frac{1}{N d}\sum_{l \ge 0} x^{2l} \langle Tr A^{2l}\rangle =
\lim_{N\to \infty} \frac{1}{N d}(N_1\langle tr T_{i}\rangle +
N_2\langle tr T_{j}\rangle) \nonumber \\
&&= \frac{t_2 f_1(x) + t_1 f_2(x)}{t_1+t_2}
\label{fxadj}
\end{eqnarray}
where $Tr$ is the $Nd$-dimensional trace,
$tr$ is the $d$-dimensional trace and 
the following g.f.s, for closed walks starting with a
node $r$ with degree $Z_a$, are defined
\begin{eqnarray}
T_{a}(x; X, \alpha) &=& \sum_{l \ge 0} x^{2l}  (A^{2l})_{r,r}
\label{muTMa}\\
f_a(x) &=& \frac{1}{d} \lim_{N\to \infty} \langle tr\, T_{a}(x; X, \alpha) \rangle
\label{fa}
\end{eqnarray}
where the arguments $X$ and $\alpha$ indicate the set of block matrices
$\{X_{i,j} \}$ and the set $\{\alpha_{i,j} \}$.
The dependence of $f$, $f_1$ and $f_2$ on $t_1$, $t_2$ and $d$ is not 
explicitly indicated to simplify the notation.

The resolvent of the Adjacency random block matrix is
\begin{equation}
r(w) = \sum_{n\ge 0} w^{-n-1} \mu_n = \frac{1}{w}f(\frac{1}{w})
\label{resolv}
\end{equation}
From it one computes the spectral density
\begin{equation}
\rho(\lambda) = -\frac{1}{\pi} \lim_{\epsilon \to 0^+} \texttt{Im} \,\, r(\lambda+i\epsilon)
\label{rhores}
\end{equation}

Locally a random BB graph with large number of nodes $N$,
with degrees $Z_1$ and $Z_2$ small respect to $N$,
looks like a BB tree with these  degrees, the edges joining
nodes of different types.
Therefore, in the large-$N$ limit, only closed walks on rooted BB trees
contribute to the moments of the spectral distribution; this contribution
factorises  in a contribution due to the $N$-node graph and
in one due to the block matrices.
Let us first discuss the $d=1$ case, in which the latter part is trivially $1$.
Consider a graph node $v$ of type $b$, i.e. of degree $Z_b$, 
belonging to a closed walk $w$ on a BB
tree graph, starting at a root of type $a$, and a walk $w_1$ of length $n$ 
from the root to $v$  along the closed walk,
that is, $w_1$ consists of the first $n$ steps of $w$.
Let $p$ be the number of distinct edges incident in $v$, belonging to $w_1$.
Starting from $v$,
there are $Z_b - p$ ways to choose a new edge as the $(n+1)$-th step in the 
walk, a single way to choose an edge already traversed in the walk.
Adding the factor $\frac{N_a}{N} = \frac{Z_{3-a}}{Z_1+Z_2}$ from 
Eq. (\ref{fxadj}),
there is a combinatorial factor for the contribution to $f$ 
associated to the closed walk
\begin{equation}
\frac{Z_{3-a}}{Z_1+Z_2} Z_a^{\underline {\delta_R}} 
\prod_i (Z_1-1)^{\underline{\delta_i-1}}
\prod_j (Z_2-1)^{\underline{\delta_j-1}}
\label{zfact}
\end{equation}
where the degree $Z_a$ of the root $R$
is determined by the type (parity) $a=1,2$ of $R$,
$\delta_i$ are the degrees of type $1$ of the nodes of the tree graph $\gamma$
covered by the closed walk, $\delta_j$ are the degrees of type $2$ of $\gamma$
(not including the root);
$n^{\underline i}$ is the falling factorial.
If $R$ has type $1$,
$\delta_R + \sum_i (\delta_i-1) = V_2$ and $\sum_j (\delta_j-1) = V_1-1$;
otherwise $\sum_i (\delta_i-1) = V_2-1$ and 
$\delta_R + \sum_j (\delta_j-1) = V_1$, where $V_a$ is the number of vertices
of $\gamma$ of type $a$ (each edge out of a node of type $a$, apart from
the one towards the root, points to a distinct node of type $3-a$).

The g.f. for the moments of the Adjacency random matrix
on BB graphs
and the corresponding spectral distribution, obtained
in \cite{gm}, are reviewed in Appendix C.A.
We checked that the moments, computed till $\mu_{14}$ with the moments method 
just described,
agree with those computed with this g.f..

Let us consider now the case of bipartite ER graphs,
with i.i.d. random variables $\alpha_{i,j}=0,1$,
where $i=1,\cdots,N_1$, $j = N_1+1, \cdots, N$, where $N=N_1+N_2$
are the indices of the nodes with average degrees $Z_1$ and $Z_2$ respectively;
one has
\begin{equation}
\langle \alpha_{i,j} \rangle = \frac{Z_1}{N_2} = \frac{Z_2}{N_1} = 
\frac{Z_1 + Z_2}{N}
\end{equation}
The g.f. $f(x)$ for the moments of the Adjacency random matrix on bipartite 
ER graphs is again given by Eq. (\ref{fxadj}); we avoided putting a label
specifying the set of graphs on which the random matrix model is defined,
to simplify the notation.
To compute the $2l$-th moment of the Adjacency matrix, one considers
the closed walks with length $n$, on a graph with $V$ vertices and
$E$ edges.
Since the averages are on i.i.d. random variables, after
separating the indices in such a way that the sums on the vertices are all
on distinct indices, the averages of the addends of a sum give all the
same contribution to $f$, with multiplicity factor 
$N_1^{\underline V_1} N_2^{\underline V_2}$; the average of $\alpha_{i,j}$
on all edges gives a factor
$\prod \langle \alpha_{i,j} \rangle = \big(\frac{Z_1+Z_2}{N} \big)^E$.
Adding the factor $\frac{1}{N}$ in Eq. (\ref{fxadj}) one gets a combinatorial
factor for the contribution of the walk to $f$
\begin{equation}
\frac{1}{N} N_1^{\underline{V_1}} N_2^{\underline{V_2}}
\big( \frac{Z_1+Z_2}{N}\big)^E
\to \frac{Z_2^{V_1} Z_1^{V_2}}{Z_1+Z_2}
\label{zfacter}
\end{equation}
for $N\to \infty$.

If the root of the closed walk is of type $1$, this gives a factor
$\frac{t_2}{t_1+t_2} Z_2^{V_1-1} Z_1^{V_2}$,
so from Eq. (\ref{fxadj}) it follows that
$f_1$ gets a factor $Z_2^{V_1-1} Z_1^{V_2} = Z_1^{E_1} Z_2^{E_2}$,
where $E_a$ is the number of edges with node of type $a$ closer to the root
and $E = E_1 + E_2$ is the total number of edges (there is a $1$-to-$1$
correspondence between the nodes of type $a$ of a tree, 
different from the root, and the edges of type $3-a$ from them towards the root.)

According to Theorem $1$ in Section III, the walks contributing
to the $2l$-th moment of the Adjacency matrix on ER graphs
are in $1$-to-$1$ correspondence with
the partitions in $P^{(2)}_2(l)$; 
the monomial $Z_1^{E_1} Z_2^{E_2}$ associated to a length-$2l$ closed walk
can be seen as the monomial associated to the corresponding
partition of $P^{(2)}_2(l)$ with $E_a$ blocks of type
$a=1,2$, using $Z_a$ as parameter to count $E_a$.

We checked that up to $\mu_{14}$ the
moments computed with the moments method just discussed and using 
the recursive equations Eq. (\ref{psijc}) agree.

For $d > 1$ there is a
remaining factor contributing to the weight of a closed walk contributing
to the moments in Eq. (\ref{fxadj}); it is
the average over the trace of the product of matrix blocks, divided by $d$
and it is the same in the bipartite ER or BB case;
it can be expressed in terms of the parameters $c_i$ in Eq. (\ref{cm}),
coming from the average on inner products of random versors associated to the
edges. $c_1=1$ is used in Appendix B.A, while the parameters $c_i$ are 
kept in it for $i > 1$. For $d=1$, $c_i$ is equal to $1$ for all $i$.
In Appendix B.A we give the first few moments of the Adjacency random 
block matrix for generic $d$
on BB graphs; since at the orders examined the only explicit
dependence on $d$ (i.e. considering the parameter $c_i$ as independent)
are due to the factors $t_a - \frac{i}{d}$ coming from the falling factorials
of $Z_a$ in Eq. (\ref{zfact}), dropping from these moments the terms
explicitly dependent from $d$ one obtains the moments on bipartite 
ER graphs for generic $d$.

Let us now consider the limit $d\to \infty$ with $t_a = \frac{Z_a}{d}$ fixed.

The leading term in Eq. (\ref{zfact}), for $Z_1$ and $Z_2$ going to infinity, 
is equal to the factor in Eq. (\ref{zfacter}), so both in
the case of BB graphs and of bipartite ER graphs
this combinatorial factor for the contribution to $f_1$  
associated to the closed walk is equal to $Z_1^{E_1} Z_2^{E_2}$ in the
limit $d\to \infty$ with $t_a = \frac{Z_a}{d}$ fixed.
In the case of a closed walk with noncrossing sequence of steps,
the average over the trace of the product of matrix blocks, divided by $d$,
gives a factor $d^{-E}$, so
in this limit the weight of such a walk is $t_1^{E_1} t_2^{E_2}$;
if the closed walk has a crossing sequence of steps, the weight 
goes to zero in this limit. This argument is the same as in \cite{PC}.
Therefore in this limit the g.f. $f_a(x)$ is the same in both models;
we will call it $f_a^{nc}$.
In the case of ER graphs, from 
Corollary $1$ in Section III it follows that in this limit
the length-$2l$ closed walks correspond to partitions in $NC^{(2)}(l)$.
$[x^{2l}t_1^{j_1}t_2^{j_2}]f_a^{nc}(x)$ is the number of closed 
walks of length $2l$,
with root of CC $a$, $j_1$ edges of CC $1$ and $j_2$ edges of CC $2$,
and with noncrossing sequence of steps.

Using the results in Section V.A for $k=2$, $z=x^2$ and $t_0 = 1$, 
Eq. (\ref{fphi0})  reads
\begin{equation}
    f_a^{nc}(x) = 1 + t_a x^2 \phi(x^2)
\label{faadj}
\end{equation}
where the superscript $nc$ refers to the fact that only noncrossing partitions
contribute to it;
Eq. (\ref{fphi}) becomes
\begin{equation}
\phi = (1 + x^2 \phi)(1+ t_1 x^2 \phi)(1+ t_2 x^2 \phi)
\label{fphi2}
\end{equation}

Express the g.f. for the moments in Eq. (\ref{fxadj}) in terms of
\begin{equation}
h(x) \equiv t_1 f_2(x) + t_2 f_1(x); \qquad f(x) = \frac{h(x)}{t_1+t_2}
\label{hf}
\end{equation}
From Eqs. (\ref{faadj}, \ref{hf})
\begin{equation}
x^2 \phi(x^2) = \frac{h(x) - t_1-t_2}{2 t_1 t_2}
\end{equation}
we get from Eq. (\ref{fphi2}) the cubic equation
\begin{equation}
h^3 + a_2 h^2 + a_1(x) h + a_0(x) = 0
\label{cubadj}
\end{equation}
with
\begin{eqnarray}
a_2 &=& 2t_1 t_2 - t_1 - t_2 \nonumber \\
a_1(x) &=& -\big((t_2-t_1)^2 + 4\frac{t_1t_2}{x^2}\big) \nonumber \\
a_0(x) &=& 4\frac{t_1t_2}{x^2}(t_1+t_2) - (t_2-t_1)^2(2t_1t_2-t_1-t_2)
\label{fa6}
\end{eqnarray}
For $t_1=t_2$ this cubic reduces to the one of the EM
distribution.

The Cardano solution of the cubic equation can be parametrized in terms of
\begin{eqnarray}
p(x) &=& a_1(x) - \frac{a_2^2(x)}{3} \nonumber \\
q(x) &=& \frac{a_1(x) a_2}{3} - a_0(x) - 2\frac{a_2^3}{27} \nonumber \\
D(x) &=& \frac{q^2(x)}{4} + \frac{p^3(x)}{27}
\label{pqD}
\end{eqnarray}
where  $D(x)$ is the discriminant of this cubic equation.
The spectral distribution is non-vanishing when there are complex roots,
that is for $D$ positive;
since $D(x) = 0$ is a cubic equation in $x^{-2}$, we can study
the discriminant $\Delta$ of this latter cubic equation in $x^{-2}$,
normalized to have the $x^{-6}$ term with unit coefficient,
$-\frac{27 D(x)}{64 t_1^3t_2^3} = 0$; we get
\begin{equation}
\Delta = -\frac{1}{1728t_1^4t_2^4}((t_1 + t_2 + t_1 t_2)^3 - 27 t_1^2 t_2^2)^3 \le 0
\end{equation}
where the inequality, holding
for $t_1, t_2 \ge 0$, is due to the arithmetic-geometric inequality;
therefore the three roots $w_i$ of $D(x) = 0$ (as cubic equation in 
$x^{-2}$) are real.

One has
\begin{equation}
D(x) = -\frac{64}{27}t_1^3t_2^3(x^{-2}-w_0)(x^{-2}-w_1)(x^{-2}-w_2)
\label{Dfact}
\end{equation}
Let us choose $w_0 \le w_1 \le w_2$.
For $x \to \infty$
\begin{equation}
D(x) \to -\frac{16}{27}((t_2 - t_1)t_1t_2(t_1 - 1)(t_2-1))^2
\label{Dlimadj}
\end{equation}
so that $D(\infty) < 0$ for $t_1 \neq t_2$, $t_a \neq 1$ and from
Eq. (\ref{Dfact}) one has $w_0 w_1 w_2 < 0$, hence
either all $w_i$ are negative, or only $w_0$ is negative.
To have $D(x) > 0$ it must be $x^{-2} < w_0$ or $w_1 < x^{-2} < w_2$;
the former case is not allowed for $x$ real; therefore the latter holds,
and $w_0 < 0$, $w_1, w_2 > 0$.

Let $\lambda = \frac{1}{x}$.
Since for $D(x) < 0$ the roots of Eq. (\ref{cubadj}) are all real, 
from Eq. (\ref{Dlimadj}) it follows
that for $t_1 \neq t_2$ and $t_a \neq 1$ the spectral distribution is
zero in the neighborhood of $\lambda=0$ (apart in
$\lambda=0$ where there is the delta function term, as discussed below).

The spectral distribution corresponding to $f_1^{nc}$ 
is, for $\sqrt w_1 < \lambda < \sqrt w_2$,
\begin{equation}
    \rho_{A,c}^{nc}(\lambda) = 
- \frac{\texttt{Im} h(\frac{1}{\lambda})}{\pi \lambda(t_1+t_2)}
\label{rhoeq}
\end{equation}
where the solution formula for the cubic gives
\begin{equation}
\texttt{Im} \, h(x) = - \frac{\sqrt{3}}{2}
\Big(
\Big( \sqrt{D(x)} + \frac{1}{2}q(x) \Big)^{\frac{1}{3}}
- \Big( -\sqrt{D(x)} + \frac{1}{2}q(x) \Big)^{\frac{1}{3}} \Big)
\label{imh}
\end{equation}

In the case $t_2=1$  one has
\begin{equation}
D = -\frac{16 t_1^2}{27 x^2} \big(4t_1x^{-4} - (8t_1^2+20t_1-1)x^{-2} +4(t_1-1)^3 \big)
\label{Dtr}
\end{equation}
To see in which region $D$ is positive, consider 
$p = 4 t_1 w^2 - (8t_1^2+20t_1-1) +4(t_1-1)^3 = 4 t_1(w-w_1)(w-w_2)$,
with $w_1 \le w_2$;
$w_1$ has the sign of $t_1 - 1$; for $w = x^{-2} = \lambda^2$ the root $w_1$
must be non negative, so the solution for $t_1 < 1$ is not acceptable.
From this it follows that if $t_1 > 1$,
$\lambda^2 \ge w_1 > 0$, so there is a gap between the negative and
the positive bands of the spectrum; if $t_1 \le 1$ one has $\lambda^2 \ge 0$,
so there is no gap.

We have thus determined the continuous part $\rho^{nc}_{A,c}(\lambda)$ of the 
spectral distribution 
$\rho^{nc}_A(\lambda) = \rho^{nc}_{A,c}(\lambda) + \rho^{nc}_{A,0}(\lambda)$ 
of the Adjacency random block matrix.

From Eqs. (\ref{cubadj}, \ref{fa6}), the cubic equation can be rewritten as
\begin{equation}
(h + 2t_1t_2 - t_1 - t_2)(h + t_1 - t_2)(h - t_1 + t_2) +
\frac{4t_1t_2}{x^2} (t_1 + t_2 - h) = 0
\end{equation}
so that from this equation and Eq. (\ref{hf}), for $x\to \infty$ one gets
$f(\infty) = c_0$, where either $c_0$ is either $1 - \frac{2t_1t_2}{t_1+t_2}$
or $\frac{|t_2-t_1|}{t_1+t_2}$.
The integral of $\rho_A(\lambda)$ around $\lambda = 0$ is $c_0$.
Assuming that there are two regions in the $t_1$,$t_2$ plane,
in which $\rho_{A,0}(\lambda)$ has these two values of $c_0$, and that reaching
a point of the border from these regions one gets the same value, one
finds that $t_2 = 1$ for $t_2 > t_1$, $t_1 = 1$ for $t_1 > t_2$.

For $N_2 \neq N_1$, the matrix $A$ has at least $|d N_2 - d N_1|$ zeros,
so that the fraction of zero eigenvalues is at least
$\frac{|N_2-N_1|}{N_1+N_2} = \frac{|t_2-t_1|}{t_1+t_2}$, which is
the second value of $c_0$.

For $Z_1 < d$ and $Z_2 < d$ the fraction of zero eigenvalues is greater.

These considerations lead us to the following guess for $\rho^{nc}_{A,0}$
\begin{equation}
    \rho^{nc}_{A,0}(\lambda) = 
(1 -\theta(1-t_1)\theta(1-t_2))\frac{|t_2-t_1|}{t_1+t_2}\delta(\lambda) +
\theta(1-t_1)\theta(1-t_2) (1 - \frac{2t_1t_2}{t_1+t_2})\delta(\lambda)
\label{rho0adj}
\end{equation}
We checked in a few cases, using multiprecision numerical integration 
with mpmath \cite{mpmath}, that the integral of $\rho^{nc}$,
gives as expected $1$,
in particular we checked a few cases in the region $0 < t_1, t_2 < 2$
and close to the line $t_1=1$ for $0 < t_2 < 10$.

On the transition line $t_1=1,\, 0 \le t_2 \le 1$ and 
$t_2=1,\, 0 \le t_1 \le 1$ the spectrum has a single band, as we saw after
Eq. (\ref{Dtr}); moving along the line $t_2 = \alpha t_1$, the fraction
of zero modes decreases with $t_1$ as $t_1$ increases inside the unit
square and it is constant outside it.

From Eq. (\ref{rho0adj}) it follows that for $t_1 > 1$ and $t_2 > 1$ one
has $\rho^{nc}_{A,0} = 0$ only for $t_1=t_2$.

We have made a few simulations with Adjacency random block matrices for small 
$d$ on random BB graphs,
to see how the spectral distribution varies with $d$.
To generate random BB graphs
we used a configuration model algorithm, based on \cite{ballobas},
consisting in attaching stubs
(half-edges) to the nodes and joining nodes of different type randomly
till all nodes are paired;
if the resulting graph is not simple another run is made; this simple
algorithm is sufficiently fast, because the graphs are sparse (the degrees are
small with respect to $N$).
To each edge a random $d$-dimensional vector with Gaussian distribution is
associated; the corresponding versor is uniformly distributed on the
sphere in $d$ dimensions. The eigenvalues of the resulting random block
matrix are computed numerically.

\begin{figure*}[h]
\begin{center}
\epsfig{file=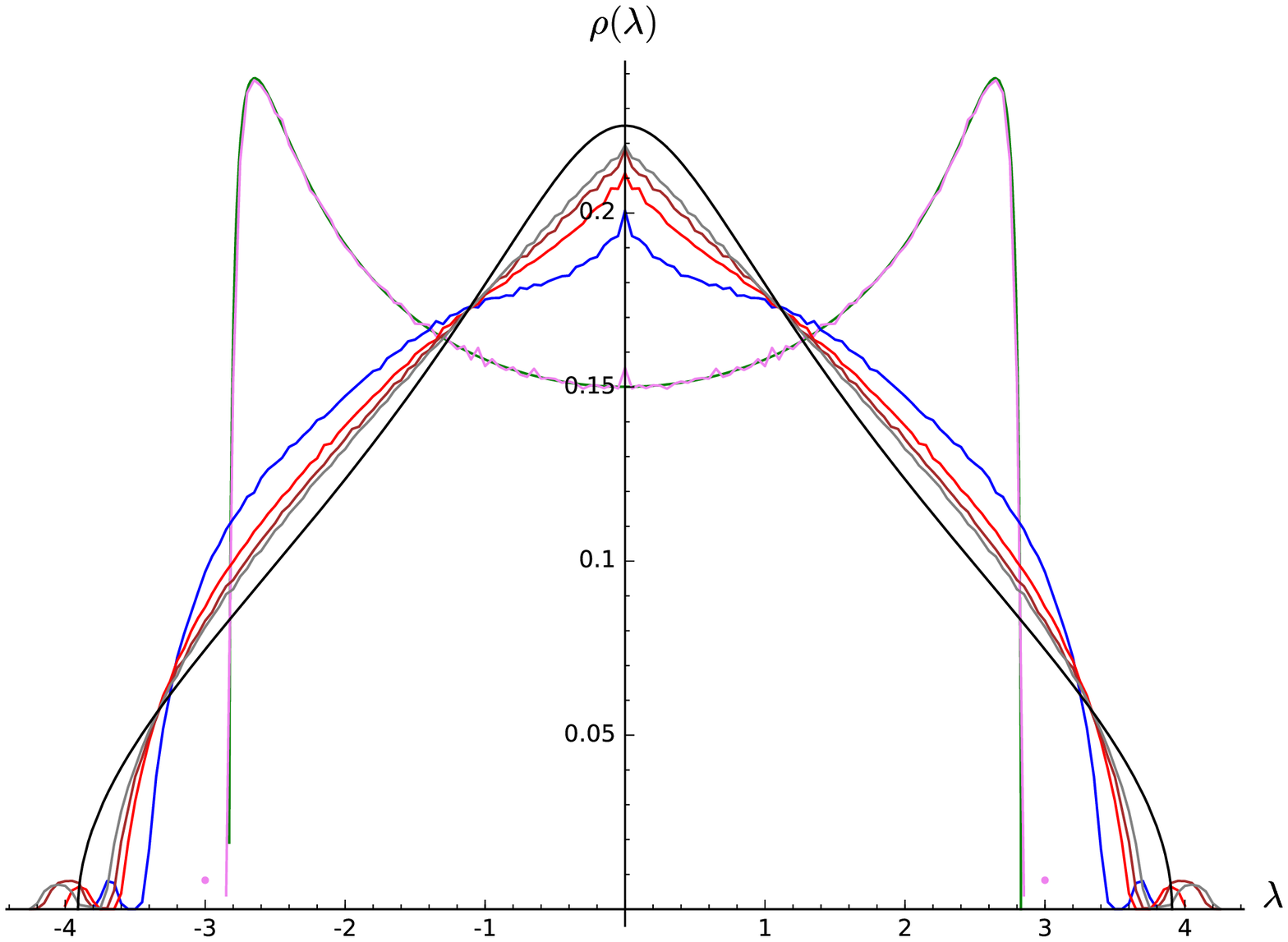, width=7.00cm}\quad
\epsfig{file=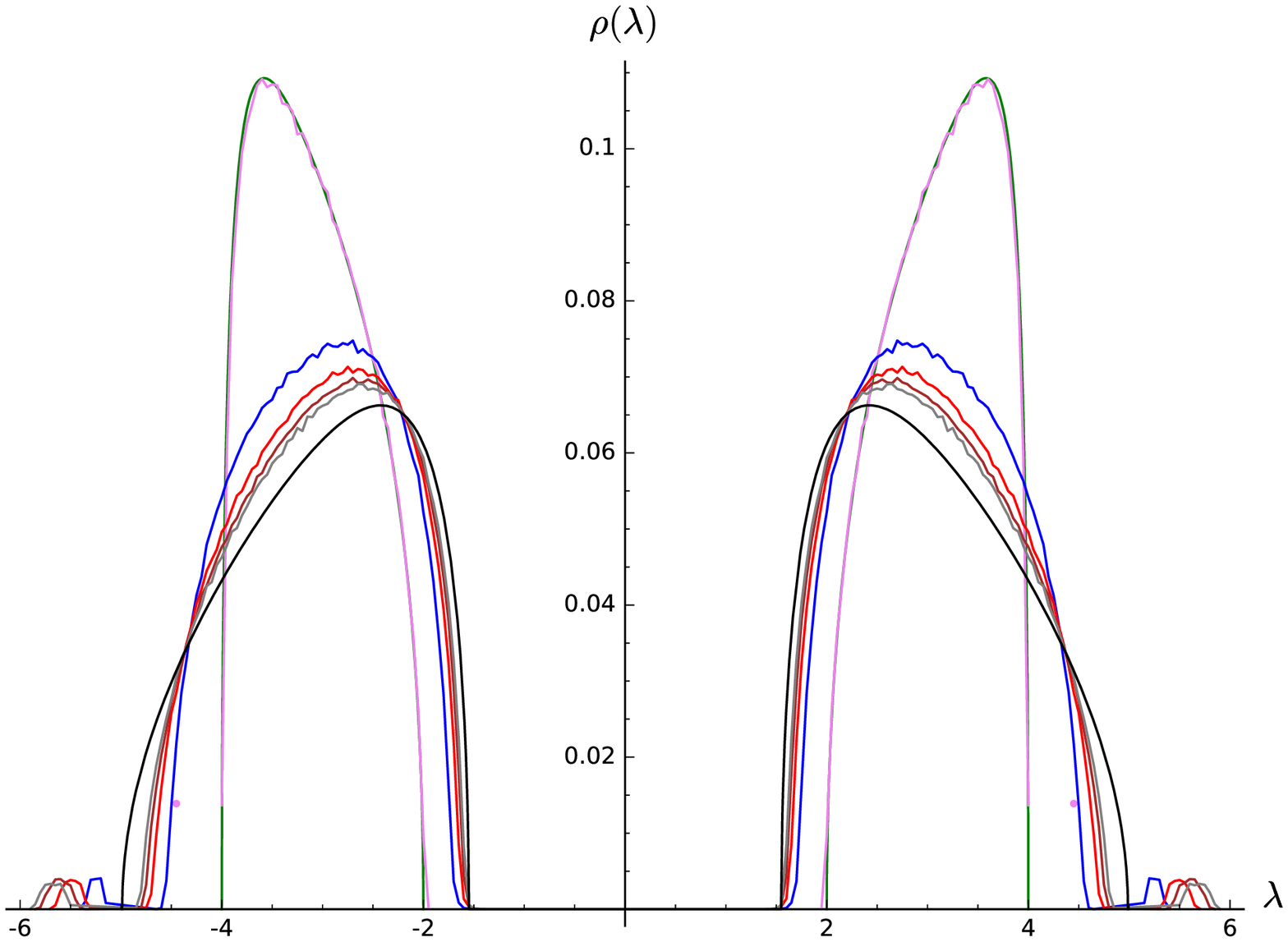, width=7.00cm}
\caption{Simulations for the spectral distribution of the Adjacency 
  random block matrix on random BB graphs.
In the left hand side figure, there are the results of a simulation with
random block matrices for the
spectral distribution of the Adjacency random block matrix in the case 
$t_1=3, t_2=3$;
the violet line is a $d=1$
simulation with $Z_1=Z_2=3$ with $100$ random graphs with $N_1=1200$, 
$N_2=1200$ nodes;
the blue line is a $d=2$ simulation with $Z_1=Z_2=6$ with $500$ random graphs
with $N_1=480$, $N_2=480$ nodes;
the red line is a $d=3$ simulation with $Z_1=Z_2=9$ with $500$ random graphs
with $N_1=480$, $N_2=480$ nodes;
the brown line is a $d=4$ simulation with $Z_1=Z_2=12$ with $500$ random graphs
with $N_1=240$, $N_2=240$ nodes;
the gray line is a $d=5$ simulation with $Z_1=Z_2=15$ with $500$ random graphs
with $N_1=300$, $N_2=300$ nodes;
the green line is the $d=1$ Kesten-McKay distribution;
the black line is the $d=\infty$ EM distribution.
In the right hand side figure, there are the results of a simulation with
random block matrices for the
spectral distribution of the Adjacency random block matrix in the case 
$t_1=2, t_2=10$;
the violet line is a $d=1$
simulation with $Z_1=2$, $Z_2=10$ with $500$ random graphs with $N_1=1200$,
$N_2=240$ nodes;
the blue line is a $d=2$ simulation with $Z_1=4$, $Z_2=20$ with $500$ random 
graphs with $N_1=1200$, $N_2=240$ nodes;
the red line is a $d=3$ simulation with $Z_1=6$, $Z_2=30$ with $500$ random 
graphs with $N_1=900$, $N_2=180$ nodes;
the brown line is a $d=4$ simulation with $Z_1=8$, $Z_2=40$ with $500$ random 
graphs with $N_1=800$, $N_2=160$ nodes;
the gray line is a $d=5$ simulation with $Z_1=10$, $Z_2=50$ with $200$ random 
graphs with $N_1=900$, $N_2=180$ nodes;
the green line is the $d=1$ exact distribution Eq. (\ref{rhod1adj});
the black line is $\rho_A^{nc}$
given by Eqs. (\ref{fa6},\ref{pqD},\ref{rhoeq},\ref{imh});
the distribution in $\lambda=0$ is not represented.
}
\label{Fig1}
\end{center}
\end{figure*}

In the left hand side figure in Fig. \ref{Fig1} are given the results of 
simulations with random block matrices in the case $t_1=t_2=3$ case.
The $d=1$ case follows the Kesten-McKay distribution \cite{kesten, MK},
which is the $Z_1=Z_2$ case of the distribution reviewed in Appendix C.A.
The $d=2,\cdots,5$ distributions approach, as $d$ increases, 
the distribution of the EM approximation.

In the right hand side figure in Fig. \ref{Fig1} are given the results of
simulations with random block matrices in the case $t_1=2$, $t_2=10$.
The $d=1$ spectral distribution Eq. (\ref{rhod1adj}) has been obtained in 
\cite{gm}; as $d$ increases, the $d=2,\cdots,5$
distributions approach the $d\to \infty$ distribution $\rho_A^{nc}$.

\subsection{Adjacency random matrix model on ER 
graphs in the $\frac{1}{s}$-expansion}

While the analytic form of g.f. of the moments and the spectral distribution
of the Adjacency random matrix 
model on random regular graphs \cite{MK} or on random BB graphs \cite{gm} 
is known, those on ER graphs and bipartite ER graphs are not. 
It is therefore interesting to
look for distributions approximating the latter ones, in some region of 
the average degree(s). One of them is the EM approximation \cite{SC}; another
is the distribution on random regular or BB graphs.

The g.f. of the moments of the Adjacency random matrix model on ER graphs
with average degree $t$ is $f(x,t) = f_1(x,t,t)$,
enumerating $P^{(2)}_2$ according to the number
of elements and of blocks of its partitions. $P^{(2)}_2$ is nc-closed,
as shown in Observation $4$.
We saw in Section II.B that from the g.f. $f(x,t)$, 
enumerating a nc-closed set of partitions according to the number of blocks,
one can obtain the g.f.
$f(x, s,\frac{t}{s}) = \psi(x,\sigma, t) = \psi_0(x,t) + \sum_{g \ge 1} \sigma^g \psi_g(x,t)$, where $\sigma = \frac{1}{s}$;
this g.f. interpolates between the g.f. enumerating the nc-closed set, 
at $\sigma=1$, and the one
enumerating the corresponding set of noncrossing partitions, for $\sigma=0$;
$\psi_g(x,t)$ is the g.f. enumerating $P^{(2)}_2$ according
to the number of elements and of blocks of its partitions
in which $g$ is the difference between the number of blocks 
and the number of irreducible components.
$\psi_0(x,t)$ is the g.f. enumerating $NC^{(2)}$ according
to the number of elements and of blocks of its partitions; 
it is equal to $f_1^{nc}(x, t, t)$ in the Adjacency random block matrix
model, in the limit $d\to \infty$ with $t = \frac{Z}{d}$
fixed; it satisfies Eq. (\ref{ban0}) for $k=2$; 
it is the EM distribution.
The $g$-th order correction around the EM approximation
$[\sigma^g] \psi(x,\sigma, t) = [\frac{1}{s^g}]f(x, s, \frac{t}{s})$
can be computed in terms of the EM approximation, via the
parameter $y = x \psi_0(x,t)$, given in Eq. (\ref{f1s2b1}),
and the generating functions enumerating partitions in $P^{(2)}_2$ 
with $h$ blocks according to the number of elements, for $h \le g+1$;
We computed the latter ones till $10$ blocks, with the
algorithm described in subsection IV.B.

The expression of the first order correction $[\sigma] \psi(x,\sigma, t)$
is given in Eq. (\ref{f1s2b}), for $k=2$. 
We computed the second and third order correction,
given in Eqs. (\ref{ntransig2},\ref{ntransig3}), where $a_{(j)}$
is expressed in terms of $f_{(i)}$ in Eq. (\ref{a12});
$f_{(i)}(x) = [t^i]f(x, t)$ for $i=1,2,3,4$ is given in
Eqs. (\ref{tfk},\ref{t2fk2},\ref{t3fk2},\ref{t4fk2}).
We checked these corrections till order $32$ in powers of $x$ using the recursive
equations Eqs. (\ref{psiH1}, \ref{psiH}) and Eq. (\ref{nctrant}).

In the case of the Adjacency random matrix model on bipartite
ER graphs with average degrees $t_1$ and $t_2$,
whose g.f. of the moments
enumerates $P^{(2)}_2$ according to the number of elements and the
CC type of its partitions, one can still define a noncrossing partition flow, 
as explained in subsection II.C; in the limit $s\to \infty$ it gives
the g.f. $f_1^{nc}(x,t_1,t_2)$, described in the above subsection, which is also
the limit, for $d\to \infty$ with $t_a = \frac{Z_a}{d}$ fixed, of the g.f.
of the moments of the Adjacency random block matrix on bipartite ER graphs.
It is the extension to the bipartite case of the EM approximation of 
the g.f. of the moments of the Adjacency random matrix model on ER graphs.
We have computed $\psi_1(x, t_1, t_2)$ using
Eq. (\ref{psik2cc}) and checked it till $x^{14}$, and for a few values of 
$t_1,t_2$ till $x^{30}$ using the recursive relations Eq. (\ref{psijc}) and 
Eq. (\ref{nctrank}).

\begin{figure*}[h]
\begin{center}
\epsfig{file=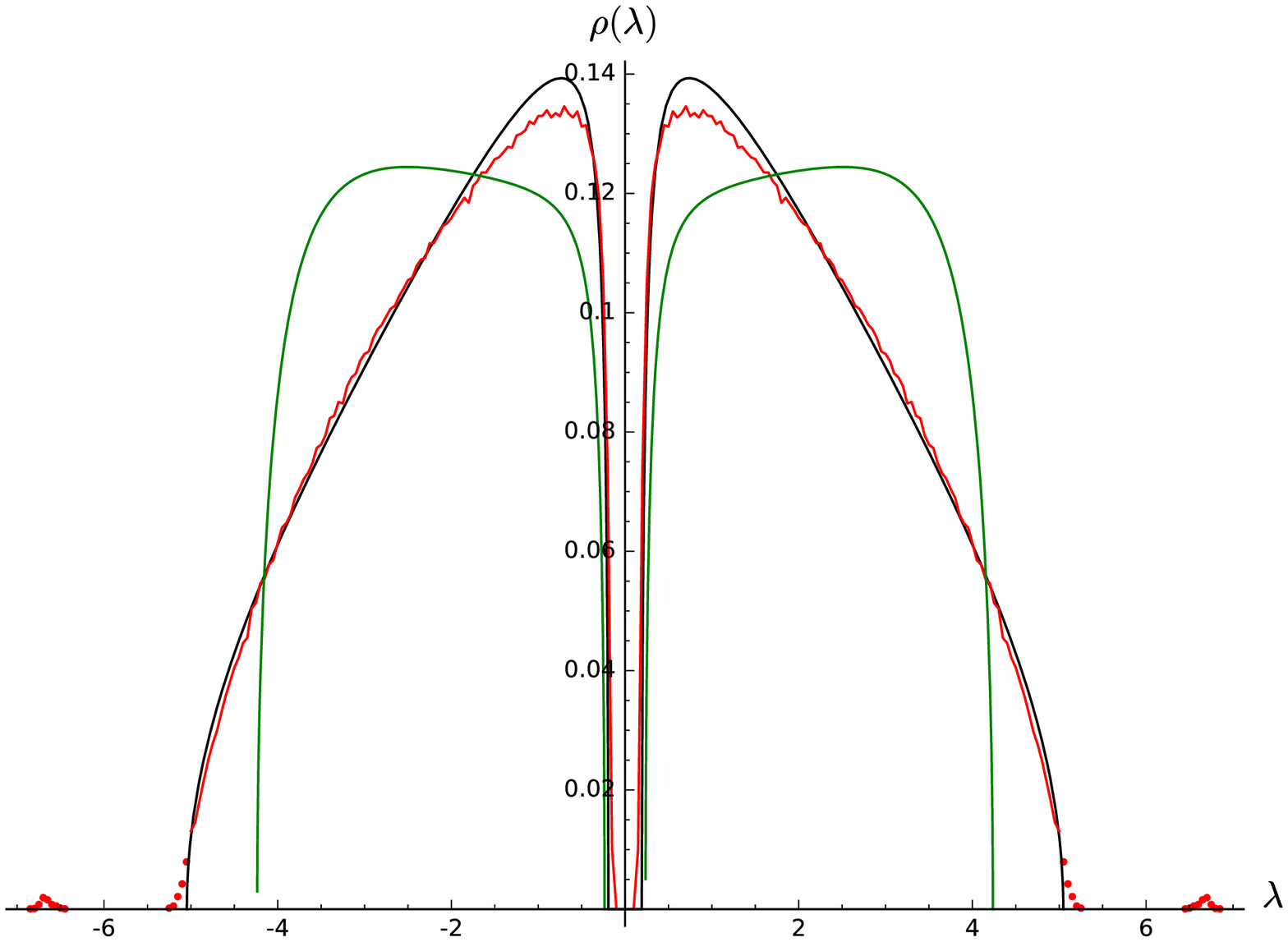, width=5.00cm } \quad
\epsfig{file=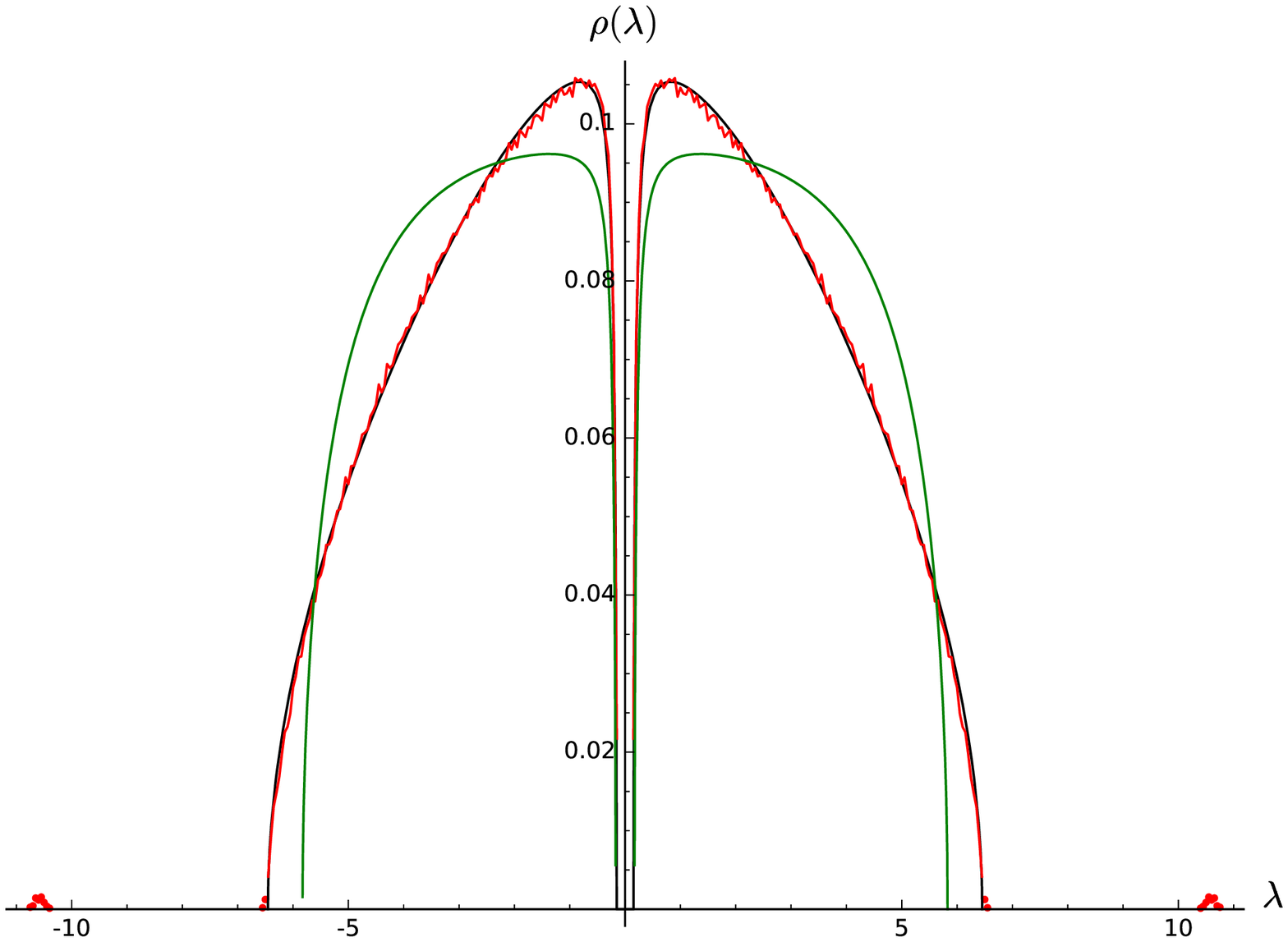, width=5.00cm  }
\epsfig{file=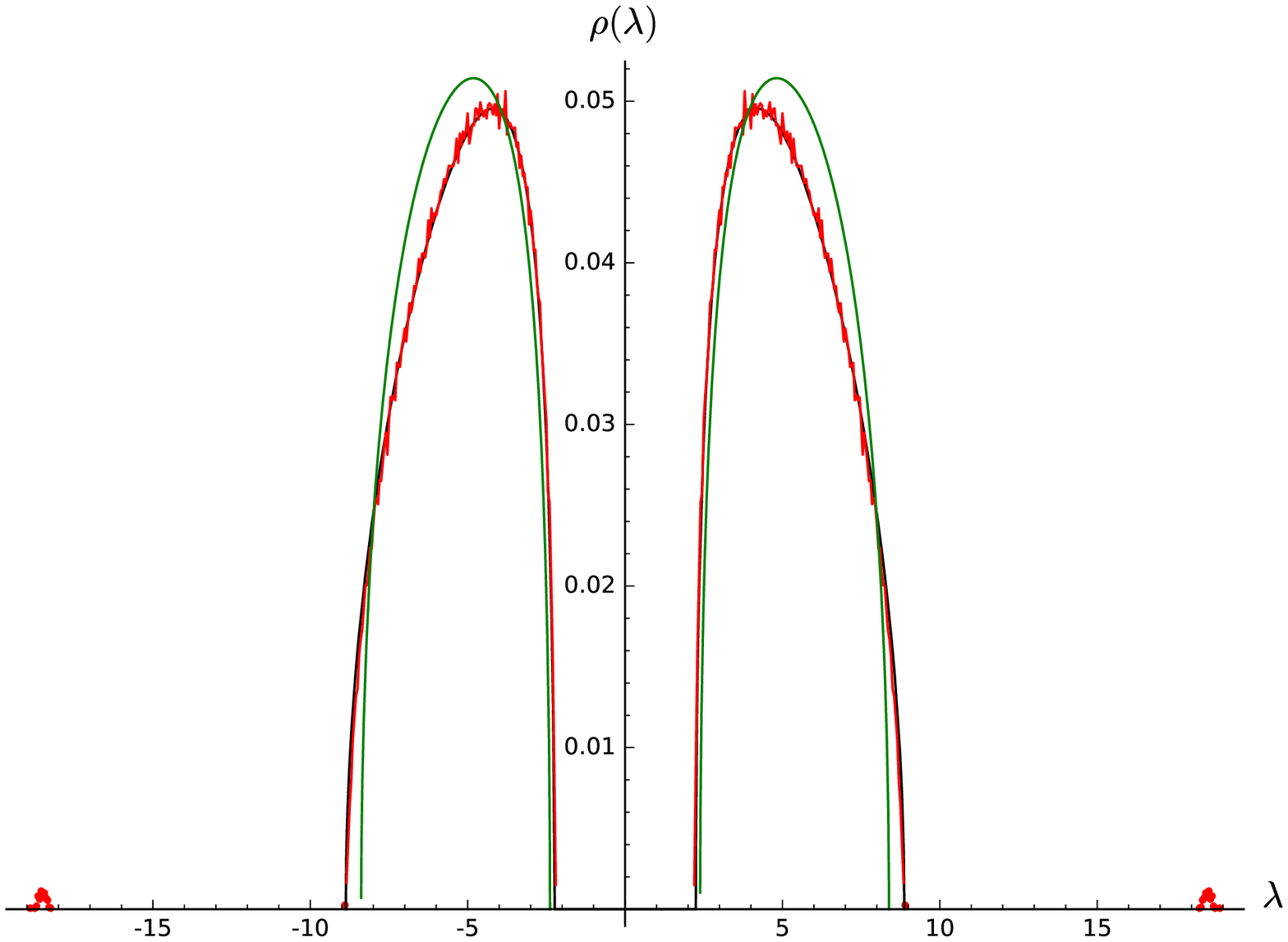, width=5.00cm  } \quad
\caption{Simulations for the spectral distribution of the Adjacency
    random matrix on bipartite ER graphs with average degrees $t_1$, $t_2$.
    Each simulation is made with $100$ Adjacency random block matrices.
    The left hand figure is for $t_1=5$ and $t_2=6$, where
    the red line is a simulation with $N_1=1800$, $N_2=1500$;
    the center figure is for $t_1=9$ and $t_2=10$, where
    the red line is a simulation with $N_1=1700$, $N_2=1530$;
    the right hand figure is for $t_1=10$ and $t_2=30$, where
    the red line is a simulation with $N_1=2400$ and $N_2=800$ nodes.
    The black line is $\rho_A^{nc}$
    given by Eqs. (\ref{fa6},\ref{pqD},\ref{rhoeq},\ref{imh}).
    For comparison, we added as
    a green line the spectral distribution of the Adjacency random matrix
    on random BB graphs given in Eq. (\ref{rhod1adj}).
    The distribution in $\lambda=0$ is not represented;
    we used bins of size $0.05$ to group eigenvalues evaluated numerically;
    within the domain of $\rho_A^{nc}$ the points, representing the
    simulated distribution on the bins, are joint by straight lines.
}
\label{figadjer}
\end{center}
\end{figure*}

Numerical simulations with Adjacency random matrices
on bipartite ER graphs indicate that $\rho_A^{nc}$ approximates well
the simulated spectral distribution, for large values of $t_1, t_2$,
as one can see in Fig. \ref{figadjer}.

Let us end this section with a remark:
in the Adjacency random matrix model on random BB graphs, the set of closed
walks contributing to the moments is equivalent to a set of partitions
${\cal P}_B$; as remarked in subsection II.B,
one can define a noncrossing partition flow from the g.f. enumerating 
${\cal P}_B$ to the one enumerating the
corresponding set of noncrossing partitions; its irreducible noncrossing 
partitions are $(1,\cdots, 2j)$, corresponding to walks on a single edge,
so the g.f. enumerating ${\cal P}_B$ flows to $NC^{(2)}$, 
like $P^{(2)}_2$. However ${\cal P}_B$
is not nc-closed, since from its irreducible partitions one can
generate partitions corresponding to walks covering rooted tree graphs
with arbitrary high degrees; for instance from the irreducible partition
$(1,2)$ one can generate the size-$2j$ partition with blocks $(2 i-1, 2 i)$,
$i=1,\cdots, j$, which corresponds to a closed walk covering a tree graph
with root of degree $j$.
Therefore one cannot compute in the case of ${\cal P}_B$ 
the non-leading terms in the $\frac{1}{s}$-expansion around $NC^{(2)}$
using the NCPT.
Of course, since the g.f. of the moments of the Adjacency random matrix on
random BB graphs is known \cite{gm} (see Appendix C.A), there is no need
of studying its $\frac{1}{s}$-expansion to find an approximation to it.

\section{Laplacian random block matrix model}

Consider a random elastic network in $d$ dimensions.
The Hessian is given by the following Laplacian random block matrix
\begin{eqnarray}
L_{i,j} &=& -\alpha_{i,j} X_{i,j}, \qquad i \neq j = 1,\cdots,N \nonumber \\
L_{i,i} &=& \sum_{j \neq i} \alpha_{i,j} X_{i,j}
\label{eqM}
\end{eqnarray}
where $\alpha_{i,j}$ is an element of the Adjacency matrix,
$X_{i,j} = |\hat v_{i,j}><\hat v_{i,j}|$, where
$\hat v_{i,j} = \frac{r_{i,j}}{|r_{i,j}|}$ and $r_{i,j}$ is the vector
from the $i$-th node to the $j$-th node.

This mechanical model describes the vibrational modes of a disordered
equilibrium configuration of soft spheres near the jamming point;
the latter system is a model for the vibrational modes of
a disordered system like glass \cite{ohern}.

The network of contacts of the soft spheres near the jamming point
is approximately a regular graph. In \cite{parisi,benetti} the case of
random regular graphs has been considered, in \cite{CZ} the
case of ER graphs with average degree $Z$.
In both cases it has been considered a mean field approximation,
replacing the $d$-dimensional random contact network with
a random regular graph and edges contact versors $\hat v_{i,j}$, which 
are considered to be independent random versors
uniformly distributed on the $d$-dimensional sphere;
it has been shown that, in the limit $N\to \infty$,
and in the limit $d\to \infty$ with $t=\frac{Z}{d}$ fixed,
the spectral distribution is the Marchenko-Pastur distribution, both
in the random regular graph case \cite{parisi} and in the ER
case \cite{CZ}, \cite{PC}.
At the isostatic point the density of states is finite at zero frequency
at the isostatic point, giving an approximate description of
the plateau found for frequency close to zero near the jamming point.

In this section we consider a variant of this model,
in which the random elastic network is BB, with 
degrees $Z_1$ and $Z_2$, or it is a bipartite ER graph
with average degrees $Z_1$ and $Z_2$.
In analogy with the case of the random regular elastic network, we consider
the mean field approximation, in which a random $d$-dimensional 
BB elastic network
is replaced by a random BB graph, where to the edges contact versors $\hat v_{i,j}$ are associated, which
are considered to be independent random versors
uniformly distributed on the $d$-dimensional sphere.

For generic $d$, the Maxwell isostatic condition for mechanical equilibrium 
\cite{maxwell} follows from equating
the number of degrees of freedom $d N = d(N_1+N_2)$ with the number of
constraints, that is the number of edges $E = N_1 Z_1 = N_2 Z_2$
(the degrees of freedom coming from translation and rotation invariance
can be neglected for large $N_1$ and $N_2$; they are considered in the study of
the rigidity of a BB mechanical framework in \cite{bolker});
so the isostatic condition is the hyperbola
\begin{equation}
t_1 t_2 - t_1 - t_2 = 0
\label{iso}
\end{equation}
or equivalently
\begin{equation}
(Z_1 - d)(Z_2 - d) = d^2
\label{iso1}
\end{equation}
Hence $Z_a - d$ is a factor of $d^2$;
taking $Z_1 \le Z_2$, the number of isostatic points for finite $d$
is equal to $\frac{1 + \tau(d^2)}{2}$, 
where $\tau(d^2)$ is the number of divisors of $d^2$.
For $d > 1$ there are always at least two isostatic points $(Z_1,Z_2)$
with $Z_1 \le Z_2$;
one is $Z_1 = Z_2 = 2d$, corresponding
to the isostatic condition $Z = 2d$ for an elastic regular network;
the other is $Z_1 = d+1$, $Z_2 = d^2 + d$.

The g.f. for the moments of $L$ is
\begin{equation}
f(x, t_1, t_2) = \sum_{n \ge 0} x^n \nu_n = 
\lim_{N\to \infty}\frac{1}{N d}\sum_{n \ge 0} x^n \langle Tr L^n\rangle =
\frac{t_2 f_1(x) + t_1 f_2(x)}{t_1+t_2}
\label{fxlap}
\end{equation}
with
\begin{eqnarray}
T_{j}(x; X, \alpha) &=& \sum_{n \ge 0} x^n  (L^n)_{j,j}
\label{muTM}\\
f_a(x, t_1,t_2) &=& \frac{1}{d} \lim_{N\to \infty} \langle 
tr T_{j}(x; X, \alpha) \rangle
\label{fl}
\end{eqnarray}
where $j$ is the index of a node of type $a$ and degree $Z_a$.

In the large-$N$ limit,
the terms in $tr (L^n)_{j_0,j_0}$ can be interpreted, similarly to
the Adjacency case, as closed walks on a tree graph rooted at $j_0$.
At the $i$-th node a walk has two kind of moves, 
one along the edge $(i,j)$ due to the elements
$L_{i,j} = -\alpha_{i,j} X_{i,j}$;
the other is due to the $j$-th term 
$\alpha_{i,j} X_{i,j}$ of $L_{i,i}$ in Eq. (\ref{eqM}); 
it corresponds to moving up and down an edge $(i,j)$; using the fact
that $\alpha_{i,j}$ and $X_{i,j}$ are symmetric and idempotent, 
it is useful to represent this
move as $\alpha_{i,j}\alpha_{j,i} X_{i,j} X_{j,i}$
\cite{bau,PC}.

The weight associated to a closed walk corresponding to a contribution
to $f_a$ is computed as in Section VI.

We have computed the first $10$ moments of the Laplacian random block matrix
for generic $d$; see Appendix B.B.

The spectral distribution in $d=1$ on random BB graphs
is given in Appendix C.B; the moments in $d=1$ on ER graphs
with average degree $Z$
can be computed with the recursion relations in \cite{bau}.
We have not studied the generalization of these recurrence relations
to bipartite ER graphs with degrees $Z_a$.

Let us describe the closed walks contributing to the moments in the Laplacian
random matrix model on ER graphs in terms of partitions. 
As in the random matrix Adjacency model, 
we generalize them to walks on $k$-gon trees, the $k = 2$ case being the
case relevant for the Laplacian random matrix model.

Let us consider two kinds of moves on a $k$-gon tree $T$; a move of the
first kind goes from a node $i$ to an adjacent node $j$ along the
edge $(i,j)$ on $T$, a move of the second kind starts at a node $i$
of a $k$-gon $G$ and loops around it once, i.e. it is a closed walk of length
$k$ on $G$ with root $i$. Let us call $W_{n,k}$ the set of closed walks
covering rooted $k$-gon trees, consisting of length-$n$ sequences of these 
moves.
In the $k=2$ case a move of the second kind corresponds to moving up and
down an edge $(i,j)$, which is due to the 
$\alpha_{i,j} X_{i,j} = \alpha_{i,j} \alpha_{j,i} X_{i,j} X_{j,i}$ term 
of $L_{i,i}$ in Eq. (\ref{eqM}); a move of the first kind corresponds to moving
along the edge $(i,j)$ with $L_{i,j}$.

From a partition $\pi$ of $[n]$ we can obtain a
bicolored partitions $p$ adding a color $g=1,2$ to each element.
We represent an element of a bicolored partition as $i_g$; $i$ is the base
element in $\pi$; let us call $\pi$ the base partition of $p$.
There are $2^n$ bicolored partitions for each partition $\pi$ of $[n]$.
In the sequential form of $p$, to an element $j_g$ of the $i$-th block of $p$
corresponds in position $j$ the element $i_g$.

Let us call the $i$-th block of the bicolored partition $p$
the subset of elements
corresponding to the $i$-th block of the base partition.
We say that a bicolored partition $p$ is noncrossing if its base partition
is noncrossing.

Define ${\cal C}^w_{(n,k)}$ as the set of bicolored partitions of $[n]$,
with the restriction that each block of a bicolored partition
has a number of elements of the first color which is multiple of $k$,
and that, given two distinct blocks $B_i$ and $B_j$, the number of elements 
of $B_j$ with the first color between two elements of $B_i$ is multiple of $k$.
${\cal C}^w_{(k)}$ is the union of ${\cal C}^w_{(n,k)}$ for all $n$.

Define the set ${\cal C}_{(n,k)}$ as the set of noncrossing
partitions in ${\cal C}^w_{(n,k)}$;
${\cal C}_{(k)}$ is the union of ${\cal C}_{(n,k)}$ for all $n$.

As an example, $p = (1_2,2_1,3_2,4_1,5_1,8_1)(6_1,7_1)$ belongs to 
${\cal C}_{(8,k)}$ ; it has sequential form 
$(1_2,1_1,1_2,1_1,1_1,2_1,2_1,1_1)$.

Let us show that ${\cal C}_{(n,k)}^w$ is isomorphic to $W_{n,k}$.

Let $w \in W_{n,k}$; expand the moves of second kind in $w$ to get a
walk $w'$, in which a move of second kind of $w$ on a $k$-gon is replaced by $k$
moves of the first kind on that $k$-gon; 
let $r$ be the number of moves of the second kind
in $w$, and let $S$ be the size-$kr$ set of the positions of the moves of the
first kind in $w'$, coming from this expansion.
By Theorem $1$ in Section III, to $w'$ corresponds a partition $p' \in P^{(k)}_k(l)$,
with $kl - r k + r = n$. Group the elements of $p'$ which belong to $S$,
starting from the smallest element, replacing $k$ elements in $S$ with
an element of color $2$; the elements not belonging to $S$ become
of color $1$; in this way one gets a bicolored partition $p$ with $n$ elements.
Since a block $B'_i$ of $p'$ has size multiple of $k$, after the grouping 
the number of elements of the first kind in the corresponding
block $B_i$ is multiple of $k$.
The number of elements of $B'_i$ between two elements of another block
$B'_j \in p'$ is multiple of $k$; after the grouping the number of
elements of the first kind in the corresponding
block $B_i$, which are between two elements of $B_j$, is multiple of $k$;
therefore $p \in {\cal C}_{(n,k)}^w$.
Viceversa, let $p \in {\cal C}_{(n,k)}^w$, with $r$ elements of color $2$.
Expand all the elements of color $2$
in the sequential form of $p$, replacing an element of color $2$
by $k$ elements of color $1$; in this way one gets the sequential form
of a partition $p'$; let $S$ be the set of the positions of the elements
of the sequential form of $p'$, which come from this expansion;
since the number of elements of color $1$ in a block
$B_i$ of $p$ is multiple of $k$, it follows that the number of elements
in the corresponding block $B_i'$ of $p'$ is multiple of $k$;
the number of elements of color $1$ in $B_i$, between two elements of
another block $B_j$ of $p$ is multiple of $k$, so that $p' \in P^{(k)}_k(l)$.
By Theorem $1$, to $p'$ corresponds a closed walk $w'$ covering a 
rooted $k$-gon tree;
starting from the beginning of the walk, regroup the steps of $w'$ with
position in $S$, replacing $k$ steps in a move of the second kind;
the steps in positions not in $S$ become moves of the first kind;
in this way one gets a walk $w \in W_{n,k}$.

One can prove that ${\cal C}^w_{(k)}$ is nc-closed as in Observation $4$.
It is easy to see that Eqs. (\ref{Aeq}-\ref{nctran}) and
Eq. (\ref{nctrant}) apply also to ${\cal C}_{(k)}$;
there is a noncrossing partition flow from the g.f. enumerating 
${\cal C}^w_{(k)}$ to the one enumerating ${\cal C}_{(k)}$.

We did not study the enumeration of the partitions in ${\cal C}^w_{(n,k)}$.
Recurrence relations to compute the moments of the Laplacian random matrix
model (i.e the $d=1$ Laplacian random block matrix model)
on ER graphs have been given in \cite{bau}, thus enumerating 
partitions in ${\cal C}^w_{(n,2)}$ with given number of blocks;
we listed the partitions of ${\cal C}^w_{(n,2)}$
till $n=10$, finding agreement with these recurrence relations.

As in the Adjacency random block matrix model, in the limit
$d\to \infty$, with $t_a = \frac{Z_a}{d}$ fixed, only the closed walks with
noncrossing sequence of moves contribute to the moments of the Laplacian
block matrix and give the same contribution in 
the case of the bipartite ER graphs and in the case of
random BB graphs.  
We will obtain the exact
g.f. of the moments in this limit; it is the same
on bipartite ER graphs or on random BB graphs.

The enumeration of ${\cal C}_{(2)}$ according to the number of blocks
gives the moments of the Laplacian random block matrix model 
on ER graphs and on the random regular graphs,
in the limit $d\to \infty$ with $t = \frac{Z}{d}$ fixed.

Let us enumerate the bicolored partitions in ${\cal C}_{(n,k)}$ 
according to the number of blocks and of elements with color $2$.
Since these partitions are noncrossing, the irreducible partitions have a 
single block and $n$ elements.
Introduce a parameter $u$ to count the number of elements with color $2$
in an irreducible partition with $n$ elements.
There are $\binom {n}{jk}$ ways to choose $jk$ elements with color $1$
so that, for $u=1$,
$a_h$ is the lacunary sum of binomial coefficients.
Following \cite{gessel} one has
\begin{equation}
1 + a(x, u) = \sum_{n\ge 0} x^n u^n \sum_{j\ge 0} \binom{n}{jk}u^{-jk} = 
\sum_{j\ge 0}u^{-jk} \sum_{n \ge jk} (ux)^n \binom{n}{jk} =
\sum_{j\ge 0}u^{-jk} \frac{(ux)^{jk}}{(1-ux)^{jk+1}}
\end{equation}
so that the g.f. enumerating the irreducible partitions of 
${\cal C}_{(k)}$ according to the number of elements 
and the number of elements with color $2$ is
\begin{equation}
a(x, u) = \frac{(1-ux)^{k-1}}{(1-ux)^k - x^k} - 1
\label{As}
\end{equation}

Using the NCPT Eq. (\ref{nctran}) and Eq. (\ref{As}) one gets, setting $s=t$,
\begin{equation}
f^{nc} = 1 + x \Big(u +  \big(\frac{x f^{nc}}{1 - u x f^{nc}}\big)^{k-1}\Big) f^{nc} (f^{nc}+t-1)
\label{egs2}
\end{equation}
For $u=0$, Eq. (\ref{egs2}) is Eq. (\ref{ban0}).
For $k=2$ Eq. (\ref{egs2}) is the g.f. of the moments
for a model in which $L_{i,i}$ has an extra factor $u = 0, 1$,
\begin{equation}
L_{i,i} = u \sum_{j \neq i} \alpha_{i,j} X_{i,j}
\label{eqMu}
\end{equation}
so that, for $u=0$, $L_{i,j}$ is the Adjacency random block matrix (with changed
sign; this does not affect its moments) and Eq. (\ref{egs2}) is
the EM approximation for the Adjacency random matrix on ER graphs \cite{SC};
for $u=1$ it is the Laplacian random block matrix;
Eq. (\ref{egs2}) is the Marchenko-Pastur distribution.
These results have been obtained, for the Laplacian and in the Adjacency
random block matrix on ER graphs, in the $d \to \infty$
limit with $t = \frac{Z}{d}$ fixed, in \cite{PC} using the decomposition of 
closed walks in primitive walks and the noncrossing sequence of moves in these
walks.

Associate a CC to each number $i=1,\cdots, n$, occurring as base element in a
sequential form $s$ of a partition in ${\cal C}^w_{(n,k)}$  in the following way:
let $m$ be the number of elements with the first color
occurring in $s$ before
the first occurrence of the base element $i$; the CC associated to
$i$ is defined to be $1+m \, mod \, k$.
The CC $a=1,\cdots, k$ of a block of $p$ is the CC of its lowest base element;
see the example in Appendix A.D.
In the case $k=2$, $p$ corresponds to a closed walk on a rooted tree;
a block  with CC $a$ corresponds to an edge,  traversed by the walk,
having parity $a$ for its node closest to the root
and contributing a factor $Z_a$ to the weight assigned
to the walk (see the discussion in the previous section).

In the case of ${\cal C}_{(k)}$ with given CC types,
corresponding in the case $k=2$ to the Laplacian random block matrix
model in the limit $d\to \infty$ with $t_a = \frac{Z_a}{d}$ fixed,
we do not have a suitable extension of the NCPT (the approach in Section II.C
cannot be used, because after decomposing a partition as
$p_1 c_1\cdots p_h c_h$, $c_i$ can have any color).
From now on we will consider the case $k=2$;
to enumerate ${\cal C}_{(n,2)}$ we will use primitive walks 
as in \cite{PC}.

We want to compute $f_a^{nc}(x, t_1, t_2)$, the g.f. enumerating
the closed walks in $\{W_{n,2}, n \ge 1\}$ with noncrossing sequence of moves,
according to the length of the walk and the CC type;
it is related to $f_1^{nc}$ by cyclic permutation of $t_1,t_2$, 
see Eq. (\ref{faf1}). The dependence on $u=0,1$ is not indicated.

Define a primitive walk as a sequence of moves starting at a node $v$ of an
edge $G$, ending at $v$ only at the end of the last move.

Any closed walk starting and ending at vertex $j_0$ has
a unique representation as concatenation of primitive walks, each one
starting and ending at vertex $j_0$.
This implies that $T_{j_0}$ in Eq. (\ref{muTM}), with $j_0$ of type $a$,
can be decomposed as
\begin{eqnarray}
T_{j_0}(x; X, \alpha) =\sum_{n\geq 0} \left( \sum_j B_{j_0,j}(x; X, \alpha) \right)^n \qquad
\label{T00A}
\end{eqnarray}
where $B_{j_0, j}(x; X, \alpha)$ is the g.f. of the matrices 
corresponding to the primitive walks starting with edge $(j_0, j)$.

Following \cite{PC} one has
\begin{equation}
\langle B_{r,s}(x; X, \alpha) \rangle_I = \alpha_{r,s} X_{r,s} 
g_a(x, t_1, t_2)
\end{equation}
where $\langle \cdots \rangle_I$ is the average of the internal random
variables $\alpha_{i,j} X_{i,j}$ corresponding to edges different from
$(r,s)$, $a$ is the type of $r$ and 
$t_a g_a$ be the g.f. of the number of primitive walks
(i.e. with a single return to the root)
with given CC types, starting at a node of type $a$ and having noncrossing 
sequence of moves. 
Using Proposition $2$ in \cite{PC}, Eqs. (\ref{muTM}, \ref{fl}) and
$\langle \alpha_{j_0,j} \rangle = t_a$, it follows that
\begin{equation}
f_a^{nc}(x,t_1,t_2) = 1 + \sum_{r\ge 1} \sum_{b=1}^r N(r, b) t_a^b 
g_a^r(x,t_1,\cdots,t_k)
\label{fN}
\end{equation}
where $t_a^b$ comes from averaging $\alpha_{j_0,j_i}$ for $b$ distinct edges
$(j_0, j_i)$.
Using the g.f. of the Narayana polynomials we get
\begin{equation}
g_a f_a^{nc} (f_a^{nc} + t_a - 1) = f_a^{nc} - 1
\label{gfb1}
\end{equation}

Let us compute $g_a(x,t_1,t_2)$ in term of $f_a^{nc}(x,t_1,t_2)$.
A primitive walk rooted at $v_1$, with type $a$, can be the move of second 
kind up and down the first edge $G=(v_1,v_2)$,
giving the term $u x$, or it can go to $v_2$ with a move of the first kind;
from there it can do an arbitrary closed walk starting from
node $v_2$ and not traversing $G$, which gives
a factor $f_{3-a}^{nc}$ (since there is one move of the first kind before it,
$v_2$ has CC is $3-a$);
after that, the walk can return to $v_1$
or one can have a move of the second kind
 $G$ with root $v_2$, contributing a factor $u x$, and after that
another arbitrary closed walk starting from
node $v_2$ and having no edge in common with $G$, and so on;
therefore at the node $v_2$ one has a factor 
$f_{3-a}^{nc} + f_{3-a}^{nc} u x f_{3-a}^{nc} + \cdots
= \frac{f_{3-a}^{nc}}{1 - f_{3-a}^{nc} u x}$.
Then it returns with a move of first kind to $v_1$.
Therefore one gets
\begin{equation}
g_a(x,t_1, t_2) = u x + x^2 \frac{f_{3-a}^{nc}(x,t_1, t_2)}{1 - f_{3-a}^{nc}(x,t_1, t_2) u x}
\label{egs}
\end{equation}

The system of equations Eqs. (\ref{gfb1}, \ref{egs}) gives a system
of algebraic equations for $f_a^{nc}$.

In the case $u=0$ one gets from Eq. (\ref{egs})
\begin{equation}
g(x,t_1, t_2) \equiv f_a^{nc}(x,t_1, t_2) g_a(x,t_1, t_2) = x^2 f_1^{nc}(x,t_1, t_2) f_2^{nc}(x,t_1, t_2)
\label{egsu0}
\end{equation}
which is the second equation in Eq. (\ref{f1nck}) for $k=2$; 
Eq. (\ref{gfb1}) gives Eq. (\ref{gfc1}); therefore we reobtain using
primitive walks these equations for the Adjacency random block matrix model.

We will set $u=1$ in the following.

From Eqs. (\ref{gfb1}, \ref{egs}) one gets
\begin{equation}
x f_a^{nc}(f_a^{nc} + t_a - 1) + (1-f_a^{nc})(1 - x f_{3-a}^{nc}) = 0
\end{equation}
from which by variable elimination one obtains
\begin{equation}
x(t_a-t_{3-a})(f_a^{nc})^3 +
    (xt_a^2 - x t_1 t_2 + 2xt_{3-a} - t_a+t_{3-a})(f_a^{nc})^2
 +(xt_1t_2 - xt_a - xt_{3-a} + t_a - 2t_{3-a})f_a^{nc} + t_{3-a} = 0
\label{lapf1a}
\end{equation}
Define 
\begin{equation}
h = t_2 f_1^{nc} + t_1 f_2^{nc}
\label{laph}
\end{equation}
Defining $\alpha = t_2 f_1^{nc}$ and $\beta = t_1 f_2^{nc}$, the two equations 
Eqs. (\ref{lapf1a})
are two cubic equations in $\alpha$ and $\beta$, $P(\alpha)=0$ and $Q(\beta)=0$
(the dependence of $P$ and $Q$ on $x$, $t_1$ and $t_2$ is not indicated 
to simplify the notation).
$h$ in Eq. (\ref{laph}) is the sum of two roots, $\alpha + \beta$ of these
cubic equations.

The polynomial
$\prod_{\alpha,\beta} (h - \alpha - \beta)$, where $\alpha$ and $\beta$
are the roots of $P(\alpha)=0$ and $Q(\beta)=0$ respectively, is given by
 the resultant of $P(\alpha)$ and of $Q(h - \beta)$, considered as a
polynomial in $\beta$. 
This resultant is a polynomial in $h$ of degree $9$,
factoring on the real field in a cubic polynomial and in a polynomial of order $6$, which we computed with Sage \cite{sage}.

Evaluating the cubic factor of the resultant 
using $h$ in Eq. (\ref{laph}) with $f_a^{nc}$,
computed till order $10$ in the Taylor expansion around $x=0$, obtained using
the moments method, one gets $0$ to this order; the same check fails for the
factor of order $6$ of the resultant; therefore $h$ satisfies the cubic equation
\begin{equation}
h^3 + a_2(x) h^2 + a_1(x) h + a_0(x) = 0
\label{cublapl}
\end{equation}
with
\begin{eqnarray}
& &a_2(x) = 2(t_1 t_2 -t_1-t_2) - (t_1+t_2)x^{-1} \nonumber \\
& &a_1(x) = (t_1 t_2-t_1-t_2)^2 + (-t_1 t_2^2 - t_2 t_1^2 + 2t_1^2 + 2t_2^2 + 
2t_1 t_2)x^{-1} + t_1 t_2 x^{-2} \nonumber \\
& &a_0(x) = (t_1^2 + t_2^2)(t_1 t_2-t_1-t_2) x^{-1} - t_1 t_2 (t_1+t_2)  x^{-2}
\label{lapf2}
\end{eqnarray}
For $x \to \infty$ this cubic equation becomes 
$h(h + t_1 t_2 - t_1 - t_2)^2 = 0$, so that
\begin{equation}
h(x=\infty) = t_1 + t_2 - t_1t_2
\label{hypo1}
\end{equation}
and the spectral distribution has a delta-function term
\begin{equation}
\rho_{L,0}^{nc}(\lambda) = \frac{t_1 + t_2 - t_1t_2}{t_1+t_2} \theta(t_1 + t_2 - t_1t_2)\delta(\lambda)
\label{hypo0}
\end{equation}
where $\lambda = \frac{1}{x}$.
This delta-function term can be understood in terms of
the interpretation of this Laplacian random block
matrix model as a mean field approximation of a BB
$d$-dimensional elastic network and of the isostatic condition
Eq. (\ref{iso}):
in the region under the hyperbola (hypostatic region) in Eq. (\ref{iso})
the number of degrees of freedom is larger than the number of constraints, so
there are $d N - E$ zero modes, where $E = N_1 Z_1 = N_2 Z_2$ is the number of
edges; the fraction of zero modes is $\frac{t_1 + t_2 - t_1t_2}{t_1+t_2}$,
and the spectral distribution has a delta-function term Eq. (\ref{hypo0}).

The continuous part of $\rho_L$
\begin{equation}
    \rho_{L,c}^{nc}(\lambda) = 
- \frac{\texttt{Im} h(\frac{1}{\lambda})}{\pi \lambda(t_1+t_2)}
\label{rhoeqL}
\end{equation}
is given by the Cardano solution 
Eqs. (\ref{pqD},\ref{imh},\ref{rhoeqL})
of the cubic equation given by (\ref{cublapl}, \ref{lapf2});
$\rho_L(\lambda) = \rho_{L,c}^{nc}(\lambda) + \rho_{0,L}^{nc}(\lambda)$.

We have checked for a few values of $t_1, t_2$,
using multi-precision numerical integration \cite{mpmath}, that the integral 
of $\rho_L$ is one.
As another check, the first $10$ moments computed with the moments
method agree with those computed from the resolvent.

In the regular graph case $t = t_1 = t_2$ the cubic equation Eq. (\ref{cublapl})
gives
\begin{equation}
{\left(h t^{2} + h^{2} - 2 \, h t - \frac{h t}{x} + \frac{2 \, t^{2}}{x}\right)} {\left(t^{2} + h - 2 \, t - \frac{t}{x}\right)} = 0
\label{cublaplreg}
\end{equation}
The first factor equal to zero gives the Marchenko-Pastur distribution
for $f = \frac{h}{2t}$.

On the isostatic line Eq. (\ref{iso}), $h$ diverges for $x\to \infty$; 
in this limit Eqs. (\ref{cublapl}, \ref{lapf2}) can be approximated to
$h^3 + a_1 h \approx 0$ leading to the fact that on the isostatic line
the resolvent and the spectral distribution diverge as
\begin{equation}
r^{nc}(w) \approx (-w)^{-\frac{1}{2}} \sqrt{\frac{t_1^2 - 2t_1 + 2}{t_1^2}};
\qquad
\rho_L^{nc}(\lambda) \approx \frac{\lambda^{-\frac{1}{2}}}{\pi} \sqrt{\frac{t_1^2 - 2t_1 + 2}{t_1^2}}
\label{isor}
\end{equation}
for $w, \lambda \to 0$.

In the case of the isostatic point $t_1=t_2=2$, this divergence
has been discussed in \cite{parisi}.

The support of $\rho^{nc}_{L,c}$ is where
the discriminant $D(x)$ of the cubic Eq. (\ref{lapf2}) is positive;
one has $D(x) = \sum_{k=2}^6 D_k x^{-k}$ with 
$D_6=-\frac{1}{108} \, {\left(t_{1} - t_{2}\right)}^{2} t_{1}^{2} t_{2}^{2} \le 0$,
negative for $t_1 \ne t_2$. 
Define $\hat D(\lambda) = D(x) x^2$, which is quartic in $\lambda = \frac{1}{x}$.
It can be positive in one or two intervals.
Varying parameters starting from a case in which the support of $\rho^{nc}$
consists of two intervals, the point in which these intervals touch has
$\hat D(\lambda_t) = \hat D'(\lambda_t) = 0$. 
From these equations one gets $\lambda_t$ as the ratio of two polynomials
in $D_k$
and one finds that at the separation between the 
one- and two-band regions of the spectral distribution
\begin{eqnarray}
& &(t_2-t_1)^6 P_1^3 P_2 = 0 \nonumber \\
& &(t_2-t_1)^8 (t_1^2 - 2 t_1 t_2 + t_2^2 + t_1) (t_1^2 - 2t_1t_2 + t_2^2 + t_2)P_1^3  P_3 = 0
\label{eqst1t2}
\end{eqnarray}
where
\begin{eqnarray}
P_1 = ((t_2-t_1)^2 - t_1 - t_2)^3 - 27(t_2-t_1)^2t_1t_2
\label{eqtr}
\end{eqnarray}
and $P_2$ and $P_3$ are polynomials with larger degrees.
The only non-trivial solution to Eqs. (\ref{eqst1t2}) for $t_1 \neq t_2$ 
is for $P_1=0$.

A parametric solution of this equation is
\begin{eqnarray}
\psi &=& y^2 - \frac{4}{27} (1-y)^3 \nonumber \\
(t_1, t_2) &=& (\frac{y}{2\psi} + \frac{1}{2\sqrt{\psi}}, 
               \frac{y}{2\psi} - \frac{1}{2\sqrt{\psi}}) \nonumber \\
(t_1, t_2) &=& (\frac{y}{2\psi} - \frac{1}{2\sqrt{\psi}},
                \frac{y}{2\psi} + \frac{1}{2\sqrt{\psi}})
\label{eqtrp}
\end{eqnarray}
with $y \in [\frac{1}{4}, 1]$. The lower branch starts in $y=1$,
$t_1=1, t_2=0$, the upper branch in $t_1=0, t_2=1$.
For $y \to \frac{1}{4}$, $\psi \to 0$, so the two transition branches
are asymptotically parallel to the $t_2=t_1$ line.

The curve $P_1(t_1, t_2) = 0$ is a line of transition points, from the
region in which the support of the continuous part of $\rho$ is a 
single band to the region in which it has two bands; let us call it 
$\gamma^{nc}$.

For $t_2 = t_1$ the resolvent satisfies the Marchenko-Pastur distribution:
the cubic polynomial  Eq. (\ref{lapf1a}) in the g.f.
$f^{nc}=f_1^{nc}=f_2^{nc}$ factorizes in a quadratic factor corresponding to the
Marchenko-Pastur distribution and a linear factor, see Eq. (\ref{cublaplreg}).
The line $t_2 = t_1$ is in the one-band region in the $t_1,t_2$ plane.

\begin{figure*}[h]
\begin{center}
\epsfig{file=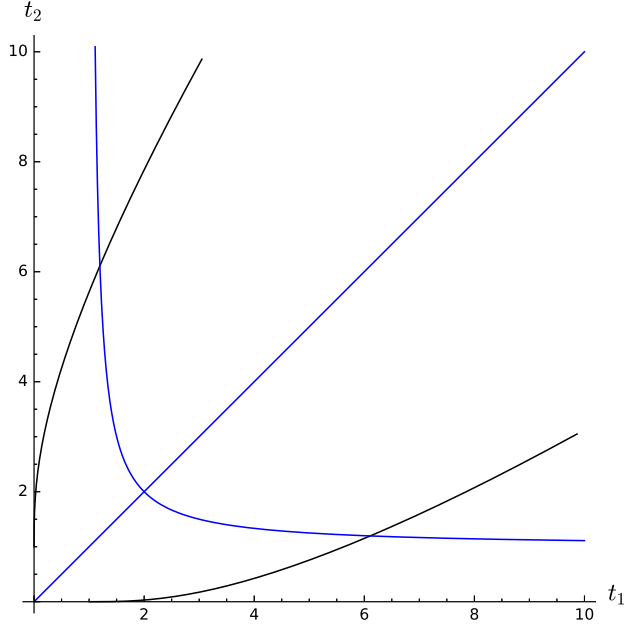, width=8.55cm  }
\caption{Diagram of the regions in the $t_1,t_2$ plane for the spectral
    distribution $\rho_L^{nc}$ in the Laplacian random block matrix model 
    for $d \to \infty$, with $t_a = \frac{Z_a}{d}$ fixed,
both in the case of bipartite ER graphs or on
BB random graphs.
The straight line $t_2 = t_1$ is the regular graph case, with the Marchenko-Pastur
distribution. $\rho_L^{nc}$ has no delta-function singularity
    above the 'isostatic' hyperbola
$t_2 = \frac{t_1}{t_1-1}$; below it ('hypostatic' region)
it has the delta-function singularity Eq. (\ref{hypo0}).
Approaching the hyperbola, $\rho_L^{nc}$ has a peak going
    to infinity on the hyperbola, see Eq. (\ref{isor}).
The two transition lines, in black, are given by
Eq. (\ref{eqtrp}); in the outer regions delimited by them, the support of
$\rho_L^{nc}$ in $\lambda > 0$
consists of two bands; in the inner region, the support consists of a
single band.
}
\label{figtr1}
\end{center}
\end{figure*}

In Figure \ref{figtr1} the isostatic line and the transition line
$\gamma^{nc}$ are drawn.

We made some simulations with random block matrices for the spectral density 
of random block Laplacian matrices with small $d$ on random BB graphs.
In the few simulations we made for $d > 1$, we find that, for small $d > 1$,
$\gamma^{nc}$ separates approximately a one-band region from a 
two-band region, with a quasi-gap between them; the separation
becomes more definite far from $\gamma^{nc}$.

\begin{figure*}[h]
\begin{center}
    \epsfig{file=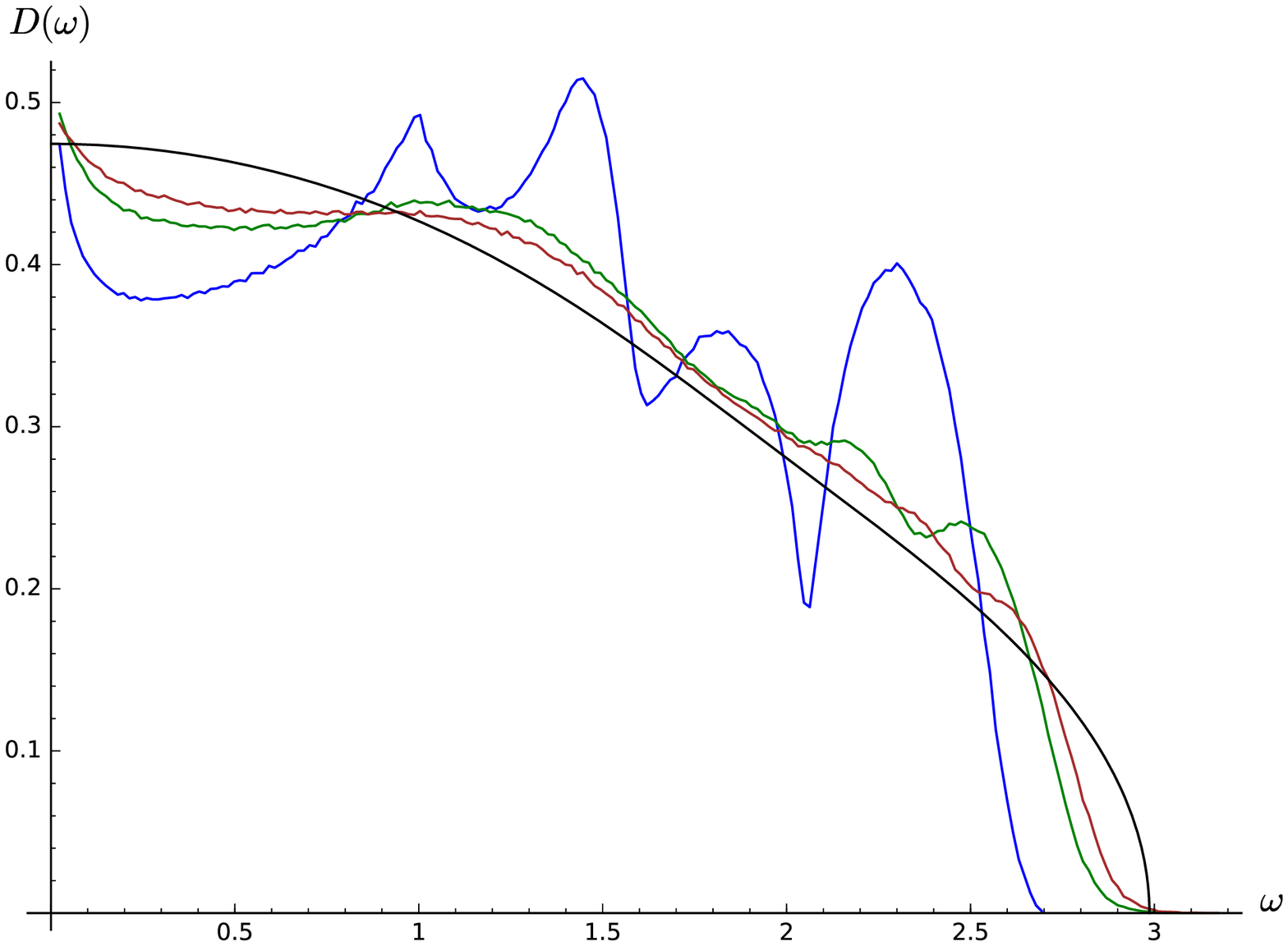, width=5.00cm  } \quad 
    \epsfig{file=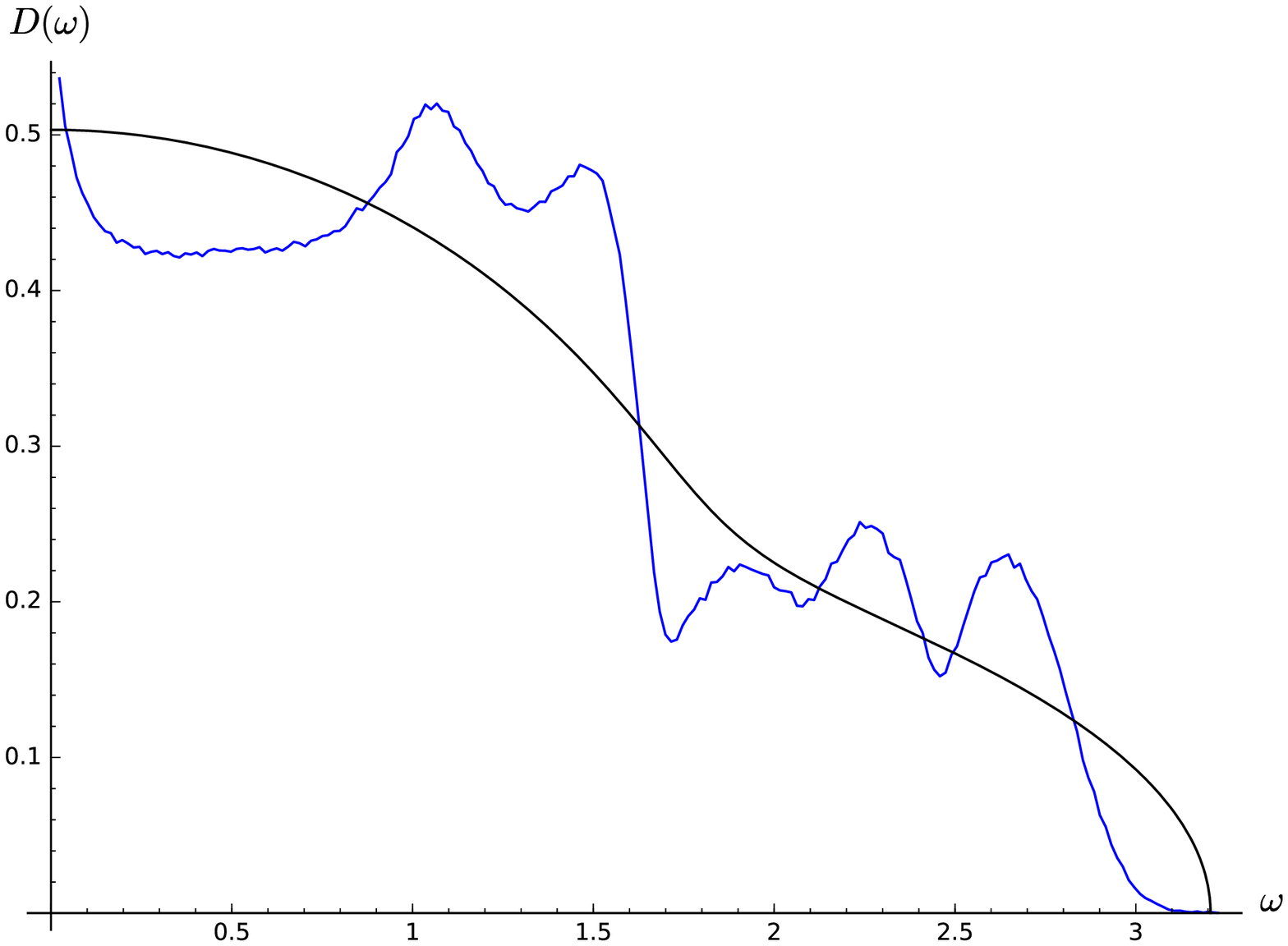, width=5.00cm} \quad  
    \epsfig{file=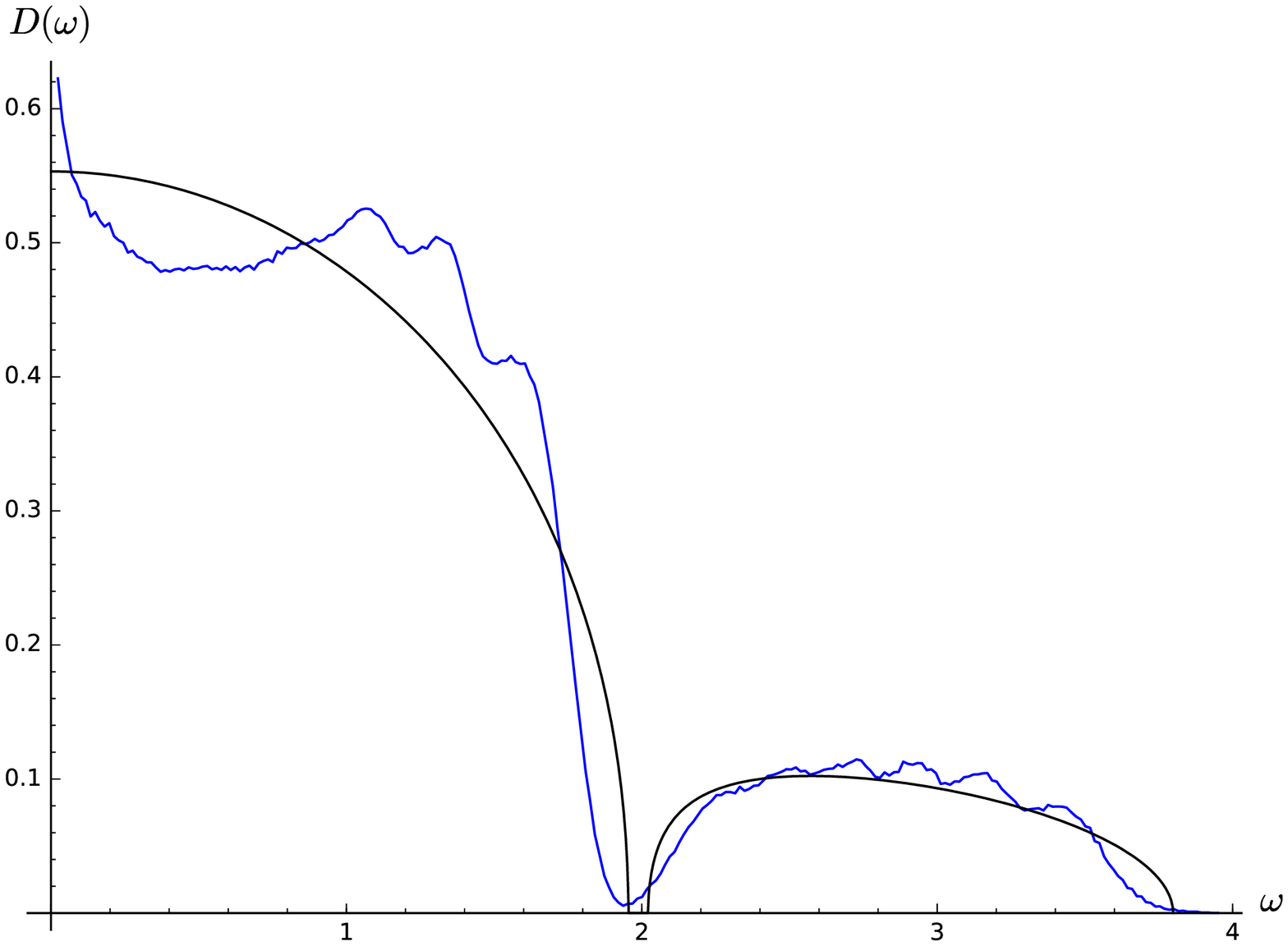, width=5.00cm}
\caption{Simulations for the density of states at isostatic points
in the Laplacian random block matrix model on random BB graphs.
    The left hand side figure shows the density of states
for $t_1=\frac{3}{2}, t_2=3$; the blue line is a $d=2$ simulation with
$Z_1=3$, $Z_2=6$ with $200$ random graphs with $N_1=1200$ and $N_2=600$ nodes;
the green line is a $d=4$ simulation with $Z_1=6$, $Z_2=12$ with $500$
random graphs with $N_1=600$ and $N_2=300$ nodes;
the brown line is a $d=6$ simulation with $Z_1=9$, $Z_2=18$ with $500$
random graphs with $N_1=360$ and $N_2=180$ nodes.
The center figure shows the density of states for
$t_1=\frac{4}{3}, t_2=4$; the blue line is a $d=3$ simulation with
$Z_1=4$, $Z_2=12$ with $200$ random graphs with $N_1=600$ and $N_2=200$ nodes.
The right hand side figure shows the density of states $t_1=\frac{7}{6}$, $t_2=7$;
the blue line is a $d=6$ simulation with $Z_1=7$, $Z_2=42$ with
$200$ random graphs with $N_1=420$ and $N_2=70$ nodes.
These line are drawn for $\lambda > 0.023$, with spacing $0.016$
The black line in these figures is the density of states
    $D(\omega) = 2\omega\rho_L^{nc}(\omega^2)$
    in the limit $d \to \infty$, $t_a = \frac{Z_a}{d}$ fixed.  }
\label{Fig3q2}
\end{center}
\end{figure*}

In Fig. \ref{Fig3q2} there are simulations for the density of states
$D(\omega) \equiv 2\omega \rho_L(\omega^2)$, where $\lambda = \omega^2$,
at three isostatic points with $t_1 \neq t_2$, 
compared with those corresponding to $\rho_L^{nc}$.
The density of states is more convenient
to represent the behavior at small frequencies than the spectral
density, going to $\frac{2}{\pi} \sqrt{\frac{t_1^2 - 2t_1 + 2}{t_1^2}}$
for $\omega \to 0$, while for $\lambda \to 0$ the spectral distribution
diverges as in Eq. (\ref{isor}).
In the left hand side figure
there is the case $t_1=\frac{3}{2}$, $t_2=3$ in $d=2,4,6$,
in particular $Z_1=3, Z_2=6$ in $d=2$.
As $d$ increases the density of states approaches the
$d=\infty$ case.
In the center figure there is the case $t_1=\frac{4}{3}$, $t_2=4$ in $d=3$,
with degrees $Z_1=4$ and $Z_2=12$.
In the right hand side figure there is the case $t_1=\frac{7}{6}$, $t_2=7$;
in $d=\infty$ there are two close bands; in $d=6$, with $Z_1=7$ and $Z_2=42$,
the density of states is close to zero near the gap between the two
bands of the $d=\infty$ case.
In these simulations one finds that for finite $d$ the density of states
around $\omega = 0$ is not flat, having a peak there,
absent in the $d\to \infty$ case.

\begin{figure*}[h]
\begin{center}
\epsfig{file=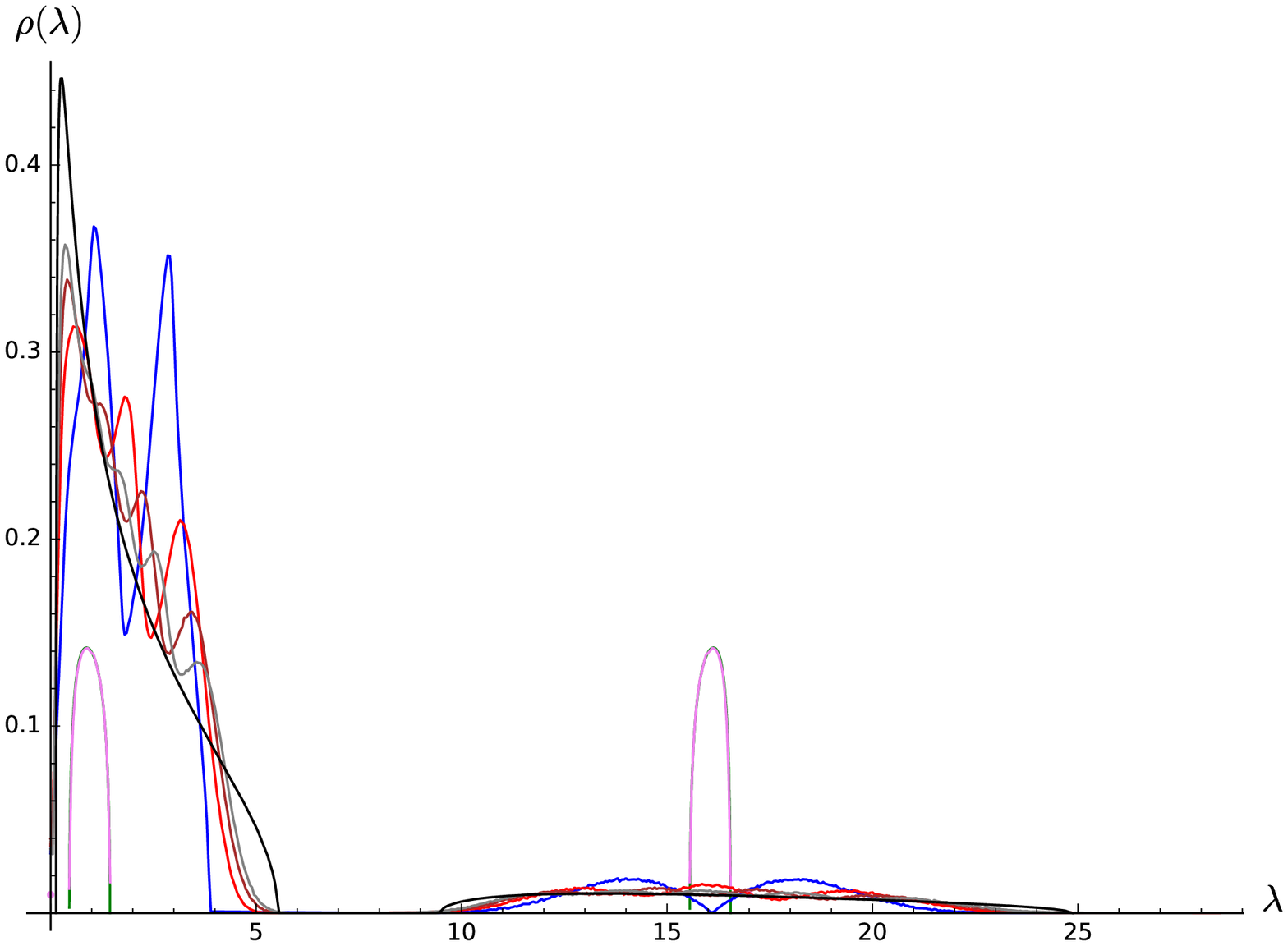, width=15.00cm  }
\caption{Spectral distribution of the Laplacian 
    random block matrix on random BB graphs
    with $t_1=\frac{Z_1}{d}=2$ and $t_2=\frac{Z_2}{d}=15$.
    The green line is the $d=1$ exact curve Eq. (\ref{rholapd1}). 
The violet line is a $d=1$
simulation with $Z_1=2$, $Z_2=15$ with $500$ random graphs with $N_1=1800$, $N_2=240$
nodes. In the $d=1$ case the delta-function in $\lambda=2$ is not indicated;
adding its contribution the area of the first band is the same as for 
    $\rho_L^{nc}$.
The blue line is a $d=2$ simulation with $Z_1=4$, $Z_2=30$ with $500$ random graphs
with $N_1=900$, $N_2=120$ nodes.
The red line is a $d=3$ simulation with $Z_1=6$, $Z_2=45$ with $500$ random graphs
with $N_1=600$, $N_2=80$ nodes.
The brown line is a $d=4$ simulation with $Z_1=8$, $Z_2=60$ with $500$ random graphs
with $N_1=480$, $N_2=64$ nodes.
The gray line is a $d=5$ simulation with $Z_1=10$, $Z_2=75$ with $500$ random graphs
with $N_1=540$, $N_2=72$ nodes.
    The black line is the $\rho_L^{nc}$ given by 
Eqs. (\ref{pqD},\ref{rhoeq},\ref{imh},\ref{cublapl},\ref{lapf2}).
We used bins of size $0.05$ to collect eigenvalues in these simulations.
}
\label{Figsp6}
\end{center}
\end{figure*}

Let us give another example in the two-band region, 
at $t_1=2, t_2=15$,  far from $\gamma^{nc}$.
The results of the simulations are given in Fig. \ref{Figsp6}; 
with the increase of $d$ the
spectral distribution approaches the exact $d\to \infty$ solution.

It appears that there is some regularity in the bands:
the second band has a number of peaks equal to $d$.

Let us make an observation about the area of the two spectral bands.

In the case $d=1$ the spectral distribution is given by 
Eqs. (\ref{rhoL0},\ref{rholapd1}) in Appendix C.B.
The two bands are placed symmetrically around $t_m \equiv \frac{\min(t_1, t_2)}{2}$ (see Eq. (\ref{ev5a}));
there is furthermore a delta-function term in $\min(t_1,t_2)$.
We have checked, using multiprecision numerical integration \cite{mpmath}, 
in about a hundred cases that the area
of the second band in the $d=1$ and the $d=\infty$ case are equal to
$\frac{\min(t_1, t_2)}{t_1 + t_2}$, which is also the area of the positive
band for $\rho_{A}^{nc}$ in the region outside the unit square,
and for $\rho_{A}$ in $d=1$.

In the cases in which there is a clear separation between the bands,
in the few simulations we made in $d > 1$, we found that
the area of the second band is approximately the same as in $d=1$ and 
$d=\infty$; 
the sum of the areas of the first and of the second band is one;
there is no delta-function term in the spectral density in $d > 1$.

\vspace{4 mm}

We have seen in the previous Section that numerical simulations indicate that 
in the case of the Adjacency random block matrix, for $t_1$ and $t_2$
large, $\rho_A^{nc}$  approaches $\rho_A$
with $d=1$. The former distribution is the leading orders in the 
$\frac{1}{s}$-expansion, 
so one can expect that at higher order the approximation improves.

In the case of the Laplacian random block matrix on bipartite ER graphs
with $t_2 > t_1$, $\rho_L^{nc}$ is
the leading order in the $\frac{1}{s}$-expansion, but we do not know
how to compute the next orders, since an extended NCPT for this case is
not available. This limits the usefulness of $\rho_L^{nc}$
as an approximation of $\rho_L$ on bipartite ER graphs;
numerical simulations indicate that $\rho_L^{nc}$ is not a good
approximation for the latter distribution; however there is some
similarity between them; the latter
is characterized by a transition from one to two bands, like in the former,
but in this case the transition is smoother.
In a few simulations with
$t_2 > t_1 \ge 2$, we find a distribution with
a single band for $t_2$ close to $t_1$; as $t_2$ increases so that
$(t_1,t_2)$ reaches $\gamma^{nc}$, there is still one 
band, but at the joining point between the two bands of the $\rho_L^{nc}$,
 one can see the beginning of a separation of the $d=1$ band
in two. In these simulations, for $t_2$ not close to $t_1$, each band
is characterized by oscillations or sub-bands,
in which the distance between the peaks is approximately $1$.

\begin{figure*}[h]
\begin{center}
\epsfig{file=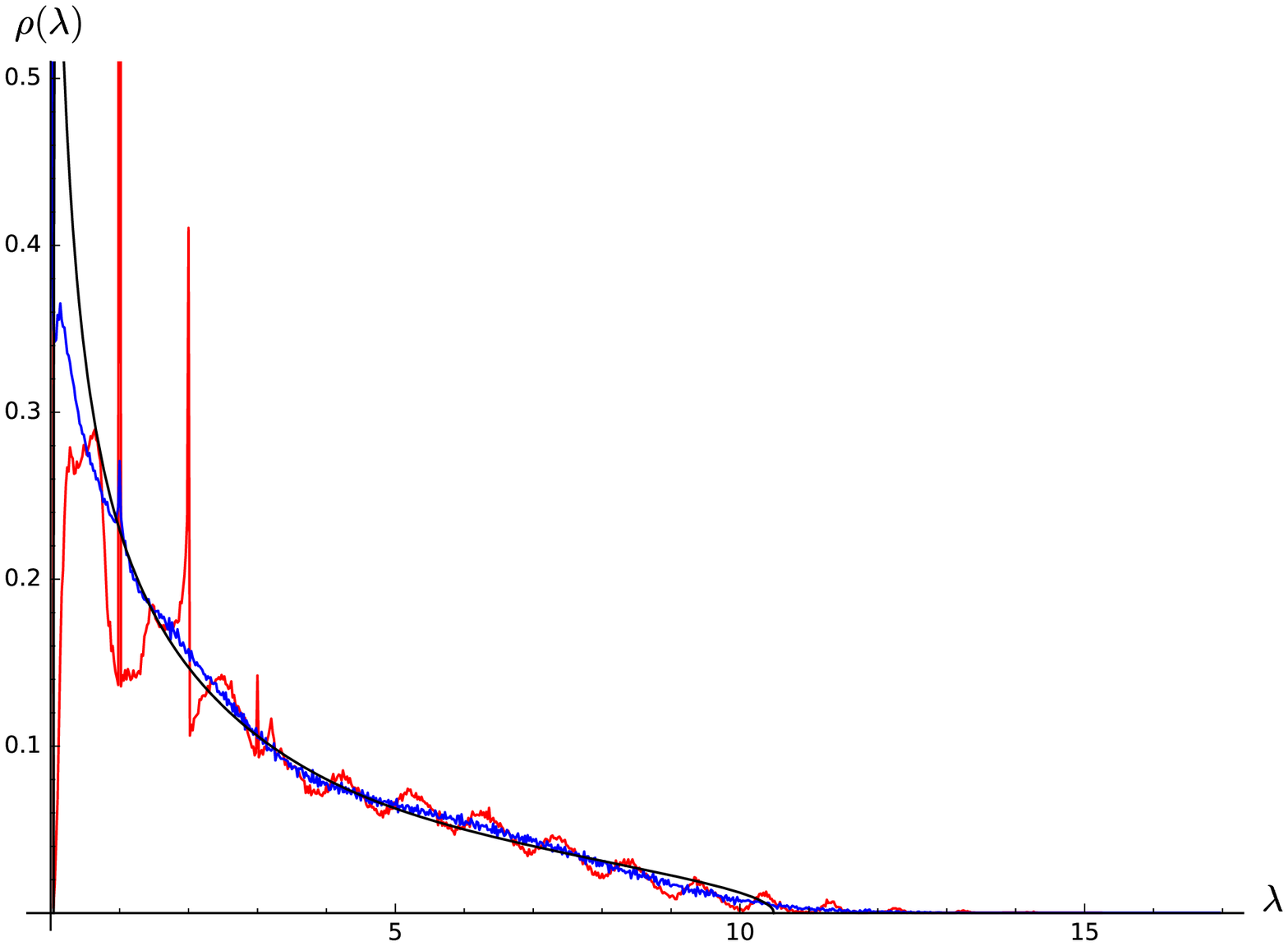, width=5.00cm  } \quad
\epsfig{file=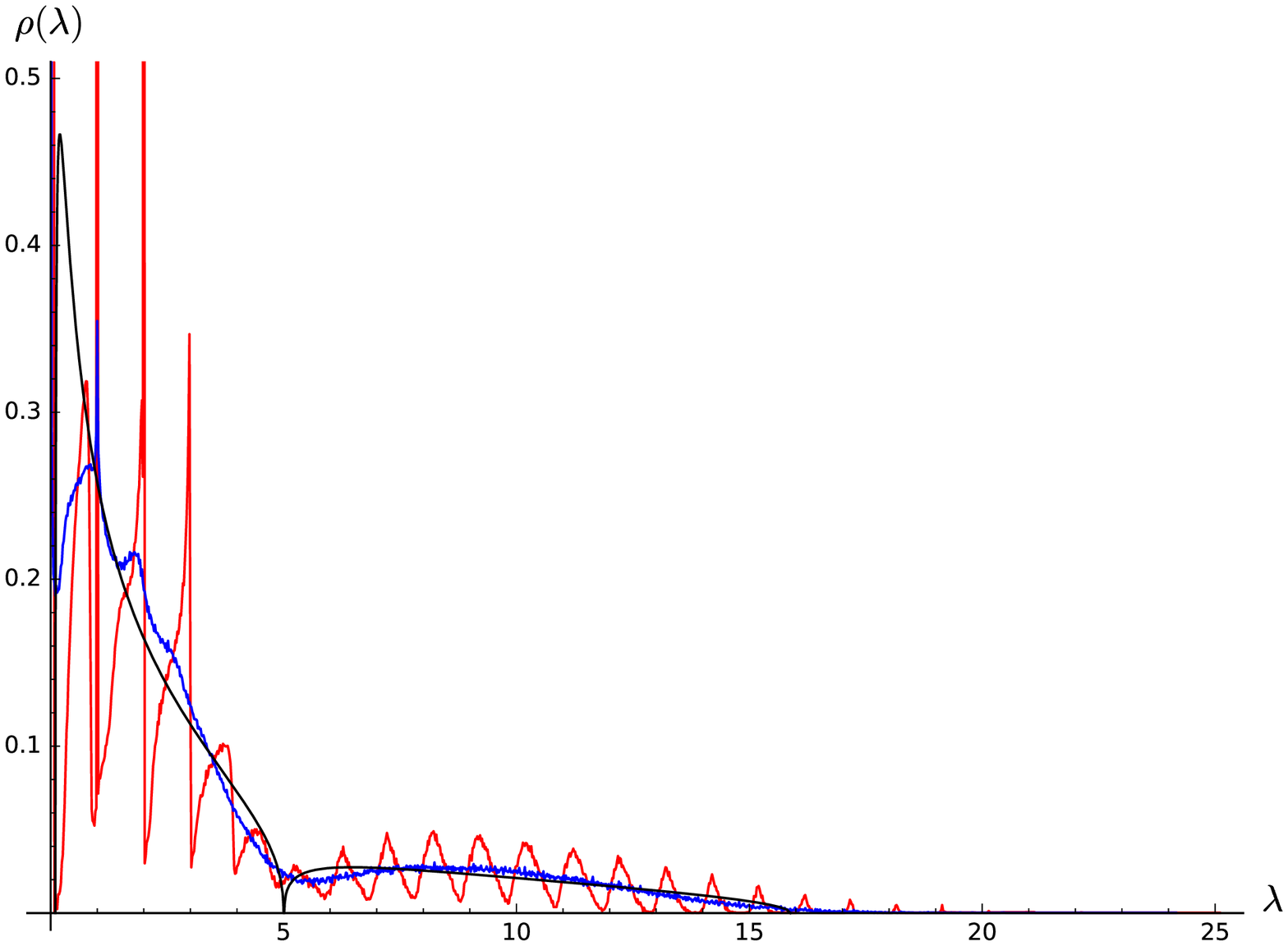, width=5.00cm  } \quad
\epsfig{file=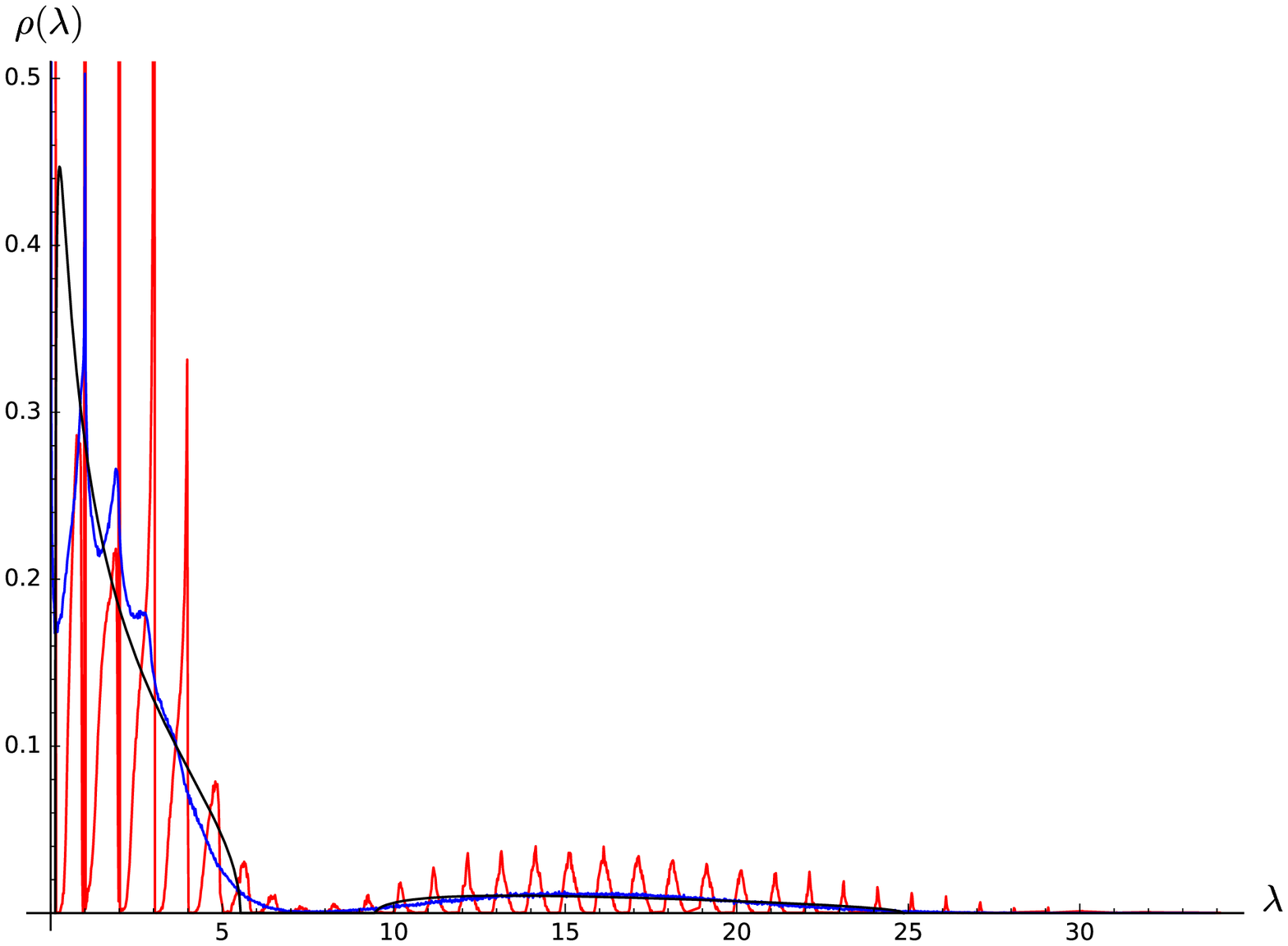, width=5.00cm  }
\caption{Spectral distribution for the Laplacian
random block matrix model on bipartite ER graphs with $t_1=2$ and
    with $t_2 = 4, 8, 15$, for $d=1, 2$ and $\infty$.
Each simulation is made with $100$ Laplacian random block matrices.
In the left hand side figure the red line is a $d=1$ simulations with 
$Z_1=2$, $Z_2=4$ and $N_1=2000$, $N_2=1000$ nodes;
the blue line is a $d=2$ simulation with $Z_1=4$, $Z_2=8$ and
$N_1=800$, $N_2=400$ nodes.
In the center figure the red line is a $d=1$ simulations with
average degrees $Z_1=2$, $Z_2=8$ and $N_1=2000$, $N_2=500$ nodes;
the blue line is a $d=2$ simulation with average degrees $Z_1=4$, $Z_2=16$
and $N_1=1280$, $N_2=320$ nodes.
In the right hand side figure the red line is a $d=1$ simulations with
average degrees $Z_1=2$, $Z_2=15$ and $Z_1=2250$, $Z_2=300$ nodes;
the blue line is a $d=2$ simulation with average degrees  $Z_1=4$, $Z_2=30$ 
and $N_1=900$, $N_2=120$ nodes.
    The black line is  $\rho_L^{nc}$ given by
Eqs. (\ref{pqD},\ref{rhoeq},\ref{imh},\ref{cublapl},\ref{lapf2}).
Bins of size $0.02$ are used to collect eigenvalues in the simulations;
the distribution around $\lambda=0$
is not well represented in these graphs; the first bin accounts approximately
for $10\%$ and $5\%$ of the area of the distributions respectively in
$d=1$ and $d=2$.
}
\label{Fig215er}
\end{center}
\end{figure*}

In Fig. \ref{Fig215er} there are simulations for $\rho_L$
on bipartite ER graphs, with $t_1=2$
and $t_2=4, 8$ and $15$. For $t_2=8$, close to the point
$(2, 7.86)$ on $\gamma^{nc}$, the band starts its separation in two bands;
for $t_2=15$ there are two non completely separate bands; in the quasi-gap
between them there are few eigenvalues, grouped in small peaks. 
There are oscillations in the
distribution, which become wider as $t_2$ increases. In this figure
there are also simulations for $\rho_L$ in $d=2$ on bipartite ER graphs;
in them the wide oscillations of the $d=1$ case are absent; the spectral
distribution in $d=2$ is closer to $\rho_L^{nc}$ than the one  in $d=1$.

\begin{figure*}[h]
\begin{center}
\epsfig{file=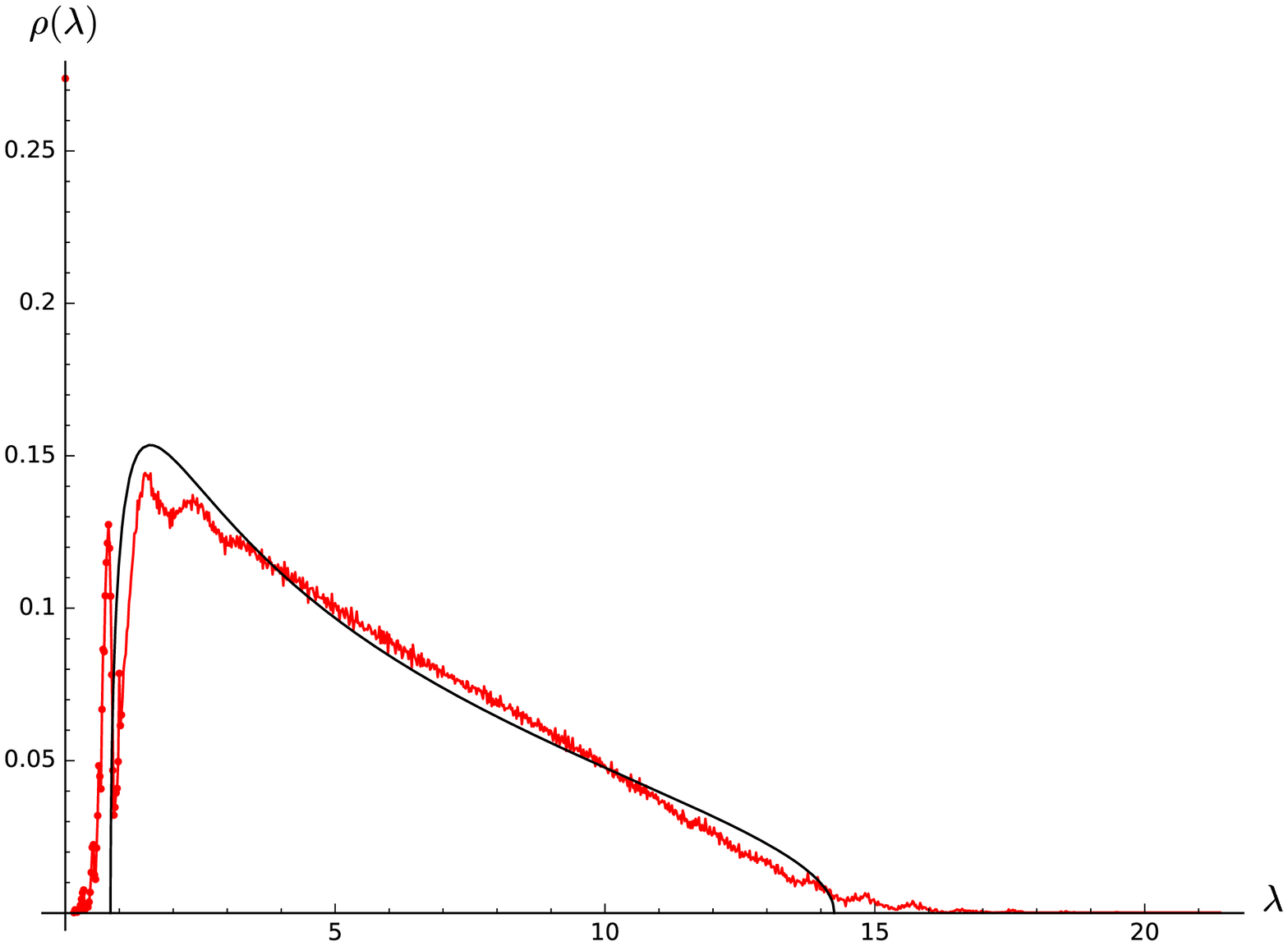, width=5.00cm  } \quad
\epsfig{file=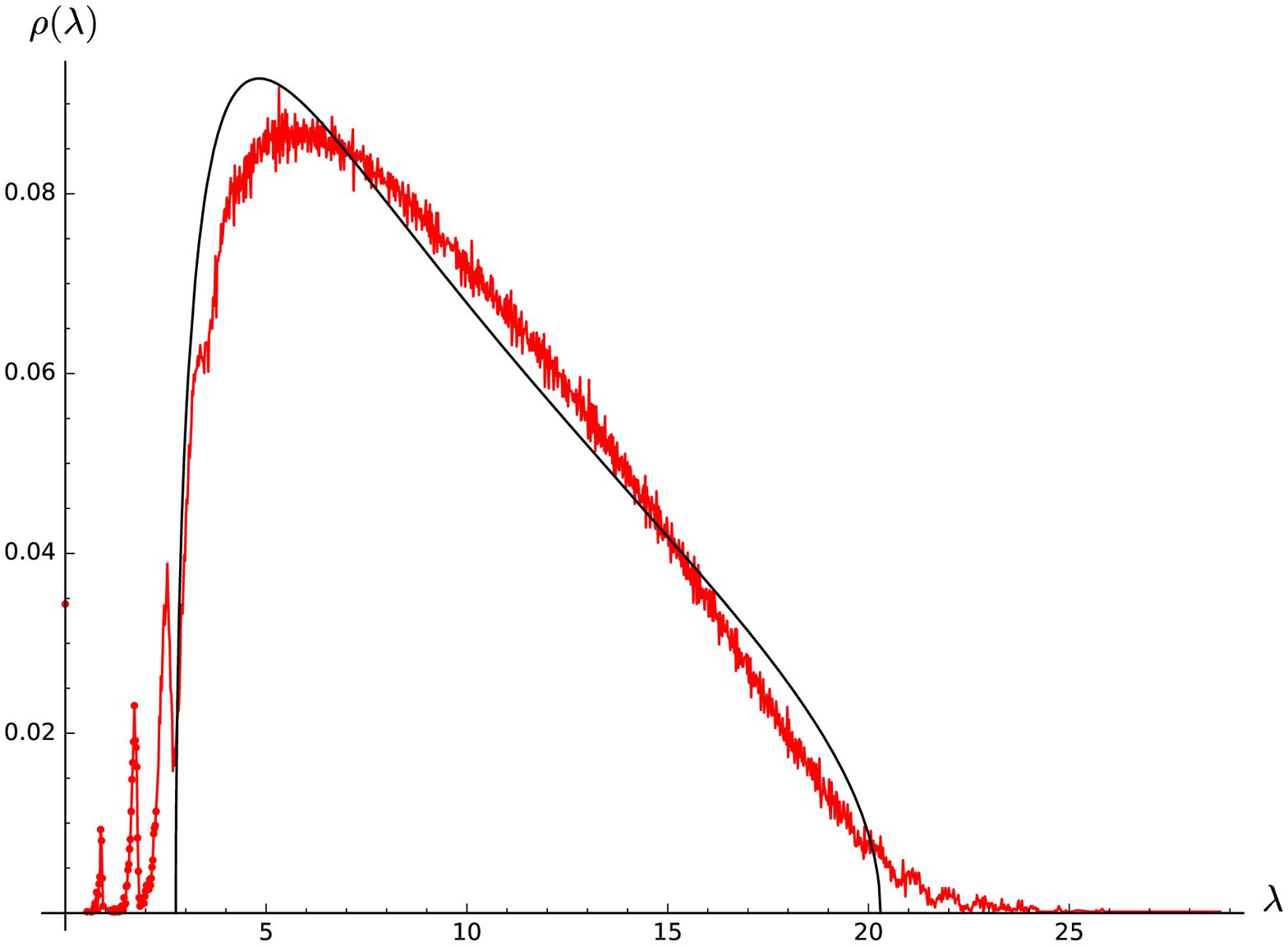, width=5.00cm  } \quad
\epsfig{file=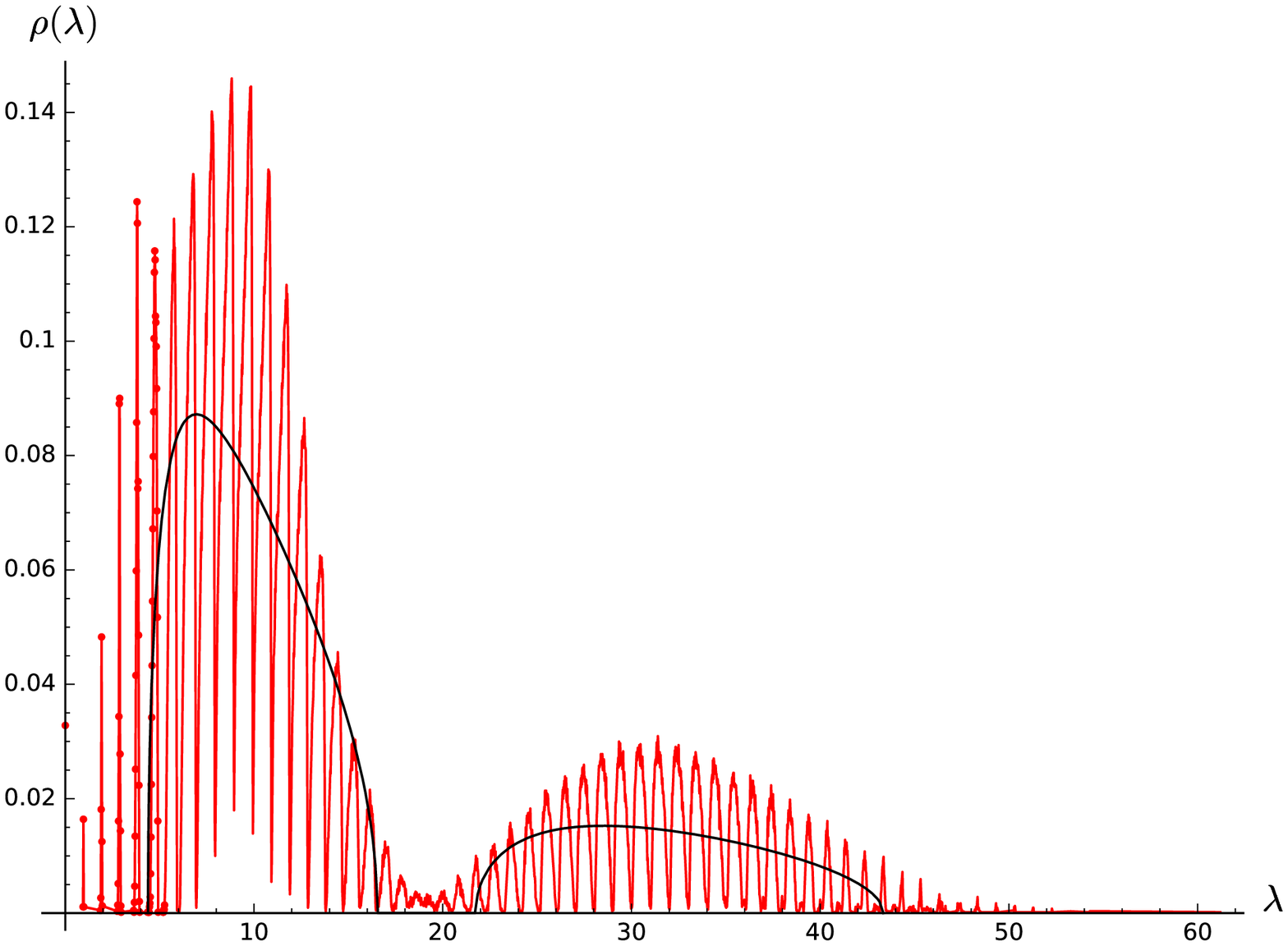, width=5.00cm  } \quad
\caption{Spectral distribution for the Laplacian
random block matrix model on bipartite ER graphs.
Each simulation is done with $100$ Laplacian random block matrices 
with ER graphs with average degrees $t_1$ and $t_2$.
In the left hand figure the red line is a simulations with
$t_1=5$, $t_2=6$ and with $N_1=1800$, $N_2=1500$ nodes.
In the center figure the red line is a simulations with
$t_1=9$, $t_2=10$ and with $N_1=1700$, $N_2=1530$ nodes.
In the right hand figure the red line is a simulations with
$t_1=10$, $t_2=30$ and with $N_1=2400$, $N_2=800$ nodes.
    The black line is  $\rho_L^{nc}$ given by
Eqs. (\ref{pqD},\ref{rhoeq},\ref{imh},\ref{cublapl},\ref{lapf2}).
The simulations are done putting eigenvalues with bin of size $0.02$;
points for $\lambda > 0$ are joined by straight lines; points for small
$\lambda$ are made visible, to clarify the graph.
}
\label{Figlap1030er}
\end{center}
\end{figure*}

In Fig. \ref{Figlap1030er} we consider the same cases as in 
Fig. \ref{figadjer},
but with Laplacian instead of Adjacency random matrices; while in the
latter case $\rho_A^{nc}$ approximates fairly well $\rho_A$
for $t_1=5, t_2=6$ and well for $t_1=9, t_2=10$ and $t_1=10, t_2=30$,
in the Laplacian case there are marked differences between $\rho_L^{nc}$ 
and $\rho_L$.
There are no oscillations when
$t_2$ is close to $t_1$, i.e. for $t_1=5, t_2=6$ and $t_1=9, t_2=10$;
in the region beyond $\gamma^{nc}$ ($t_1=10, t_2=20.9$ is on $\gamma^{nc}$)
there are wide oscillations, with spacing around $1$ between the peaks,
see the right hand side figure with the case $t_1=10, t_2=30$.

\subsection{Discussion}
The Laplacian random block matrix model on random BB
graphs studied in this section can be interpreted as a mean field approximation 
of a random BB $d$-dimensional elastic network; the fact
that, in the limit $d \to \infty$ and $\frac{Z_a}{d}$ fixed,
the isostatic condition Eq. (\ref{iso1}) is satisfied in the limit
of zero frequency can be considered as a check of the validity of this
mean field approximation; numerical simulations for small $d > 1$ give
similar results. Another check is the fact that at the isostatic points
the spectral distribution diverges for $\lambda \to 0$, as expected
in a continuous constraint satisfaction problem at the SAT-UNSAT transition
\cite{fpuz}. For $d \ge 2$ there are at least two isostatic points;
if $d$ has many divisors, there are many isostatic points.

A random BB $d$-dimensional
elastic network can model a closely packed disordered system of particle
with a short range central repulsive force and with a bipartite contact network.
We do not know if there is a disordered solid
with these characteristics.
In such a system, the particle sizes should be chosen in such a way that
crystallization be avoided \cite{ohern}; in particular
in a bidimensional system, at the isostatic point with $Z_1=3$, $Z_2=6$
the particles could form a dice lattice; this lattice has been studied
in a nearest-neighbor tight-binding model \cite{wang}.

If such a jammed system can be realized, the corresponding random elastic
network should have an excluded volume constraint.
Numerical simulations in the Laplacian random block matrix model 
on random regular graphs
in $d=3$ indicate \cite{sphrep} that the peak in 
the density of states $\omega = 0$
is due to the region in which the angle between two contact versors
$\hat v_{i,j}$ and $\hat v_{i,k}$ is small.
Such a region exists because in the mean field approximation the
contact versors are independent versors randomly distributed on
the $d$-dimensional sphere.
In the interpretation of the random network as the contact network
of soft spheres near the jamming point, such a region is
excluded due to sphere repulsion.
Putting a cutoff on the small angles between contact versors, in
\cite{sphrep} no peak is found in the density of states around zero;
the density of states around zero is lower than without this cutoff, and
fits well with the molecular dynamics simulations in \cite{ohern}
and \cite{MZ}.

The peak in the case $d=3$ Laplacian random block matrix model
on random BB graphs with degrees $Z_1=4$ and $Z_2=12$ in
Fig. \ref{Fig3q2} is higher than in the case of simulations with
random block matrices on random regular graphs with $Z=6$ \cite{sphrep}.
We argue that this is due to the fact that in this case small angles between 
versors starting from a node of degree $Z_2$ are distributed more densely
than in the regular graph case,
being $Z_2 = 12$ twice the degree in the latter case.

In a simulation of a random BB elastic
network in finite dimension, near one of the isostatic points, 
we expect that the density of 
states has a plateau around $\omega=0$ which is lower than the density of 
states in the simulations with the random block matrices considered here.

\newpage

\section{Conclusion}
We have studied the moments and the spectral distribution of the Adjacency 
and Laplacian
block matrix in a sparse random block matrix model on bipartite biregular (BB)
random graphs or on bipartite Erd\"os-Renyi (ER) graphs 
with coordination degrees 
$Z_1, Z_2$. The blocks are projectors on independent random $d$-dimensional 
versors, uniformly distributed on the sphere in $d$ dimensions.
In $d=1$ these models are the corresponding
Adjacency and Laplacian random matrix models.

In the limit $d \to \infty$ with $t_a = \frac{Z_a}{d}$ fixed,
the computation of the moments of the spectral distribution reduces to
an enumeration problem of closed walks on rooted trees, in which the sequence 
of edges is noncrossing; as in \cite{PC},
exact solutions for these spectral distributions can be obtained
decomposing closed walks in primitive walks;
these solutions are the same on BB random graphs or on bipartite ER graphs.
In the case $t_1=t_2$, the spectral distribution  of the Adjacency 
and Laplacian random block matrix reduces to the effective medium (EM)
approximation \cite{SC} and the Marchenko-Pastur distributions respectively,
as found in the case of ER graphs in \cite{PC}.

In this limit, the spectral distribution
of the Laplacian random block matrix, representing the Hessian of
a BB random elastic network, is characterized by two lines
in the $t_1$-$t_2$ plane:

i) the isostatic line $t_1 t_2 = t_1 + t_2$, at which the
spectral density diverges as $\lambda^{-\frac{1}{2}}$ for $\lambda \to 0$; 

ii) the transition line, separating a region with a single spectral band
from one with two spectral bands.

For finite $d$, the isostatic condition reads
$(Z_1-d)(Z_2-d) = d^2$, so it is determined by the divisors of $d^2$.
Numerical simulations with Laplacian random block matrices on random BB
graphs, with $d > 1$ small, indicate that the spectral distribution diverges
at the isostatic points, as expected at the SAT-UNSAT transition
of a continuous constraint satisfaction problem \cite{fpuz}.

Numerical simulations in the case of BB graphs indicate that for $d > 1$ 
the transition 
line is approximately the same as the one in the $d\to \infty$ limit.

In $d=1$, in the case of BB graphs there are two bands for any $Z_1 \ne Z_2$;
in the case of ER graphs, numerical simulations suggest that
there is a transition from the one-band
to the two-bands region, with a quasi-gap between them.

We show that
the $l$-th moment of the Adjacency random block matrix model on ER graphs
is, in the limit $d\to \infty$ with $t = \frac{Z}{d}$ fixed, the
generating function (g.f.)
enumerating $NC^{(2)}(l)$ according to the number of blocks; 
the corresponding spectral distribution is the EM approximation
of the Adjacency random matrix model on ER graphs \cite{CZ,PC}.

The $l$-th moment of the Adjacency random matrix model on ER 
graphs is the g.f. enumerating a set of partitions, recently
investigated in \cite{bose}, which we call $P^{(2)}_2(l)$.
The Adjacency random block matrix model on ER graphs interpolates
between $P^{(2)}_2$, for $d=1$, and $NC^{(2)}$, for $d\to \infty$.
A similar phenomenon occurs for other sets of partitions; 
we define a flow from the g.f. enumerating a set of partitions, for $s=1$, 
to one enumerating the
corresponding set of noncrossing partitions, for $s \to \infty$.
In a class of sets of partitions, first studied
in \cite{beiss}, here called nc-closed sets of partitions,
we find expressions for the three non-leading terms in the 
$\frac{1}{s}$-expansion.
In particular the g.f. enumerating $P^{(2)}_2$ flows to the one enumerating
$NC^{(2)}$; the leading order in the
$\frac{1}{s}$-expansion is the EM approximation; we compute the next three
orders of this expansion.
We study similarly the flow from the g.f. enumerating $P^{(k)}_k$, 
a generalization of $P^{(2)}_2$,
to the g.f. enumerating $NC^{(k)}$, the set of $k$-divisible noncrossing partitions.

The Adjacency random matrix model on bipartite ER graphs leads to 
define congruence class (CC) types for $k$-divisible partitions.
We compute the g.f. enumerating $NC^{(k)}$
according to the CC types.
There is a noncrossing partition flow from the g.f. enumerating 
$P^{(k)}_k$ according to the CC types and the one enumerating $NC^{(k)}$
according to the CC types.
For $k=2$ the leading order in the $\frac{1}{s}$-expansion
is the $d\to \infty$ limit of the g.f. of the moments
of the Adjacency random block matrix model on bipartite ER graphs.
Numerical simulations indicate that, for large average degrees
$t_1$ and $t_2$,
the spectral distribution at the leading order in $\frac{1}{s}$
approximates well the spectral distribution of the Adjacency
random matrix model on bipartite ER graphs.

The $l$-th moment of a Wishart-type random matrix model with
the product of $k$  Gaussian random matrices \cite{lecz, leczsa}
is simply expressed in terms of the g.f. enumerating $NC^{(k)}(l)$ 
according to the rank and the CC type of its partitions.
We give a $1$-to-$1$ correspondence between partitions in $NC^{(k)}(l)$
and the pair partitions contributing to the $l$-th moments in this
Wishart-type random matrix model,
mapping partitions with given CC type and rank to pair partitions with given
pairing type.

Let us list a few open problems:

(i) To study the non-leading order approximation of the
spectral distribution of the Adjacency random matrix on ER graphs, 
knowing its moments at the same order.

(ii) Numerical simulations indicate that
the spectral distribution of the Laplacian random matrix on bipartite
ER graphs, close to the transition line or beyond it, has
wide oscillations with distance around $1$ between peaks; 
it would be interesting to understand their origin.

(iii) Find a disordered solid whose contact network can be modeled
by a random BB elastic network, with degrees $Z_1 \neq Z_2$ at an
isostatic point,
and hence by the Laplacian random block matrix on random BB graphs;
a possible candidate in $d=2$ would have $Z_1=3$ and $Z_2=6$, with
the dice lattice \cite{wang} as the corresponding crystal form.

\vspace{10 mm}
Acknowledgments.

I thank Gianni Cicuta for suggesting this problem, numerous discussions
and for computing the resolvent of the Laplacian random matrix model
on BB random graphs, reported in Appendix C.

\newpage

\section{Appendix A: Examples of partitions with CC types}
\subsection{Example of extended NCPT for partitions of size multiple of $2$}
Consider first $NC_2(l)$; the only irreducible
partition is $(1,2)$; the g.f. enumerating $NC_2(l)$ with given CC types is
\begin{equation}
a^{(2)}(x, t_1, t_2) = x^2 t_1
\nonumber
\end{equation}
Solving the linear system of equations
Eqs. (\ref{faf1}, \ref{nctrank}) one gets, for $s=1$, the quadratic equation
\begin{equation}
t_2 x^2 f_1^{(2)}(x, t_1, t_2)^2 + f_1^{(2)}(x, t_1, t_2)(t_1 x^2 - t_2 x^2 - 1) + 1 = 0 
\nonumber
\end{equation}
For $t_1=t_2=1$, $[x^{2i}]f_1^{(2)}(x) = C_i$ is the $i$-th Catalan number.

Here is $f_1^{(2)}$ till order $8$
\begin{equation}
f_1^{(2)}(x, t_1, t_2) = {\left(t_{1}^{4} + 6 \, t_{1}^{3} t_{2} + 6 \, t_{1}^{2} t_{2}^{2} + t_{1} t_{2}^{3}\right)} x^{8} + {\left(t_{1}^{3} + 3 \, t_{1}^{2} t_{2} + t_{1} t_{2}^{2}\right)} x^{6} + {\left(t_{1}^{2} + t_{1} t_{2}\right)} x^{4} + t_{1} x^{2} + 1
\nonumber
\end{equation}
Let us give the partitions to this order and the corresponding monomials in 
$t_1,t_2$
\begin{equation}
\begin{aligned}
    (1, 2) && t_{1} && \quad
(1, 2)(3, 4) && t_{1}^{2} \\
(1, 4)(2, 3) && t_{1} t_{2} && \quad
(1, 2)(3, 4)(5, 6) && t_{1}^{3} \\
(1, 2)(3, 6)(4, 5) && t_{1}^{2} t_{2} && \quad
(1, 4)(2, 3)(5, 6) && t_{1}^{2} t_{2}  \\
(1, 6)(2, 3)(4, 5) && t_{1} t_{2}^{2}&& \quad
(1, 6)(2, 5)(3, 4) && t_{1}^{2} t_{2}\\
(1, 2)(3, 4)(5, 6)(7, 8) && t_{1}^{4}&& \quad
(1, 2)(3, 4)(5, 8)(6, 7) && t_{1}^{3} t_{2}\\
(1, 2)(3, 6)(4, 5)(7, 8) && t_{1}^{3} t_{2}&& \quad
(1, 2)(3, 8)(4, 5)(6, 7) && t_{1}^{2} t_{2}^{2}\\
(1, 2)(3, 8)(4, 7)(5, 6) && t_{1}^{3} t_{2}&& \quad
(1, 4)(2, 3)(5, 6)(7, 8) && t_{1}^{3} t_{2}\\
(1, 4)(2, 3)(5, 8)(6, 7) && t_{1}^{2} t_{2}^{2}&& \quad
(1, 6)(2, 3)(4, 5)(7, 8) && t_{1}^{2} t_{2}^{2}\\
(1, 8)(2, 3)(4, 5)(6, 7) && t_{1} t_{2}^{3}&& \quad
(1, 8)(2, 3)(4, 7)(5, 6) && t_{1}^{2} t_{2}^{2}\\
(1, 6)(2, 5)(3, 4)(7, 8) && t_{1}^{3} t_{2}&& \quad
(1, 8)(2, 5)(3, 4)(6, 7) && t_{1}^{2} t_{2}^{2}\\
(1, 8)(2, 7)(3, 4)(5, 6) && t_{1}^{3} t_{2}&& \quad
(1, 8)(2, 7)(3, 6)(4, 5) && t_{1}^{2} t_{2}^{2}
\nonumber
\end{aligned}
\end{equation}
Let us now consider the set of partitions generated by
$(1,2)$ and $(1,2,5)(3,4,6)$. The second of these partitions has size multiple
of $2$, but it is not $2$-divisible.
The g.f. of the number of
irreducible partitions with given CC types is
\begin{equation}
a(x, t_1, t_2) = x^2 t_1 + x^6 t_1^2
\nonumber
\end{equation}
From the Eqs. (\ref{nctrank}) one gets 
\begin{eqnarray}
&&f_1(x, t_1, t_2) = 1 + x^2 t_1 f_1 f_2 + x^6 t_1^2 f_1^3 f_2^3 \nonumber \\
&&f_2(x, t_1, t_2) = 1 + x^2 t_2 f_1 f_2 + x^6 t_2^2 f_1^3 f_2^3
\nonumber
\end{eqnarray}
from which, eliminating $f_2$, one gets a 6th-order equation in $f_1$.
Let us give $f_1 - f_1^{(2)}$ till order $8$
\begin{equation}
f_1(x, t_1, t_2) - f_1^{(2)}(x, t_1, t_2) = {\left(4 \, t_{1}^{3} + 3 \, t_{1}^{2} t_{2} + t_{1} t_{2}^{2}\right)} x^{8} + t_{1}^{2} x^{6}
\nonumber
\end{equation}
corresponding to the partitions
\begin{equation}
\begin{aligned}
(1, 2, 5) (3, 4, 6) && t_1^2 && \quad
(1, 2, 5) (3, 4, 6) (7, 8) && t_1^3 \\
(1, 2, 5)(3, 4, 8)(6, 7) && t_1^2 t_2 && \quad
(1, 2, 7)(3, 4, 8)(5, 6) && t_1^3 \\
(1, 2, 7)(3, 4)(5, 6, 8) && t_1^3 && \quad
(1, 2)(3, 4, 7)(5, 6, 8) && t_1^3 \\
(1, 2, 7)(3, 6, 8)(4, 5) && t_1^2 t_2 && \quad
(1, 4, 7)(2, 3)(5, 6, 8) && t_1^2 t_2 \\
(1, 8)(2, 3, 6)(4, 5, 7) && t_1 t_2^2
\nonumber
\end{aligned}
\end{equation}

\subsection{$3$-divisible noncrossing partitions}
Let us compute the g.f. $f_1$ enumerating partitions
in $NC^{(3)}(l)$ for $l \le 3$ according to the CC types.

For $l=1$ there is only one $3$-divisible noncrossing partition,
$(1, 2, 3)$; one has
\begin{equation}
[x^3] f_1 = t_1
\nonumber
\end{equation}

For $l=2$ the $3$-divisible partitions and the corresponding monomials are
\begin{equation}
\begin{aligned}
(1, 2, 3, 4, 5, 6) && t_1 &&  \quad && 
(1, 2, 3)(4, 5, 6) && t_1^2 && \quad &&
(1, 2, 6)(3, 4, 5) && t_1 t_3 && \quad &&
(1, 5, 6)(2, 3, 4) && t_1 t_2 \\
\nonumber
\end{aligned}
\end{equation}
From these terms one gets
\begin{equation}
[x^6] f_1 = t_1^2 + t_1 t_2 + t_1 t_3 + t_1
\nonumber
\end{equation}

At order $l=3$ the $3$-divisible noncrossing partitions and the corresponding
monomials are
\begin{equation}
\begin{aligned}
(1, 2, 3, 4, 5, 6, 7, 8, 9) && t_{1} && \quad
(1, 2, 3, 4, 5, 6)(7, 8, 9) && t_{1}^{2} && \quad
(1, 2, 3, 4, 5, 9)(6, 7, 8) && t_{1} t_{3} \\
(1, 2, 3, 4, 8, 9)(5, 6, 7) && t_{1} t_{2} && \quad
(1, 2, 3, 7, 8, 9)(4, 5, 6) && t_{1}^{2} && \quad
(1, 2, 3)(4, 5, 6, 7, 8, 9) && t_{1}^{2} \\
(1, 2, 3)(4, 5, 6)(7, 8, 9) && t_{1}^{3} && \quad
(1, 2, 3)(4, 5, 9)(6, 7, 8) && t_{1}^{2} t_{3} && \quad
(1, 2, 3)(4, 8, 9)(5, 6, 7) && t_{1}^{2} t_{2} \\
(1, 2, 6, 7, 8, 9)(3, 4, 5) && t_{1} t_{3} && \quad
(1, 2, 6)(3, 4, 5)(7, 8, 9) && t_{1}^{2} t_{3} && \quad
(1, 2, 9)(3, 4, 5, 6, 7, 8) && t_{1} t_{3} \\
(1, 2, 9)(3, 4, 5)(6, 7, 8) && t_{1} t_{3}^{2} && \quad
(1, 2, 9)(3, 4, 8)(5, 6, 7) && t_{1} t_{2} t_{3} && \quad
(1, 2, 9)(3, 7, 8)(4, 5, 6) && t_{1}^{2} t_{3} \\
(1, 5, 6, 7, 8, 9)(2, 3, 4) && t_{1} t_{2} && \quad
(1, 5, 6)(2, 3, 4)(7, 8, 9) && t_{1}^{2} t_{2} && \quad
(1, 5, 9)(2, 3, 4)(6, 7, 8) && t_{1} t_{2} t_{3} \\
(1, 8, 9)(2, 3, 4, 5, 6, 7) && t_{1} t_{2} && \quad
(1, 8, 9)(2, 3, 4)(5, 6, 7) && t_{1} t_{2}^{2} && \quad
(1, 8, 9)(2, 3, 7)(4, 5, 6) && t_{1}^{2} t_{2} \\
(1, 8, 9)(2, 6, 7)(3, 4, 5) && t_{1} t_{2} t_{3} \\
\nonumber
\end{aligned}
\end{equation}

from which one obtains
\begin{equation}
[x^9] f_1 = t_{1}^{3} + 3 t_{1}^{2} t_{2} + t_{1} t_{2}^{2} + 3 t_{1}^{2} t_{3} + 3 t_{1} t_{2} t_{3} + t_{1} t_{3}^{2} + 3 t_{1}^{2} + 3 t_{1} t_{2} + 3 t_{1} t_{3} + t_{1}
\nonumber
\end{equation}

The above listed $[x^n]f_1$ terms are in agreement with Eq. (\ref{fad})
for $k=3$ and $t_0=1$.

\subsection{Example with $NC^{(2)}(l) \equiv NC^2_{4l}(W_1^l)$}
Let us illustrate the isomorphism between $NC^2_{4l}(W_1^l)$ with $NC^{(2)}(l)$
for $l=1,2,3,4$.

There is a single partition $(1,2)$ in $NC^{(2)}(1)$.
The first element of the block has CC $1$, so the monomial is $t_1$.
The list representation is $a = [1,2]$
so the intermediate string representation is 
$s' = \langle 1^* 1 2 2^* \rangle$. There is no substring $1 2 2^* 1^*$,
so the string representation is $s = s'$
and the pair partition is $\pi = (1,2)(3,4)$.
There is a $1$ on the right in the pairings, so $r_1(\pi)=1$; 
the number of $2$'s on the right in the pairings is $r_2(\pi)=0$;
from $r_0(\pi)=l=1, r_3(\pi)=0$ one gets $j_0(\pi) = r_0(\pi)-r_1(\pi)=0$, 
$j_1(\pi)=r_1(\pi)-r_2(\pi)=1$
and $j_2 = r_2(\pi)-r_3(\pi)=0$, so we get the monomial 
$t_0^{j_0} t_1^{j_1} t_2^{j_2} = t_1$
as before.

In the following tables there is the partition of $NC^{(2)}(l)$ in 
the first column;
in the second column the monomial $t_0^{j_0} t_1^{j_1} t_2^{j_2}$,
in the third column the string representation,
in the fourth column the partition in $NC^2_{4l}(W_1^l)$.

For $l=1$ we have just seen that
\begin{equation}
\begin{aligned}
(1,2) && t_1 && \langle 1^* 1 2 2^* \rangle && (1,2)(3,4)
\nonumber
\end{aligned}
\end{equation}
For $l=2$,
let us illustrate the algorithm in the case of $p = (1,4)(2,3)$.
The list representation is $a = [1,[2,1],2]$;
replacing square brackets with angular brackets,
$1$ with $1^* 1 2$ and $2$ with $2^*$ one gets
$s = \langle 1^* 1 2 \langle 2^* 1^* 1 2\rangle 2^*\rangle$, 
which determines the contractions; the pair partition indicates the positions
of the contractions, so $\pi = (1,2)(3,8)(4,7)(5,6)$.

\begin{equation}
\begin{aligned}
(1,2,3,4) && t_0 t_1 && \langle 1^* \langle 1 2 2^* 1^*\rangle 1 2 2^* \rangle && (1,6)(2,5)(3,4)(7,8) \\
(1,2)(3,4) && t_1^2  && \langle 1^* 1 2 2^* \rangle \langle 1^* 1 2 2^* \rangle&& (1,2)(3,4)(5,6)(7,8) \\
(1,4)(2,3) && t_1 t_2 && \langle 1^* 1 2 \langle 2^* 1^* 1 2 \rangle 2^* \rangle && (1,2)(3,8)(4,7)(5,6) \\
\nonumber
\end{aligned}
\end{equation}
giving $[x^4]f_1 = t_1^2 + t_1t_2 + t_0t_1$

For $l=3$
\begin{equation}
\begin{aligned}
(1,2,3,4,5,6) && t_0^2 t_1 && 
\langle 1^* \langle 1 2 2^* 1^* \rangle \langle 1 2 2^* 1^* \rangle 1 2 2^* \rangle && (1,10)(2,5)(3,4)(6,9)(7,8)(11,12) \\
(1,2,3,4)(5,6) &&t_0t_1^2 && 
\langle 1^* \langle 1 2 2^* 1^* \rangle 1 2 2^*\rangle \langle 1^* 1 2 2^* \rangle && (1,6)(2,5)(3,4)(7,8)(9,10)(11,12) \\
(1,2,3,6)(4,5) &&t_0t_1t_2 && 
\langle 1^* \langle 1 2 2^* 1^* \rangle 1 2 \langle 2^* 1^* 1 2  \rangle 2^* \rangle && (1,6)(2,5)(3,4)(7,12)(8,11)(9,10) \\
(1,2,5,6)(3,4) &&t_0t_1^2 && 
\langle 1^* \langle 1 2 2^* \langle 1^* 1 2 2^* \rangle 1^* \rangle 1 2 2^* \rangle  && (1,10)(2,9)(3,4)(5,6)(7,8)(11,12) \\
(1,4,5,6)(2,3) &&t_0t_1t_2&& 
\langle 1^* \langle 1 2 \langle 2^* 1^* 1 2 \rangle 2^* 1^* \rangle 1 2 2^* \rangle && (1,10)(2,9)(3,8)(4,7)(5,6)(11,12) \\
(1,6)(2,3,4,5) &&t_0t_1t_2&& 
\langle 1^* 1 2 \langle 2^* 1^* \langle 1 2 2^* 1^* \rangle 1 2 \rangle 2^* \rangle && (1,2)(3,12)(4,11)(5,10)(6,9)(7,8) \\
(1,2)(3,4,5,6) && t_0t_1^2 && 
\langle 1^* 1 2 2^*\rangle \langle 1^* \langle 1 2 2^* 1^* \rangle 1 2 2^* \rangle && (1,2)(3,4)(5,10)(6,9)(7,8)(11,12) \\
(1,2)(3,4)(5,6) &&t_1^3&& 
\langle 1^* 1 2 2^* \rangle \langle 1^* 1 2 2^* \rangle \langle 1^* 1 2 2^* \rangle  && (1,2)(3,4)(5,6)(7,8)(9,10)(11,12) \\
(1,2)(3,6)(4,5) &&t_1^2 t_2&& 
\langle 1^* 1 2 2^* \rangle \langle 1^* 1 2 \langle 2^* 1^* 1 2\rangle 2^* \rangle&& (1,2)(3,4)(5,6)(7,12)(8,11)(9,10) \\
(1,4)(2,3)(5,6) && t_1^2 t_2 && 
\langle 1^* 1 2 \langle 2^* 1^* 1 2  \rangle 2^*  \rangle\langle 1^* 1 2 2^* \rangle && (1,2)(3,8)(4,7)(5,6)(9,10)(11,12) \\
(1,6)(2,3)(4,5) &&t_1 t_2^2 && 
\langle 1^* 1 2 \langle 2^* 1^* 1 2 \rangle\langle 2^* 1^* 1 2 \rangle 2^* \rangle && (1,2)(3,12)(4,7)(5,6)(8,11)(9,10) \\
(1,6)(2,5)(3,4) &&t_1^2t_2 && 
\langle 1^* 1 2 \langle 2^* \langle 1^* 1 2 2^* \rangle 1^* 1 2 \rangle 2^* \rangle && (1,2)(3,12)(4,11)(5,6)(7,8)(9,10)
\nonumber
\end{aligned}
\end{equation}
giving
$[x^6]f_1 = t_1^3 + 3t_1^2t_2 + 3t_0t_1^2 + t_1t_2^2 + 3t_0t_1t_2 + t_0^2t_1$.

Let us illustrate the case $(1,2,5,6)(3,4)$; it has two blocks of $CC$ 1,
so that, adding $t_0$ to get a monomial of degree $l=3$, one gets $t_0 t_1^2$.
The list representation is $[1,2,[1,2],1,2]$.
To this list corresponds the intermediate string representation 
$\langle 1^* 1 2 2^* \langle 1^* 1 2 2^* \rangle 1^* 1 2 2^* \rangle$.
Insert the brackets for the $1 2 2^* 1^*$ term to get
$s = \langle 1^* \langle 1 2 2^* \langle 1^* 1 2 2^* \rangle 1^* \rangle 1 2 2^* \rangle$; now the contractions are determined,
so one obtains the corresponding pair partition.
The length-$4$ strings within brackets, excluding nested sub-brackets,
are one $S_0$ and two $S_1$, so that from Eq. (\ref{Ws}) one gets again 
the monomial $t_0 t_1^2$.

In the case $l=4$ we give a representative $p$ for each orbit under cyclic 
permutation of $NC^{(2)}(4)$;
in the first column there is $p$, in the second column the corresponding string
representation encoding the partition of $NC^2_{16}(W_1^4)$,
in the third column the polynomial associated to the partitions in the orbit.
As one can see from the previous examples, the pair partition $\pi$ is 
easily read from the string representation of $p$.

\begin{equation}
\begin{aligned}
&&(1,2,3,4,5,6,7,8) && 
\langle 1^* \langle 1 2 2^* 1^* \rangle \langle 1 2 2^* 1^* \rangle \langle 1 2 2^* 1^* \rangle  1 2 2^* \rangle && t_0^3 t_1\\
&&(1,2,3,4,5,6)(7,8) && 
\langle 1^* \langle 1 2 2^* 1^* \rangle \langle 1 2 2^* 1^* \rangle  1 2 2^* \rangle \langle 1^* 1 2 2^* \rangle && 4 t_0^2 t_1(t_1+t_2) \\
&&(1,2,3,4)(5,6,7,8) &&
\langle 1^* \langle 1 2 2^* 1^* \rangle 1 2 2^* \rangle \langle 1^* \langle 1 2 2^* 1^* \rangle 1 2 2^* \rangle
    && 2 t_0^2 t_1(t_1+t_2) \\
&&(1,2,3,4)(5,6)(7,8) &&
\langle 1^* \langle 1 2 2^* 1^* \rangle 1 2 2^* \rangle \langle 1^* 1 2 2^* \rangle \langle 1^* 1 2 2^* \rangle&& 4 t_0t_1(t_1^2 + t_2^2) \\
&&(1,2,3,4)(5,8)(6,7) && 
\langle 1^* \langle 1 2 2^* 1^* \rangle 1 2 2^* \rangle \langle 1^* 1 2 \langle 2^* 1^* 1 2 \rangle 2^* \rangle && 8t_0 t_1^2t_2 \\
&&(1,2,3,6)(4,5)(7,8) && 
\langle 1^* \langle 1 2 2^* 1^* \rangle 1 2 \langle 2^* 1^* 1 2 \rangle 2^* \rangle \langle 1^* 1 2 2^* \rangle && 8 t_0t_1^2 t_2\\
&&(1,2,5,6)(3,4)(7,8) && 
\langle 1^* \langle 1 2 2^* \langle 1^* 1 2 2^* \rangle 1^* \rangle 1 2 2^* \rangle \langle 1^* 1 2 2^* \rangle && 2t_0t_1(t_1^2+t_2^2) \\
&&(1,2)(3,4)(5,6)(7,8) && 
\langle 1^* 1 2 2^* \rangle \langle 1^* 1 2 2^* \rangle \langle 1^* 1 2 2^* \rangle \langle 1^* 1 2 2^* \rangle && t_1(t_1^3+t_2^3) \\
&&(1,2)(3,4)(5,8)(6,7) && 
\langle 1^* 1 2 2^* \rangle \langle 1^* 1 2 2^* \rangle \langle 1^* 1 2 \langle 2^* 1^* 1 2 \rangle 2^* \rangle && 4t_1^2t_2 (t_1+t_2)\\
&&(1,2)(3,8)(4,7)(5,6) &&
\langle 1^* 1 2 2^* \rangle \langle 1^* 1 2 \langle 2^* \langle 1^* 1 2 2^* \rangle 1^* 1 2 \rangle 2^* \rangle && 2t_1^2t_2(t_1+t_2)
\nonumber
\end{aligned}
\end{equation}
giving
$[x^8]f_1 = t_1^4 + 6 t_1^3t_2 + 6 t_0t_1^3 + 6 t_1^2t_2^2 + 16 t_0t_1^2t_2 + 6 t_0^2t_1^2 + t_1 t_2^3 + 6t_0t_1t_2^2 + 6t_0^2t_1t_2 + t_0^3t_1$.
The above listed $[x^n]f_1$ terms are in agreement with Eq. (\ref{fad})
for $k=2$.

\subsection{Example with ${\cal C}^w_{(n,2)}$ }
Let us consider list the bicolored partitions of ${\cal C}^w_{(n,2)}$ at the 
first $4$ orders. We will write them in sequential form.

For $n=1$ there is only $(1_2)$, with one element of CC $1$,  so 
that $[x]f_1 = t_1$.

For $n=2$ one has
\begin{equation}
\begin{aligned}
(1_1,1_1) &&  t_1 && \quad && (1_2,1_2)  && t_1 && \quad && 
(1_2,2_2) &&  t_1^2
\nonumber
\end{aligned}
\end{equation}
so that $[x^2]f_1 = t_{1}^{2} + 2 \, t_{1}$

For $n=3$ one has
\begin{equation}
\begin{aligned}
(1_1,1_1,1_2) &&  t_{1} && \quad && (1_1,1_2,1_1) &&  t_{1} && \quad &&
(1_2,1_1,1_1) &&  t_{1} && \quad &&  (1_2,1_2,1_2) &&  t_{1} \\
(1_1,1_1,2_2) &&  t_{1}^{2} && \quad &&
(1_2,1_2,2_2) &&  t_{1}^{2} && \quad &&
(1_1,2_2,1_1) &&  t_{1} t_{2} && \quad &&
(1_2,2_2,1_2) &&  t_{1}^{2} && \quad && \\
(1_2,2_1,2_1) &&  t_{1}^{2} && \quad &&
(1_2,2_2,2_2) &&  t_{1}^{2} && \quad &&
(1_2,2_2,3_2) &&  t_{1}^{3} && \quad &&
\nonumber
\end{aligned}
\end{equation}
so $[x^3] f_1 = t_{1}^{3} + 5 \, t_{1}^{2} + t_{1} t_{2} + 4 \, t_{1}$.

For $n=4$
\begin{equation}
\begin{aligned}
(1_1,1_1,1_1,1_1) &&  t_{1} && \quad &&
(1_1,1_1,1_2,1_2) &&  t_{1} && \quad &&
(1_1,1_2,1_1,1_2) &&  t_{1} && \quad &&
(1_1,1_2,1_2,1_1) &&  t_{1} \\
(1_2,1_1,1_1,1_2) &&  t_{1} && \quad &&
(1_2,1_1,1_2,1_1) &&  t_{1} && \quad &&
(1_2,1_2,1_1,1_1) &&  t_{1} && \quad &&
(1_2,1_2,1_2,1_2) &&  t_{1} \\
(1_1,1_1,1_2,2_2) &&  t_{1}^{2} && \quad &&
(1_1,1_2,1_1,2_2) &&  t_{1}^{2} && \quad &&
(1_2,1_1,1_1,2_2) &&  t_{1}^{2} && \quad &&
(1_2,1_2,1_2,2_2) &&  t_{1}^{2} \\
(1_1,1_1,2_2,1_2) &&  t_{1}^{2} && \quad &&
(1_1,1_2,2_2,1_1) &&  t_{1} t_{2} && \quad &&
(1_2,1_1,2_2,1_1) &&  t_{1} t_{2} && \quad &&
(1_2,1_2,2_2,1_2) &&  t_{1}^{2} \\
(1_1,1_1,2_1,2_1) &&  t_{1}^{2} && \quad &&
(1_1,1_1,2_2,2_2) &&  t_{1}^{2} && \quad &&
(1_2,1_2,2_1,2_1) &&  t_{1}^{2} && \quad &&
(1_2,1_2,2_2,2_2) &&  t_{1}^{2} \\
(1_1,1_1,2_2,3_2) &&  t_{1}^{3} && \quad &&
(1_2,1_2,2_2,3_2) &&  t_{1}^{3} && \quad &&
(1_1,2_2,1_1,1_2) &&  t_{1} t_{2} && \quad &&
(1_1,2_2,1_2,1_1) &&  t_{1} t_{2} \\
(1_2,2_2,1_1,1_1) &&  t_{1}^{2} && \quad &&
(1_2,2_2,1_2,1_2) &&  t_{1}^{2} && \quad &&
(1_1,2_2,1_1,3_2) &&  t_{1}^{2} t_{2} && \quad &&
(1_2,2_2,1_2,3_2) &&  t_{1}^{3} \\
(1_1,2_1,2_1,1_1) &&  t_{1} t_{2} && \quad &&
(1_1,2_2,2_2,1_1) &&  t_{1} t_{2} && \quad &&
(1_2,2_1,2_1,1_2) &&  t_{1}^{2} && \quad &&
(1_2,2_2,2_2,1_2) &&  t_{1}^{2} \\
(1_2,2_1,2_1,2_2) &&  t_{1}^{2} && \quad &&
(1_2,2_1,2_2,2_1) &&  t_{1}^{2} && \quad &&
(1_2,2_2,2_1,2_1) &&  t_{1}^{2} && \quad &&
(1_2,2_2,2_2,2_2) &&  t_{1}^{2} \\
(1_2,2_1,2_1,3_2) &&  t_{1}^{3} && \quad &&
(1_2,2_2,2_2,3_2) &&  t_{1}^{3} && \quad &&
(1_1,2_2,3_2,1_1) &&  t_{1} t_{2}^{2} && \quad &&
(1_2,2_2,3_2,1_2) &&  t_{1}^{3} \\
(1_2,2_1,3_2,2_1) &&  t_{1}^{2} t_{2} && \quad &&
(1_2,2_2,3_2,2_2) &&  t_{1}^{3} && \quad &&
(1_2,2_2,3_1,3_1) &&  t_{1}^{3} && \quad &&
(1_2,2_2,3_2,3_2) &&  t_{1}^{3} \\
(1_2,2_2,3_2,4_2) &&  t_{1}^{4} && \quad &&
(1_2,2_2,1_2,2_2) &&  t_{1}^{2}
\label{c42}
\end{aligned}
\end{equation}
where the last partition is the only one which is not noncrossing; one gets
$[x^4]f_1 = t_{1}^{4} + 9 \, t_{1}^{3} + 2 \, t_{1}^{2} t_{2} + t_{1} t_{2}^{2} + 19 \, t_{1}^{2} + 6 \, t_{1} t_{2} + 8 \, t_{1}$

At the first three orders all the bicolored partitions are noncrossing, so the
generating function $f_1$ enumerating $C^w_{(2)}$ and $f_1^{nc}$ enumerating
$C_{(2)}$ satisfy
$f_1 = f_1^{nc} + x^4 t_1^2 + O(x^5)$.

$f_1^{nc} + O(x^5)$ satisfies Eq. (\ref{lapf1a}) modulo $O(x^5)$.

\section{Appendix B: Moments of the random block matrix}
We give here the analytic evaluation of a few moments of the Adjacency
and Laplacian random block matrices on BB graphs.
The moments are displayed in terms of the variables 
$t_1=\frac{Z_1}{d}$, $t_2=\frac{Z_2}{d}$, $d$ and the parameters $c_i$,
coming from the integration over random unit vectors
\cite{CZ,PC}
\begin{eqnarray}
\frac{c_m}{d}&=&
\frac{ (2m-1)!!}{2^m} \frac{\Gamma \left(\frac{d}{2}\right)}{\Gamma \left(m+\frac{d}{2}\right)} \quad , \label{cdef} \nonumber \\
c_2 &=& \frac{3}{(d+2)}, \quad \cdots
\label{cm}
\end{eqnarray}

\subsection{Moments of the Adjacency random block matrix}
The g.f. of the moments of the Adjacency random block matrix is 
expressed by Eq. (\ref{fxadj}) in terms of $f_a$, with $a=1,2$,  
where $f_a$ is defined in Eq. (\ref{fa}); $f_2$ is obtained
from $f_1$ exchanging $t_1$ with $t_2$.

Define
\begin{equation}
f_{1,2i} = [x^{2i}]f_1
\end{equation}
while $[x^{2i+1}]f_1 = 0$.

The moments of the spectral distribution are given by
\begin{equation}
\mu_i = \frac{t_2 f_{1,i} + t_1 f_{2,i}}{t_1+t_2}
\end{equation}

We have computed the moments through $i=14$ on BB graphs; 
here we give them through $i=10$.
\begin{eqnarray}
f_{1,2} &=& t_{1} \nonumber\\
f_{1,4} &=& t_{1}^{2} + t_{1} t_{2} + t_{1} - \frac{2 \, t_{1}}{d} \nonumber\\
f_{1,6} &=& t_{1}^{3} + 3 \, t_{1}^{2} t_{2} + t_{1} t_{2}^{2} + 3 \, t_{1}^{2} + 3 \, t_{1} t_{2} + t_{1} - \frac{6 \, {\left(t_{1}^{2} + t_{1} t_{2} + t_{1}\right)}}{d} + \frac{7 \, t_{1}}{d^{2}} \nonumber\\
f_{1,8} &=& t_{1}^{4} + 6 \, t_{1}^{3} t_{2} + 6 \, t_{1}^{2} t_{2}^{2} + t_{1} t_{2}^{3} + c_{2} t_{1}^{2} + 6 \, t_{1}^{3} + c_{2} t_{1} t_{2} + 16 \, t_{1}^{2} t_{2} + 6 \, t_{1} t_{2}^{2} + 6 \, t_{1}^{2} + 6 \, t_{1} t_{2} + t_{1} -
\nonumber\\
&&    \frac{2 \, {\left(6 \, t_{1}^{3} + 16 \, t_{1}^{2} t_{2} + 6 \, t_{1} t_{2}^{2} + c_{2} t_{1} + 17 \, t_{1}^{2} + 17 \, t_{1} t_{2} + 6 \, t_{1}\right)}}{d} + \frac{37 \, t_{1}^{2} + 37 \, t_{1} t_{2} + 40 \, t_{1}}{d^{2}} - \frac{32 \, t_{1}}{d^{3}} \nonumber\\
f_{1,10} &=& 
t_{1}^{5} + 10 \, t_{1}^{4} t_{2} + 20 \, t_{1}^{3} t_{2}^{2} + 10 \, t_{1}^{2} t_{2}^{3} + t_{1} t_{2}^{4} + 5 \, c_{2} t_{1}^{3} + 10 \, t_{1}^{4} + 10 \, c_{2} t_{1}^{2} t_{2} + 50 \, t_{1}^{3} t_{2} + 5 \, c_{2} t_{1} t_{2}^{2} + 50 \, t_{1}^{2} t_{2}^{2} + 10 \, t_{1} t_{2}^{3} + 5 \, c_{2} t_{1}^{2} + 
\nonumber\\
&&20 \, t_{1}^{3} + 5 \, c_{2} t_{1} t_{2} + 50 \, t_{1}^{2} t_{2} + 20 \, t_{1} t_{2}^{2} + 10 \, t_{1}^{2} + 10 \, t_{1} t_{2} + t_{1} - 
\nonumber\\
&&\frac{5 \, {\left(4 \, t_{1}^{4} + 20 \, t_{1}^{3} t_{2} + 20 \, t_{1}^{2} t_{2}^{2} + 4 \, t_{1} t_{2}^{3} + 5 \, c_{2} t_{1}^{2} + 22 \, t_{1}^{3} + 5 \, c_{2} t_{1} t_{2} + 54 \, t_{1}^{2} t_{2} + 22 \, t_{1} t_{2}^{2} + 2 \, c_{2} t_{1} + 22 \, t_{1}^{2} + 22 \, t_{1} t_{2} + 4 \, t_{1}\right)}}{d} + 
\nonumber\\
&&\frac{5 \, {\left(23 \, t_{1}^{3} + 57 \, t_{1}^{2} t_{2} + 23 \, t_{1} t_{2}^{2} + 6 \, c_{2} t_{1} + 66 \, t_{1}^{2} + 66 \, t_{1} t_{2} + 26 \, t_{1}\right)}}{d^{2}} - \frac{5 \, {\left(49 \, t_{1}^{2} + 49 \, t_{1} t_{2} + 58 \, t_{1}\right)}}{d^{3}} + \frac{173 \, t_{1}}{d^{4}}
\label{momER}
\end{eqnarray}
For $d=1$ these moments agree with those computed using Appendix C.A; for 
$d\to \infty$, with $t_a=\frac{Z_a}{d}$ fixed,
$[x^{2i}]f_1$ is given by $[z^{i}]f_1$ in Eq. (\ref{fad}), for $k=2$
and $t_0=1$.

The moments in the ER case can be obtained neglecting the
terms containing explicitly $d$ (but keeping $c_i$), as explained in Section VI.

For $d=1$ the moments in the random matrix model on bipartite ER
graphs agree with those computed enumerating partitions of $P^{(2)}_2$, 
see subsection IV.C for $k=2$.

Let us make a comparison between the first $4$ moments computed here with
 Appendix A.C, setting $t_0=1$ there.
In the limit $d\to \infty$ with $t_a=\frac{Z_a}{d}$ fixed the moments
are computed enumerating partitions in $NC^{(2)}$ with given CC types.
For $l < 4$, $P^{(2)}_2(l) = NC^{(2)}(l)$, so the moments are the same
in this limit and in $d=1$, as can be seen in Eq. (\ref{momER})
and in $[x^{2l}]f_1$ in Appendix A.C.
$[x^8]f_1$ in Eq. (\ref{momER}) contains the terms
$c_2 t_1^2$ and $c_2 t_1 t_2$, corresponding in $P^{(2)}_2(4)$ respectively
to the crossing partitions $(1,2,5,6)(3,4,7,8)$ and $(1,4,5,8)(2,3,6,7)$;
in the limit $d\to \infty$, $c_2$ vanishes so one gets the
same as enumerating $NC^{(2)}(4)$, see Appendix A.C.

$P^{(2)}_2(5)$ consists of the partitions in $NC^{(2)}(5)$ and 
in $30$ crossing partitions: $10$ are irreducible partitions,
belonging to the orbit of $(1,2,3,4,7,8)(5,6,9,10)$, mentioned 
in Sections III and IV, which correspond to the contribution
$5 c_2 t_1^2 + 5 c_2 t_1 t_2$ in $f_{1,10}$;
there are $10$ reducible partitions, with two irreducible components,
in the orbit of
$(1,2,5,6)(3,4,7,8)(9,10)$, corresponding to the contribution
$5 c_2 t_1^3 + 5 c_2 t_1 t_2^2$ and $10$ reducible partitions in the orbit of
$(1,2,5,6)(3,4,7,10)(8,9)$, corresponding to $10 c_2 t_1^2 t_2$.
All the crossing partitions of $P^{(2)}_2(4)$ and $P^{(2)}_2(5)$
have $b(p) - c(p) = 1$, where $b(p)$ and $c(p)$
are respectively the number of blocks and the number of irreducible
components of the partition $p$, so they contribute to the $\frac{1}{s}$ 
term of the $\frac{1}{s}$-expansion, as can be checked in Eq. (\ref{fsk2bc}).

\subsection{Moments of the Laplacian random block matrix}
Define
\begin{equation}
f_{1,i} = [x^{i}]f_1(x)
\end{equation}
where $f_a$ is defined in Eq. (\ref{fl}); $f_2$ is obtained
from $f_1$ exchanging $t_1$ with $t_2$; the moments of the spectral
distribution are given by Eq. (\ref{fxlap}),
\begin{equation}
\nu_i = \frac{t_2 f_{1,i} + t_1 f_{2,i}}{t_1+t_2}
\end{equation}
and for $d\to \infty$
with $t_a = \frac{Z_a}{d}$ fixed they agree with the cubic equations
Eq. (\ref{lapf1a}).

We have computed the first 10 moments on BB graphs.
Here we list the first $5$ moments.

\begin{eqnarray}
f_{1,1} &=& t_1 \nonumber\\
f_{1,2} &=& t_{1}^{2} + 2 \, t_{1} - \frac{t_{1}}{d} \nonumber\\
f_{1,3} &=& t_{1}^{3} + 5 \, t_{1}^{2} + t_{1} t_{2} + 4 \, t_{1} - \frac{3 \, {\left(t_{1}^{2} + 2 \, t_{1}\right)}}{d} + \frac{2 \, t_{1}}{d^{2}} \nonumber\\
f_{1.4} &=& t_{1}^{4} + c_{2} t_{1}^{2} + 9 \, t_{1}^{3} + 2 \, t_{1}^{2} t_{2} + t_{1} t_{2}^{2} + 18 \, t_{1}^{2} + 6 \, t_{1} t_{2} + 8 \, t_{1} - \frac{6 \, t_{1}^{3} + c_{2} t_{1} + 29 \, t_{1}^{2} + 5 \, t_{1} t_{2} + 24 \, t_{1}}{d} + \frac{11 \, {\left(t_{1}^{2} + 2 \, t_{1}\right)}}{d^{2}} - \frac{6 \, t_{1}}{d^{3}} \nonumber\\
f_{1,5} &=& t_{1}^{5} + 5 \, c_{2} t_{1}^{3} + 14 \, t_{1}^{4} + 3 \, t_{1}^{3} t_{2} + 2 \, t_{1}^{2} t_{2}^{2} + t_{1} t_{2}^{3} + 9 \, c_{2} t_{1}^{2} + 50 \, t_{1}^{3} + c_{2} t_{1} t_{2} + 20 \, t_{1}^{2} t_{2} + 10 \, t_{1} t_{2}^{2} + 56 \, t_{1}^{2} + 24 \, t_{1} t_{2} + 16 \, t_{1} - \nonumber\\
&&\frac{10 \, t_{1}^{4} + 15 \, c_{2} t_{1}^{2} + 87 \, t_{1}^{3} + 15 \, t_{1}^{2} t_{2} + 8 \, t_{1} t_{2}^{2} + 10 \, c_{2} t_{1} + 170 \, t_{1}^{2} + 50 \, t_{1} t_{2} + 80 \, t_{1}}{d} + \nonumber\\
&&\frac{35 \, t_{1}^{3} + 10 \, c_{2} t_{1} + 167 \, t_{1}^{2} + 23 \, t_{1} t_{2} + 140 \, t_{1}}{d^{2}} -
\frac{50 \, {\left(t_{1}^{2} + 2 \, t_{1}\right)}}{d^{3}} + \frac{24 \, t_{1}}{d^{4}}
\label{lapdmom}
\end{eqnarray}

For $d=1$ these moments agree with those computed using Appendix C.B;
for $d \to \infty$ with $t_a=\frac{Z_a}{d}$ fixed, the moments agree
with those computed enumerating the partitions in ${\cal C}_{(n,2)}$,
see Eq. (\ref{lapf1a}).

The moments in the ER case can be obtained neglecting the
terms containing explicitly $d$ (but keeping $c_i$).

Let us make a comparison between the moments computed here with those computed 
in Appendix A.D by listing partitions in ${\cal C}^w_{(n,2)}$ till $n=4$.
In $d=1$ the moments are obtained enumerating ${\cal C}^w_{(n,2)}$,
so in Eq. (\ref{lapdmom}) one obtains the same results as in Appendix A.D.
In the limit $d \to \infty$ with $t_a=\frac{Z_a}{d}$ fixed,
the moments are computed enumerating partitions in ${\cal C}_{(n,2)}$,
the noncrossing subset of ${\cal C}^w_{(n,2)}$.
For $n=1,2,3$, ${\cal C}^w_{(n,2)} = {\cal C}_{(n,2)}$; 
in ${\cal C}^w_{(4,2)}$ there is one noncrossing partition, the
last term in Table (\ref{c42}), corresponding in $f_{1,4}$ in
Eq. (\ref{lapdmom}) to the $c_2 t_1^2$ term.

\section{Appendix C: Spectral distribution of the random matrix models on 
BB graphs.}
In the first subsection we
review the derivation of the g.f. of the moments
of the Adjacency random matrix on random BB graphs;
in the second subsection
we report the spectral distribution of the Laplacian random matrix model on
BB graphs.

\subsection{Adjacency random matrix model on random BB graphs}
The spectral distribution of the Adjacency random matrix model 
on random BB graphs has been obtained in \cite{gm}.

We give here a derivation of it, similar to the one in the case
of regular graphs \cite{wanless}.

Let $T_{a,s}(x,Z_1,Z_2)$, $a=1,2$ be the g.f. enumerating the
closed walks on an infinite, nearly BB rooted tree;
this tree is BB with degrees $Z_1, Z_2$, apart from
the root $R_a$ of degree $s = Z_a-1, Z_a$; 
the nodes which are neighbors of $R_a$ have degree $Z_{3-a}$;
$f_a(x,Z_1,Z_2) = T_{a,Z_a}(x,Z_1,Z_2)$ is the g.f. defined
in Eq. (\ref{fa}) in the case $d=1$.
Assume $Z_1 \ge Z_2 \ge 2$.

A closed walk decomposes in a length-$1$ walk $w_1$ moving from $R_a$ 
to a neighbor node $v$, a closed walk $w_2$ starting from $v$ and avoiding 
$R_a$, the length-$1$ walk $w_3$ from $v$ to $R_a$,
a closed walk $w_4$ starting with $R_a$.
There are $s$ choices for $w_1$; the first step in $w_2$ can go to one
of $Z_{3-a}-1$ neighbors, so one has
\begin{equation}
T_{a,s}(x, Z_1,Z_2) - 1 = x^2 s T_{3-a,Z_{3-a}-1}(x, Z_1,Z_2) T_{a,s}(x,Z_1,Z_2)
\label{Tsbip}
\end{equation}
Taking $s=Z_a-1$ in these two equations and solving the linear system
with the unknown $T_{a,Z_1-1}(x,Z_1,Z_2)$ and $T_{a,Z_2-1}(x,Z_1,Z_2)$ 
one gets a second-order equation for $T_{2,Z_2-1}(x,Z_1,Z_2)$ with solution
\begin{equation}
T_{2; Z_2-1}(x,Z_1,Z_2) =
\frac{1 - x^2(Z_2-Z_1) -
\sqrt{(1-x^2(Z_2-Z_1))^2 - 4x^2(Z_1-1)}}{2x^2(Z_1-1)}
\nonumber
\end{equation}
Using this equation in Eq. (\ref{Tsbip}) with $a=1$ and $s=Z_1$ one
gets an expression for $f_1(x,Z_1,Z_2) = T_{1; Z_1}(x,Z_1,Z_2)$ and for
$f_2(x,Z_1,Z_2) = f_1(x,Z_2,Z_1)$
\begin{eqnarray}
f_a(x,Z_1,Z_2)&=& \frac{1}{1-x^2Z_aT_{3-a,Z_{3-a}-1}(x,Z_1,Z_2)} = \nonumber \\
&&\frac{2 - Z_a +x^2Z_a(Z_a-Z_{3-a}) + Z_a\sqrt{(x^2B^2-1)(x^2A^2-1)}}{2(1-x^2Z_1Z_2)}
\label{Tsbip2}
\end{eqnarray}
where
\begin{equation}
A = \sqrt{Z_1-1} - \sqrt{Z_2-1}; \qquad B = \sqrt{Z_1-1} + \sqrt{Z_2-1}
\label{Tsbip2a}
\end{equation}
We checked till $n=14$ with Eq. (\ref{Tsbip2}) the computation of 
$[x^n]f_1(x,Z_1,Z_2)$ 
with the method of moments described in subsection VI.A.
The spectral density is \cite{gm}
\begin{eqnarray}
&&\rho_A(\lambda) = \frac{|Z_1-Z_2|}{Z_1+Z_2}\delta(\lambda) + \rho_{A,c}(\lambda)
\nonumber \\
&&\rho_{A,c}(\lambda) =
\frac{Z_1Z_2}{\pi |\lambda| (Z_1+Z_2)} \frac{\sqrt{(B^2-\lambda^2)(\lambda^2-A^2)}}{Z_1Z_2-\lambda^2}, \qquad A \le |\lambda| \le B
\label{rhod1adj}
\end{eqnarray}

\subsection{Laplacian random matrix model on random BB graphs}
The Laplacian spectral distribution can be obtained from the
Adjacency spectral distribution using the observation that
$A^2$ commutes with $L$.

For the Adjacency matrix the eigenvalue equation can be written as
\begin{equation}
X v_2 = \lambda v_1; \qquad X^T v_1 = \lambda v_2
\label{ev1}
\end{equation}
If $N_2 > N_1$, i.e $Z_2 < Z_1$,
there is an eigenvalue $\lambda=0$ with multiplicity
$N_2 - N_1$; the above equation reduces to $X v_2 = 0$.
The eigenvalue equation for $L$ is $L v = Z_2 v$, so that the spectral
distribution $\rho_L(\nu)$ has a factor
\begin{equation}
\rho_{L,0}(\nu) = \frac{|Z_2-Z_1|}{Z_2 + Z_1} \delta (\nu - Z_2)
\label{rhoL0}
\end{equation}
Let us now consider the case $\lambda \neq 0$.
$A^2$ has the eigenvalue $\lambda^2$, with eigenvectors
$v^T = (v_1^T, v_2^T)^T$ and $\tilde v^T = (v_1^T, -v_2^T)^T$.
Since $A^2$ commutes with $L$, the eigenvectors of $L$ must be a linear
combination of the eigenvectors of $A^2$,
\begin{equation}
(L - \nu) (\alpha v + \beta \tilde v) = 0
\label{ev2}
\end{equation}
so that
\begin{eqnarray}
& &\alpha (Z_1 - \lambda - \nu) + \beta (Z_1 + \lambda - \nu) = 0 \nonumber \\
& &\alpha (Z_2 - \lambda - \nu) + \beta (-Z_2 - \lambda + \nu) = 0
\nonumber
\end{eqnarray}
so that $\nu(\lambda)$ is a solution of the quadratic equation
\begin{equation}
\nu^2 - \nu (Z_1+Z_2) + Z_1 Z_2 - \lambda^2 = 0
\label{ev3a}
\end{equation}
One has
\begin{equation}
\rho_{L,c}(\nu) d \nu = \rho_{A,c}(\lambda) d \lambda
\nonumber
\end{equation}
where $\rho_{L,c}$ is the non-singular part of 
$\rho_L(\nu) = \rho_{L,0}(\nu) + \rho_{L,c}(\nu)$;
using Eq. (\ref{ev3a}),
\begin{eqnarray}
&&\rho_{L,c}(\nu) =
\rho_{A,c}(\lambda(\nu))|\frac{d \lambda(\nu))}{d \nu} |
\nonumber \\
&&\lambda(\nu) \equiv \sqrt{\nu^2 - \nu (Z_1+Z_2) + Z_1 Z_2}
\label{LA2}
\end{eqnarray}
From Eqs. (\ref{rhod1adj},\ref{LA2}) one gets
\begin{equation}
\rho_{L,c}(\nu) =
\frac{Z_1 Z_2 |Z_1 + Z_2 - 2 \nu|}{2\pi (Z_1 + Z_2)}
\frac{\sqrt{[-(p-1)^2 - \nu^2 + \nu (Z_1+Z_2)]
[\nu^2 - \nu (Z_1+Z_2) + (p+1)^2]}}{\nu (Z_1+Z_2-\nu)
[\nu^2 - \nu (Z_1+Z_2) + Z_1 Z_2]}
\label{rholapd1}
\end{equation}
where
\begin{eqnarray}
p = \sqrt{(Z_1-1)(Z_2-1)}
\label{ev5b}
\end{eqnarray}

$\rho_{L,c}$ has support in the intervals $(\nu_0,\nu_1)$
and $(\nu_2,\nu_3)$
\begin{eqnarray}
\nu_0 &=& \frac{Z_1+Z_2}{2} - \frac{1}{2}\sqrt{(Z_1+Z_2)^2 - 4(p-1)^2} \nonumber \\
\nu_1 &=& \frac{Z_1+Z_2}{2} - \frac{1}{2}\sqrt{(Z_1+Z_2)^2 - 4(p+1)^2} \nonumber \\
\nu_2 &=& \frac{Z_1+Z_2}{2} + \frac{1}{2}\sqrt{(Z_1+Z_2)^2 - 4(p+1)^2} \nonumber \\
\nu_3 &=& \frac{Z_1+Z_2}{2} + \frac{1}{2}\sqrt{(Z_1+Z_2)^2 - 4(p-1)^2}
\label{ev5a}
\end{eqnarray}

We checked numerically for few values of $Z_1, Z_2$ that
$\int \rho_L(\nu) d\nu = 1$ and that the first $10$ moments 
$\nu_i = \int \lambda^i \rho_L(\lambda) d\lambda$
agree, for $d=1$,  with those computed for any $d$ with the moments method 
(see Appendix B.B). The moments can be computed analytically
from the moments of the Adjacency random matrix expanding
$\nu_i = \frac{1}{N} \langle Tr L^i \rangle$ in terms of $Z_1$, $Z_2$ and
$\mu_j$, with $j \le i$.


\begin{thebibliography}{}
\bibitem{wigner} E. P. Wigner, \textsl{On the distribution of the roots of 
certain symmetric matrices}, Ann. of Math. {\bf 67}, 325  (1958).
\bibitem{dyson} F. J. Dyson, \textsl{The Dynamics of a Disordered Linear Chain},
    Phys. Rev. {\bf 92}, 1331 (1953).
\bibitem{Erd} P. Erd\"os and A. Renyi, \textsl{On the evolution of random graphs},
Magyar Tud. Akad. Kut. Int. Kozl. {\bf 5} (1960) 17.
\bibitem{bau} M. Bauer, O. Golinelli, \textit{Random incidence matrices:
moments of the spectral density}, J. Stat. Phys. 103, 301-337
(2001).
\bibitem{khor} A. Khorunzhy and V. Vangerovsky,
\textit{On Asymptotic Solvability of Random Graph's Laplacians}, arxiv:math-ph/0009028 (2000); O. Khorunzhy, M.
Shcherbina, and V. Vengerovsky, \textit{Eigenvalue distribution
of large weighted random graphs}, J. Math. Phys. 45, 1648 (2004).
\bibitem{SC} G. Semerjian and L. F. Cugliandolo, 
\textsl{Sparse random matrices: the eigenvalue spectrum revisited},
I. Phys. {\bf A 35}, 4837, (2002).
\bibitem{parisi} G. Parisi,
\textsl{Soft modes in jammed hard spheres (I): Mean
field theory of the isostatic transition}, arxiv:1401.4413
(2014).
\bibitem{maxwell} J.C. Maxwell, Philos. Mag. {\bf 27}, 294 (1864).
\bibitem{ohern} C.S. O'Hern, L.E. Silbert, A.J. Liu, S.R. Nagel,
\textsl{Jamming at zero temperature and zero applied stress: The epitome
of disorder}, Phys. Rev. {\bf E68}, 011306 (2003).
\bibitem{wnw} M. Wyart, S. R. Nagel and T.A. Witten, \textsl{Geometrical origin of
excess low-frequency vibrational modes in weakly-connected amorphous solids},
Europhys. Lett. {\bf 72}, 486 (2005).
\bibitem{fpuz} S. Franz, G. Parisi, P. Urbani and F. Zamponi,
    \textsl{Universal spectrum of normal modes in low-temperature glasses},
PNAS {\bf 112} 14439 (2015).
\bibitem{marcpas} V.A. Marchenko,  and  L.A. Pastur,
\textsl{Distribution of eigenvalues for some sets of random matrices},
        Mat. Sb. {\bf 72}, 507 (1967); Math. USSR Sbornik {\bf 1} 457 (1967).
\bibitem{benetti} F.P.C. Benetti, G. Parisi, F. Pietracaprina, G. Sicuro,
\textsl{Mean-field model for the density of states of jammed soft
spheres}, Phys. Rev. {\bf E 97}, 062157 (2018).
\bibitem{CZ} G. M. Cicuta, J. Krausser, R. Milkus, A. Zaccone,
\textsl{Unifying model for random matrix theory in arbitrary space dimension},
Phis. Rev. {\bf E 97}, 032113 (2018).
\bibitem{PC} M. Pernici and G. M. Cicuta,
\textsl{Proof of a conjecture on the infinite dimension limit of a unifying model for random matrix theory}, J. Stat. Phys. {\bf 175}, 384 (2019).
\bibitem{dkl} A. Dembczak-Kolodziejczyk, A. Lytova, 
\textsl{On the empirical spectral distribution for certain models
related to sample covariance matrices with different correlations},
arXiv:math/2103.03204 (2021).
\bibitem{bose} A. Bose, K. Saha, A. Sen and P. Sen, 
\textsl{Random matrices with independent entries: beyond non-crossing partitions}
arXiv:math/2103.09443 (2021).
\bibitem{speicher} R. Speicher, \textsl{Free probability theory and 
non-crossing partitions}, Sem. Lothar. Combin. {\bf 39}, B39c (1997).
\bibitem{voic91} D. Voiculescu, \textsl{Limit laws for random matrices
    and free products}, Invent. math. {\bf 104}, 201 (1991).
\bibitem{bani} T. Banica, S.T. Belinschi, M. Capitaine and B. Collins,
\textsl{Free Bessel laws}, Canad. J. Math. {\bf 63}, 3 (2011).
\bibitem{bani2} T. Banica, J. Bichon and B. Collins,
\textsl{The hyperoctahedral quantum group}, 
        J. Ramanujan Math. Soc. {\bf 22}, 345 (2007).
\bibitem{voi} D.V. Voiculescu, K.J. Dykema and A. Nica, 
\textsl{Free random variables}, AMS (1992).
\bibitem{alex} N. Alexeev, F. G\"otze and A.N. Tikhomirov,
\textsl{On the asymptotic distribution of singular values of products of large rectangular random matrices} J. Math. Sci. {\bf 408}, 9 (2012).
\bibitem{lecz} R. Lenczewski, \textsl{Limit distributions of random matrices},
    Adv. Math. {\bf 263}, 253 (2014).
\bibitem{leczsa} R. Lenczewski and R. Salapata,
    \textsl{Multivariate Fuss-Narayana polynomials and their application to random matrices}, Electron. J. Combin. {\bf 20}, 41 (2013).
\bibitem{wishart} J. Wishart, \textsl{The generalized product moment 
distribution in samples from a normal multivariate population},
Biometrika {\bf 20A}, 32 (1928).
\bibitem{leczsa2} R. Lenczewski and R. Salapata,
\textsl{Asymptotic distributions of Wishart type products of random matrices},
Colloquium Mathematicum, {\bf 155}, 67 (2019).
\bibitem{krew} G. Kreweras, \textsl{Sur les partitions non croisees d'un cycle},
    Discr. Math. {\bf 1}, 279 (1972).
\bibitem{proding} H. Prodinger, \textsl{A correspondence between ordered trees
and noncrossing partitions}, Discrete Math. {\bf 46}, 205 (1983).
\bibitem{edel} P. H. Edelman, \textsl{Chain enumeration and non-crossing partitions}, Discr. Math. {\bf 31}, 171 (1980).
\bibitem{speich94} R. Speicher,
    \textsl{Multiplicative functions on the lattice of 
non-crossing partitions and free convolution}, Math. Ann. {\bf 298} 611 (1994).
\bibitem{beiss} J.S. Beissinger, \textsl{The enumeration of irreducible 
combinatorial objects}, Journ. of Combinatorial Theory {\bf A38}, 143 (1985).
\bibitem{callan} D. Callan, \textsl{Sets, Lists and Noncrossing Partitions},
Journ. of. Integer Sequences, {\bf 11}, 08.1.3 (2008).
\bibitem{voi86} D.V. Voiculescu, \textsl{Addition of certain non-commuting
random variables}, J. of Functional Analysis, {\bf 66}, 323 (1986).
\bibitem{stein} P. R. Stein, \textsl{On a class of linked
    diagrams. I. Enumeration}, J. of Combinatorial Theory, Series A
    {\bf 24}, 357 (1978).
\bibitem{stein2} P. R. Stein and C.J. Everett, \textsl{On a class of linked
    diagrams. II. Asymptotics}, Discrete Math. {\bf 21}, 309 (1978).
\bibitem{sim} R. Simion, \textsl{Noncrossing partitions}, 
Discr. Math. {\bf 217}, 367 (2000).
\bibitem{arm} D. Armstrong, \textsl{Generalized noncrossing partitions and combinatorics of Coxeter groups}, arxiv:math/0611106.
\bibitem{klazar} M. Klazar, On $abab$-free and $abba$-free Set Partitions,
Europ. J. Combinatorics {\bf 17} (1996), 53-68.
\bibitem{flanoy} P. Flajolet, M. Noy, \textsl{Analytic combinatorics of
chord diagrams}, in \textsl{Formal power series and algebraic combinatorics}, Sp 191,  Springer (2000).
\bibitem{gm} C.D. Godsil and B. Mohar,
\textsl{Walk generating functions and spectral measures of infinite graphs},
Linear Algebra and its applications, {\bf 107}, 191 (1988).
\bibitem{mpmath} F. Johansson and others, \textsl{mpmath: a Python library for arbitrary-precision floating-point arithmetic} (version 0.18), (2013), http://mpmath.org/.
\bibitem{kesten} H. Kesten, \textsl{Symmetric random walks on groups},
    Trans. Amer. Math. Soc., 92:336 (1959).
\bibitem{MK} B.D. McKay, \textsl{The expected eigenvalue distribution
    of a large regular graph}, Linear Algebra Appl. {\bf 40}, 203 (1981).
\bibitem{sage}
\emph{SageMath, the Sage Mathematics Software System (Version
  7.3)}, The Sage Developers, 2016, {\tt https://www.sagemath.org}.
\bibitem{gil} J.B. Gil, P.R.W. McNamara, J.O. Tirrell and M.D. Weiner,
    \textsl{From Dyck paths to standard Young tableaux}, 
    Ann. of Combinatorics {\bf 24}, 69 (2020). 
\bibitem{klei} D. Kleitman, \textsl{Proportions of irreducible diagrams},
    Studies in Appl. Math. {\bf 49}, 297 (1970).
\bibitem{bolker} E.D. Bolker and B. Roth, \textsl{When is a bipartite graph
    a rigid framework?}, Pacific Journal of Mathematics, {\bf 90}, 1 (1980).
\bibitem{ballobas} B. Ballobas, \textsl{A probabilistic proof of an asymptotic 
formula for the number of labelled regular graphs}, 
European Journal of Combinatorics, {\bf 1}, 311 (1980).
\bibitem{wanless} I.M. Wanless, 
\textsl{Counting Matchings and Tree-Like Walks in Regular Graphs},
Combinatorics, Probability and Computing {\bf 19}, 463 (2010).
\bibitem{gessel} I.M. Gessel and T. Lengyel, \textsl{On the order of 
    Stirling numbers and alternating binomial coefficient sums}, 
 Fibonacci Quarterly {\bf 39}, 444 (2001).
\bibitem{sphrep} M. Pernici,
\textsl{Mean-field density of states of a small-world model and a jammed soft spheres model}, arxiv:2001.02622 (2020).
\bibitem{MZ} R. Milkus, A. Zaccone, \textsl{Local inversion-symmetry breaking
    controls the boson peak in glasses and crystals}, Phys. Rev. B{\bf 93},
094204 (2016).
\bibitem{wang} F. Wang and Y. Ran, \textsl{Nearly flat band with Chern number 
$C=2$ on the dice lattice}, Phys. Rev. {\bf B84}, 241103(R) (2011).
\end{thebibliography}
\end{document}